\documentclass[pdflatex,sn-nature]{sn-jnl}
\usepackage{astjnlabbrev-jh}
\usepackage{amsmath,amssymb}
\usepackage{tabularx}
\usepackage{textcomp}
\usepackage{xcolor}

\begin{document}

\title[Zieba --- LHS 3844\,b with JWST]{The dark and featureless surface of rocky exoplanet LHS 3844\,b from JWST mid-infrared spectroscopy}

\author*[1]{Sebastian Zieba}\email{sebastian.zieba@cfa.harvard.edu}

\author[2]{Laura Kreidberg}
\author[3]{Brandon P. Coy}
\author[4]{Aaron Bello-Arufe}
\author[5]{Kimberly Paragas}
\author[6]{Xintong Lyu}

\author[7,8,9,4]{Renyu Hu}

\author[10]{Aishwarya Iyer}
\author[3]{Edwin S. Kite}
\author[6]{Daniel D.B. Koll}
\author[11]{Kay Wohlfarth}

\author[12]{Emerson Whittaker}
\author[5]{Heather Knutson}
\author[13,14]{Robin Wordsworth}

\author[15]{Caroline Morley}
\author[16]{Laura Schaefer}

\affil[1]{Center for Astrophysics, Harvard \& Smithsonian, 60 Garden Street, Cambridge, MA 02138, USA}
\affil[2]{Max-Planck-Institut f\"ur Astronomie, K\"onigstuhl 17, D-69117 Heidelberg, Germany}
\affil[3]{Department of the Geophysical Sciences, University of Chicago, Chicago, IL 60637, USA}
\affil[4]{Jet Propulsion Laboratory, California Institute of Technology, Pasadena, CA 91109, USA}
\affil[5]{Division of Geological and Planetary Sciences, California Institute of Technology, 1200 E California Blvd, Pasadena, CA 91125, USA}

\affil[6]{Peking University, Beijing, People’s Republic of China}

\affil[7]{Department of Astronomy \& Astrophysics, The Pennsylvania State University, University Park, PA 16802, USA}
\affil[8]{Center for Exoplanets and Habitable Worlds, The Pennsylvania State University, University Park, PA 16802, USA}
\affil[9]{Institute for Computational and Data Science, The Pennsylvania State University, University Park, PA 16802, USA}

\affil[10]{NASA Goddard Space Flight Center, Greenbelt, MD, USA}
\affil[11]{Arbeitsgebiet Bildsignalverarbeitung, TU Dortmund University, Otto-Hahn-Str. 4, D-44227 Dortmund, Germany}

\affil[12]{Department of Earth, Planetary, and Space Sciences, University of California, Los Angeles, Los Angeles, CA 90095, USA}

\affil[13]{School of Engineering and Applied Sciences, Harvard University, Cambridge, MA 02138, USA}
\affil[14]{Department of Earth and Planetary Sciences, Harvard University, Cambridge, MA 02138, USA}

\affil[15]{Department of Astronomy, University of Texas at Austin, 2515 Speedway, Austin, TX 78722, USA}
\affil[16]{Department of Earth \& Planetary Sciences, Stanford University, Stanford, CA 94305, USA}

\maketitle

\textbf{JWST has opened a new era in the study of rocky exoplanets, enabling direct characterization of their surfaces with mid-infrared spectroscopy.
Different types of rock have distinct spectral features that are diagnostic of the chemical composition and other physical properties like surface texture. Measurements of these features can provide valuable clues about a planet’s geologic history and interior processes.
Here we report a JWST 5--12 \textmu m thermal emission spectrum for the rocky exoplanet LHS 3844\,b. It is best matched by a dark, low-silica surface, such as basalt or other olivine-rich materials. The spectrum rules out fresh powder surfaces; however, space weathering can darken the powders and make them more consistent with the data. The data also disfavor trace concentrations of CO$_2$ or SO$_2$ gas (with 5$\sigma$ and 3$\sigma$ upper limits of 100 mbar and 10 \textmu bar, respectively). Taken together, these results are well fit by an old, space-weathered surface with no evidence of accumulated volcanic gases.
}

\section{Main}
\subsection{Introduction}

Since the launch of JWST, thermal emission spectra have been measured for several rocky planets smaller than 1.5 Earth radii where the temperature at the substellar point is expected to be lower than the melting point of ultramafic rocky materials ($\sim$ 1500 K) (see Ref. \cite{ParkCoy2024, Kreidberg2025}).
The dayside temperatures for most of these planets are close to the theoretical maximum, indicating poor day-night heat redistribution \cite{Kreidberg2025}. While possible atmosphere scenarios are still being explored for individual planets, the overall ensemble of data suggests low surface pressures ($\lesssim10$ bar), with most planets consistent with no heat redistribution at all (and thus likely no atmosphere) \cite{ParkCoy2024}. For these cases, it is possible to observe the surface directly and characterize its composition \cite{hu2012}.

The hot ($T_{\rm{eq}} = 805$ K), small (1.3 $\rm{R}_\oplus$) exoplanet LHS 3844\,b \cite{Vanderspek2019} is one of the strongest candidates for an atmosphere-free rocky planet. Previous Spitzer data revealed a deep secondary eclipse and a phase curve with no measurable nightside flux or phase offset in the 4--5 \textmu m range \cite{Kreidberg2019}. The results were fit well with a bare rock surface model with a low albedo, potentially basaltic or ultramafic in composition. However, one broad wavelength band was not sufficient to constrain the full diversity of possible rocky surfaces \cite{whittaker22, lyu2024}. 

In this work, we observed LHS 3844\,b using JWST  (GO program \#1846; PI Laura Kreidberg, co-PI Renyu Hu) with the goal of measuring the planet's permanent dayside emission spectrum from 5--12 \textmu m to better determine the surface properties. Of all transiting rocky planets, LHS 3844\,b has the highest expected signal-to-noise ratio for thermal emission spectroscopy, making it an ideal target for surface characterization. We observed three eclipses of the planet by its M5 host star using JWST's MIRI/LRS. Each observation spanned 2.58 hours, using an identical setup (see ``JWST MIRI/LRS Observations'' in Methods).

\subsection{Results}\label{sec2}

\begin{figure}
    \centering
    \includegraphics[width=0.95\linewidth]{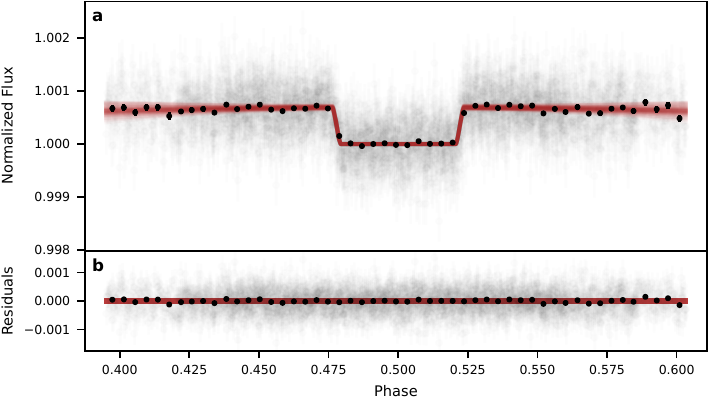}
    \caption{\textbf{Phase folded JWST MIRI/LRS white light curve of LHS 3844\,b.} \textbf{a,} The white light curve covers wavelengths from 5.060 to 12.368 \textmu m and combines the data from all three JWST observations. We measure a final eclipse depth of $696 \pm 18$ ppm, resulting in a 39$\sigma$ detection of the planet's eclipse. We show 100 random draws from the SZ reduction posterior of the astrophysical model, which consists of a planetary eclipse and a phase curve, in red, while the observed data are plotted in grey. The binned observations are also shown, with each bin containing approximately 100 integrations when all three observations contribute to the phase bin. \textbf{b,} Residuals from the best-fit model. The error bars correspond to 1$\sigma$ and are smaller than the marker for most bins.}
    \label{fig:white_lc_phasefolded}
\end{figure}

We performed two reductions of the data using the open-source pipeline \texttt{Eureka!} \cite{Bell2022}. The two reductions were carried out blindly with respect to each other and differed in the settings assumed during various stages of the pipeline. We first fit the planet's white light curve (5.060--12.368 \textmu m) alongside five available TESS sectors of the star to refine the planet's ephemeris and other planetary parameters. Our best-fit phase-folded white light curve is shown in Figure \ref{fig:white_lc_phasefolded}. We find a white light curve eclipse depth of $f_p/f_s = 696 \pm 18$ ppm. As a first step in assessing the surface properties, we converted the eclipse depth to an observed brightness temperature of $T_{\rm{obs}} = 985^{+16}_{-17}$ K, which is consistent (within 1.3$\sigma$) with the Spitzer value of $T = 1040 \pm 40$ K reported by Ref. \cite{Kreidberg2019}. We then compared our observed temperature to the maximum value expected for a bare rock planet without heat redistribution, $T_{\rm{max}} = T_{\rm{eff}}/\sqrt{a/r_s}\left(\frac{2}{3}\right)^{1/4}$, where $T_{\rm{eff}}$ is the stellar effective temperature and $a/r_s$ is the semi-major axis to star radius ratio. We obtained a brightness temperature ratio $\mathcal{R}=0.93^{+0.04}_{-0.05}$ (see ``Planetary albedo and brightness temperature ratio'' in Methods). 
Due to the limited phase coverage of the planet, we cannot constrain the phase curve amplitude. We fixed it to the eclipse depth, consistent with the findings from the Spitzer observations. In addition, by analyzing the transits observed by TESS and the eclipses observed by JWST, we placed upper limits on the eccentricity of $e < 0.017$ ($2\sigma$) and $e < 0.039$ ($3\sigma$) and find that the planet's orbit is consistent with circular. 

We then derived the planet's thermal emission spectrum by separating the white light curves from each observation into 12 spectroscopic bins from 5.060--12.368 \textmu m, each with the same full width of 0.609 \textmu m. We fit for one eclipse depth across all three observations and then calculated an average between the two independent data reductions, following the approach from previous works \cite{Greene2023, Zieba2023}. The two reductions agree well within 1$\sigma$, introducing negligible additional systematic error relative to the photon noise (see ``Final combined planetary spectrum'' in Methods).  
To calculate the absolute planet flux in each spectral bin, we used the flux-calibrated stellar spectrum from the MIRI/LRS in-eclipse data and from Spitzer. This avoids biases in the results from systematic errors in stellar models \cite{fauchez2025stellar}. See ``Data Reduction'' and ``Data Analysis'' in the Methods section for further details of the reductions and light curve fits.

\begin{figure}
    \centering
    \includegraphics[width=0.95\linewidth]{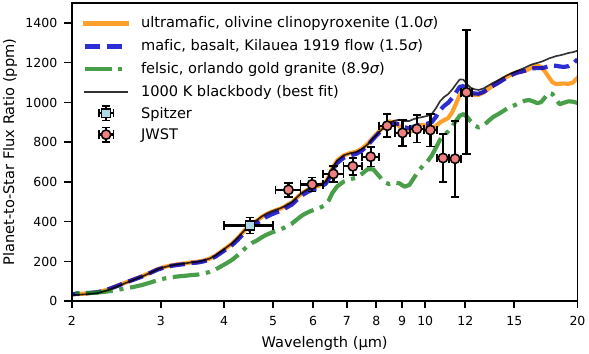}
    \caption{\textbf{The measured planet-to-star flux ratio as a function of wavelength compared to a range of solid slab surfaces.} The observations were taken using Spitzer (blue square-shaped marker) and JWST (red circle-shaped markers). The error bars on the data are 1$\sigma$. A selection of models, taken from Ref. \cite{Paragas2025}, represent solid slab surfaces with varying compositions. Solid orange line: an ultramafic, olivine clinopyroxenite EG-19-63 sample ($\sim$ 47 wt\% SiO$_2$) recovered from Emigrant Gap, California, USA, which shows good agreement with the observations. Dashed blue line: a mafic, basaltic sample from the 1919 Kilauea eruption in Hawaii, USA ($\sim51$ wt\% SiO$_2$) that is also consistent with the data. Dashed, dotted green line: a felsic sample representing the highest SiO$_2$ content model ($\sim73$ wt\% SiO$_2$) published in Ref. \cite{Paragas2025}. The legend lists the rejection significances of these models. With a solid black line, we also show a 1000 K black body curve for the planet, which is the best fit value using the combined Spitzer and JWST dataset.}
    \label{fig:surface}
\end{figure}

The joint JWST and Spitzer spectrum is well-fit by a blackbody with a temperature of $1000^{+15}_{-14}$ K, with no significant spectral features. 
From this, we obtain a brightness temperature ratio $\mathcal{R}=0.96^{+0.03}_{-0.04}$. This value is close to $R = 1$ and, therefore, indicative of a low Bond albedo, which we calculate at $A_B = 0.14^{+0.13}_{-0.14}$, assuming no heat redistribution. 
To interpret which surfaces could produce such a spectrum, we compared the data to several suites of rocky surface models, starting with a newly published spectral library by Ref. \cite{Paragas2025}, which includes a broader range of compositions than earlier datasets \cite{hu2012}, as well as three distinct surface textures (solid slabs, coarsely crushed materials, and fine powders).
Figure \ref{fig:surface} shows a selection of the slab surface models from Ref. \cite{Paragas2025}. We show the overall best-fit surface from the entire database, olivine clinopyroxenite (sample EG-19-63 from Ref. \cite{Paragas2025}), an ultramafic rock, which is consistent with the data at the $\sim1.0\sigma$ level. Mafic surfaces, such as a basalt sample from the 1919 Kilauea flow, also fit the data well ($\sim1.5\sigma$). For comparison, we also show the worst-fitting model, a light-colored granite (``the orlando gold'' granite sample in Ref. \cite{Paragas2025}), which is ruled out at the $\sim8.9\sigma$ level. 

\begin{figure}
    \centering
    \includegraphics[width=0.8\linewidth]{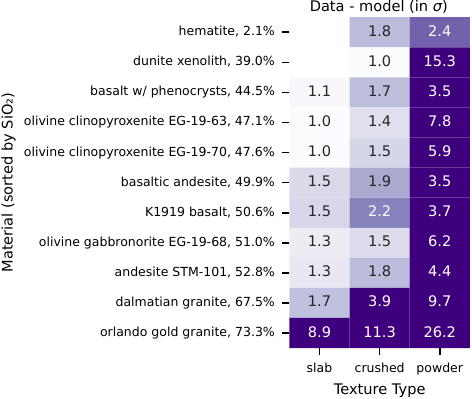}
    \caption{\textbf{Comparison between the measured emission spectrum and a wide range of surface compositions and textures}. The models in this figure were published in Ref. \cite{Paragas2025}. The materials are ordered from top to bottom by increasing SiO$_2$ content. Each column represents a different surface texture: solid slabs (left), coarsely crushed (middle), and powdered (right). The number in each grid cell indicates the difference between the measured spectrum (Spitzer and JWST) and the model prediction for a given material and texture, expressed in terms of $\sigma$. Therefore, a higher number corresponds to a greater disagreement between the model and observation. There were no slab samples available for hematite and dunite xenolith in Ref. \cite{Paragas2025}.}
    \label{fig:textures_vs_sio2}
\end{figure}

To systematically explore the suite of models from Ref. \cite{Paragas2025}, we sort them by surface texture and SiO$_2$ content. The goodness of fit across this parameter space is shown in Figure \ref{fig:textures_vs_sio2}. We find that the slab-textured surfaces fit the data best (generally consistent to within $2\sigma$). Crushed surfaces have a marginally worse fit due to their higher reflectivity, but are still broadly consistent with the data. Powders, on the other hand, have the highest reflectivity and yield cooler surfaces than we observe, due to increased multiple scattering with decreasing grain size \cite{Paragas2025}. All powders but hematite are ruled out at $>3\sigma$.  In addition to the powders, we also rule out the surface with the highest SiO$_2$ content, the orlando gold granite slab. This rock has a strong Si-O stretching feature between 8 and 10 \textmu m that is not seen in our data (see Fig. \ref{fig:surface}).
This result is consistent with expectations, as water, which is a key factor in the widespread formation of granite, is highly unlikely on a hot, bare-rock planet like LHS 3844\,b \cite{Campbell1983}.
In the Solar System, granite is abundant only on Earth. This highlights the planet's unique crustal composition resulting from geological processes such as heating and partial melting, driven by plate tectonics and the incorporation of water \cite{Taylor1989, hu2012, mansfield2019, hammond25, Paragas2025}. 
Although abundant granite is usually taken to be a proxy for water, other high-SiO$_2$ rocks can be formed without requiring liquid water. Such compositions can also arise from processes like remelting of thick basaltic crusts \cite{Lee2025, Phillips2025}, and this has been hypothesized to occur on Venus. Our spectrum indicates that such processes (to be tested by upcoming Venus exploration missions such as VEM on VERITAS and VenSpec-M on EnVision, see Ref. \cite{Widemann2023} and references therein) do not set the composition of LHS 3844\,b's dayside.

While all fresh powders, except the hematite, are ruled out, it is possible that darkening agents that have been observed in the solar system, such as space-weathering–induced nanometer-sized iron particles (npFe) or elemental carbon, could lower reflectance and increase emission and bring them into closer agreement with the data (see Section “Bare rock surfaces --- darkening agents and space weathering” in Methods). 
Iron particles of several to hundreds of nanometers in size produced by micrometeorite impacts and solar wind irradiation, is the principal darkening agent on the Moon and has been inferred on Mercury and other airless bodies \cite{cassidy1975,Pieters2000,hapke2001,pieters2016}. For a short-period planet such as LHS 3844\,b (orbital period $\sim 11$ h), weathering-induced darkening is expected to occur within $10^2$--$10^3$ years \cite{Zieba2023,lyu2024}. 

We assess the impact of npFe using the powder spectra of Ref. \cite{hu2012} and the weathering parameterization of Ref. \cite{lyu2024}, with volume fractions of 0.1--0.5\% consistent with lunar soils \cite{hapke2001,Noble2007}, and an extreme upper bound of 5\%, equivalent to the total iron content of Earth’s crust redistributed in form of npFe to the surface \cite{Fleischer1954}. In the spectral range of MIRI, it is a reasonable approximation to use only the smallest size fraction of iron, as it only affects the spectral brightness and not the slope. For carbon, we test fractions up to 5\%, matching the darkest regions of Mercury \cite{Klima2018}.

The overall effect of the weathering is to darken surfaces, causing higher brightness temperatures in the wavelengths covered by MIRI/LRS. We find that lunar-like weathering can sufficiently darken the surfaces to make them consistent with our measurements; for example, basalt with 0.5\% npFe agrees with the data at 1.9$\sigma$ compared to 2.5$\sigma$ without any weathering. However, we find the large Si-O features from granitoid surfaces are still ruled out at 5.1$\sigma$ even under the most extreme weathering scenario (5\% npFe). Thus, we conclude that granite surfaces can be excluded for LHS 3844\,b even if they have been severely space-weathered (see ``Bare rock surfaces --- darkening agents and space weathering'' in Methods). 

Finally, we consider the possibility of a tenuous atmosphere. The proximity of LHS 3844\,b to its host star leads to extreme stellar irradiation received by the planet, making a stable, thick atmosphere unlikely due to rapid atmospheric escape \cite{diamond21}. Based on Spitzer observations, Ref. \cite{Kreidberg2019} ruled out atmospheres with surface pressures greater than 10 bar, regardless of the atmosphere's composition. However, recent models also suggest that N$_2$/O$_2$-rich atmospheres could survive intense XUV flux  \cite{nakayama22,chatterjee2024}. In addition, transient outgassing events might still generate thin, detectable atmospheres. JWST’s MIRI/LRS is sensitive to key volcanic gases like SO$_2$, which is a promising tracer due to its strong spectral features in the near infrared wavelengths, particularly between 7--8 \textmu m \cite{Hu2013, whittaker22, LICHTENBERG2025}. 
SO$_2$ is, however, chemically short-lived in terrestrial exoplanet atmospheres. Therefore, requiring a substantial, and/or continuous surface outgassing rate to maintain a detectable level of SO$_2$ in the atmosphere \cite{Kaltenegger2010, Hu2013}.
Here, we perform grid forward modeling of potential atmospheric scenarios for LHS 3844\,b using the radiative-convective equilibrium code \texttt{HELIOS
3.0} \cite{malik2017helios,Malik2019,whittaker22}. We focus on two scenarios indicative of recent outgassing events; trace SO$_2$ with a CO$_2$-dominated background and trace SO$_2$ with an O$_2$-dominated background.

\begin{figure}
    \centering
    \includegraphics[width=0.999\linewidth]{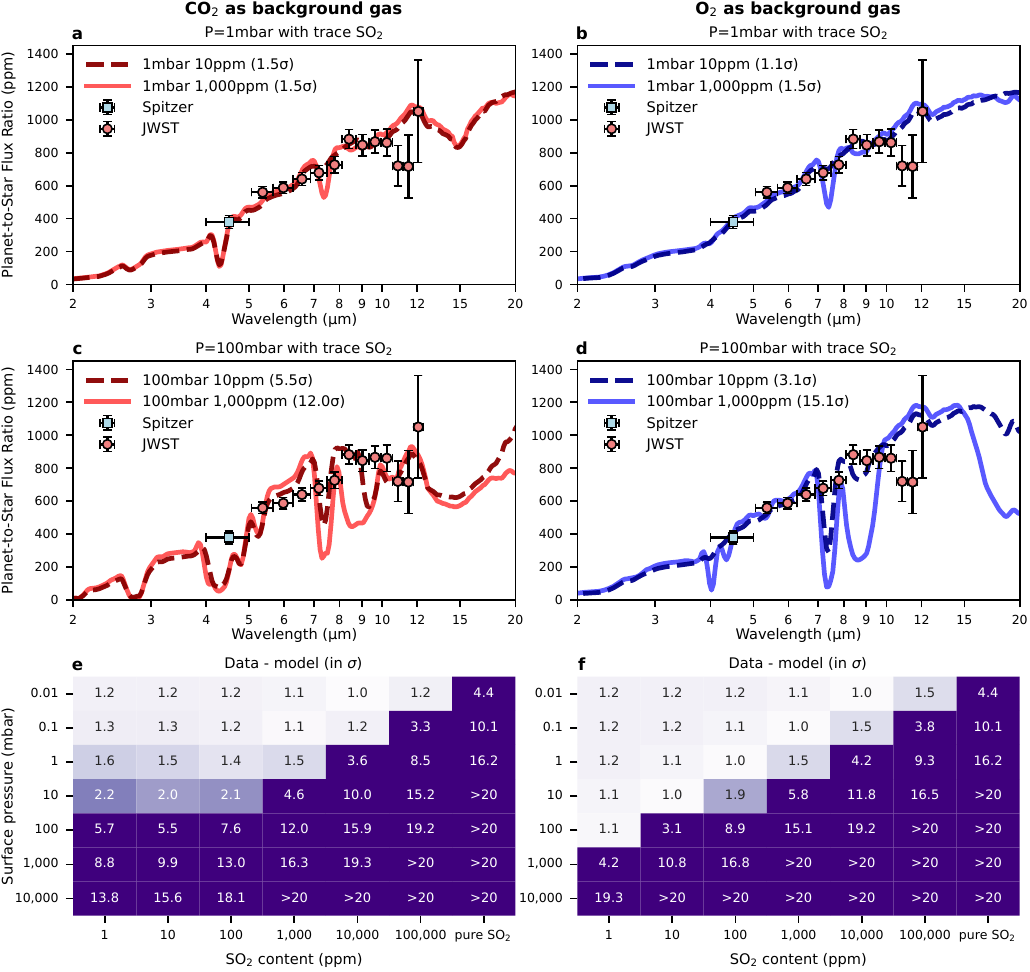}
    \caption{\textbf{A comparison between the observations and a range of atmospheric models for LHS 3844\,b}. The observations were taken using Spitzer (blue square-shaped marker) and JWST (red circle-shaped markers). We consider two different background gases: CO$_2$ (panels \textbf{a}, \textbf{c}, and \textbf{e}), which has strong features in the infrared, and O$_2$ (panels \textbf{b}, \textbf{d}, and \textbf{f}), which is generally featureless in the observed wavelength range. We show a selection of different surface pressures (1 mbar, panels a and b; 100 mbar, panels \textbf{c} and \textbf{d}) and SO$_2$ abundances (10 ppm, dashed lines; 1,000 ppm, solid lines). The bottom row (panels \textbf{e} and \textbf{f}) shows the goodness of fit for the full grid of surface pressures and SO$_2$ content we considered. As in Fig.~\ref{fig:textures_vs_sio2}, the number in each grid cell indicates the difference between the data and the model in units of $\sigma$. For CO$_2$-dominated atmospheres, we exclude surface pressures $\geq 100\,$mbar at a $5\sigma$ level. In both CO$_2$ and O$_2$-dominated atmospheres, SO$_2$ partial pressures $\geq 10\,$µbar are also ruled out ($3\sigma$).}
    \label{fig:atmosphere}
\end{figure}

As shown in Figure \ref{fig:atmosphere}, by combining Spitzer and JWST observations, we rule out atmospheres with surface pressures $\geq 1$ bar, regardless of composition (specifically, surface pressures of $\geq 100\,$mbar for CO$_2$-dominated atmospheres at 5$\sigma$ and $\geq 1\,$bar for O$_2$-dominated atmospheres at 4$\sigma$). Due to the pronounced SO$_2$ absorption, we further rule out SO$_2$ partial pressures of $\geq 10\,$µbar for both CO$_2$ and O$_2$ dominated atmospheres at $3\sigma$ ($\geq 100\,$µbar at $8\sigma$).
An important caveat for these atmosphere results is the assumption of radiative-convective equilibrium and no clouds. Changes in the temperature structure, for example, due to aerosol radiative feedback, could affect the amplitude of spectral features \cite[e.g.,][]{mansfield2019, Powell2024}. We leave a detailed exploration of these effects to future work. 

To put these atmospheric constraints in context, we compare them to known SO$_2$ concentrations in the solar system for Earth, Venus, and Io. 
For all these bodies, SO$_2$ points towards current or past volcanic activity.

With our data, we can rule out ($>12\sigma$) Venus-like SO$_2$ concentrations, i.e., a CO$_2$ dominated atmosphere with a surface pressure of 92 bar and SO$_2$ abundances of about $\sim180$ ppm \cite{Barker1979, Encrenaz2018, Oyama1979, Bezard1993, Taylor2014}. 
The Earth has low SO$_2$ abundances in the order of 10 ppt (0.00001 ppm) with 1 bar of surface pressure \cite{dePater2010, Seinfeld2006, Brimblecombe1989}. This low abundance is due to the Earth's oceans acting as a sink for SO$_2$, preventing the accumulation of this volcanic gas, as in the case of Venus \cite{Loftus2019}.
For an Io-like case, a locally variable SO$_2$ dominated atmosphere with surface pressures around a few nanobar \cite{Lellouch1990, Lellouch1992, Tsang2016, Lopes2023}, the surface pressures and abundances of SO$_2$ are too low to create significant features in the emission spectrum of LHS 3844\,b to be detectable by JWST  \cite{Pearl1979}.
For our observations, even if the surface pressure is as low as 0.1 mbar, we can rule out ($>3\sigma$) SO$_2$ contents of 10\% in the atmosphere regardless of the background gas. 
We can therefore disfavor accumulated volcanic gases for LHS 3844\,b. These observations will enable future studies to better constrain the amount of outgassing for the planet, given the observed lack of spectral planetary features.
We, however, emphasize that the spectral consistency of slab surfaces might also be consistent with ongoing volcanic resurfacing if any outgassed volatiles, such as SO$_2$ or CO$_2$, would quickly collapse into cold traps on the planet’s nightside \cite{Wordsworth2015}. 

\subsection{Discussion}\label{sec3}
 
Many rocks have spectral features in the 8--12 \textmu m wavelength range, including the Christiansen feature and the Restrahlen bands, which are sensitive to the chemical composition, crystal structure, and grain size. To explore which rock compositions fit this part of the spectrum best, we compare our data with models from the RELAB spectral database\footnote{\url{https://sites.brown.edu/relab/relab-spectral-database/}}.
We identify a qualitatively similar spectrum (i.e., an Si-O stretching feature between 8--12 \textmu m, leading to an apparent dip around 11 \textmu m) among multiple olivine-rich surfaces, including fayalite ($0.6\sigma$, Sample ID: DD-MDD-101) and an olivine-rich lunar basalt sample ($1.0\sigma$, Sample ID: LR-CMP-169) (see ``Bare rock surfaces --- RELAB comparison'' in Methods). 
The slight preference for olivine-clinopyroxenite over the Kilauea 1919 flow surface shown in Figure \ref{fig:surface}, based on models from Ref. \cite{Paragas2025}, is driven by the higher amounts of olivine in the olivine-clinopyroxenite model, leading to a better agreement between the olivine-rich surface and the data. If subsequent observations confirm the presence of an Si-O stretching feature between 8--12 \textmu m with the same shape as tentatively appears in our data, it would therefore likely indicate an olivine-dominated mineralogy on the planetary surface. This can be tested with the higher precision spectrum expected from JWST GO 7953 (PI: Kimberly Paragas; \cite{Paragas2025_JWST7953}), which will observe an additional 8 MIRI/LRS eclipses and a NIRSpec G395H eclipse.

Taken together, this suite of model comparison points towards two possible explanations for the very dark and featureless surface observed for LHS 3844\,b. One is a fresh slab, formed recently, without time for micrometeorite impacts to pulverize the rock. However, such a young surface would suggest ongoing geologic activity, and we find no evidence for CO$_2$ and SO$_2$, both expected products of volcanic outgassing. Alternatively, we also find good fits to the data for an old, powdery, and space-weathered surface, much like Earth's moon. Again, additional observations of LHS 3844\,b will help distinguish between these possibilities: a thermal phase curve was recently observed (JWST Program GO 4008, PI: Sebastian Zieba; \cite{Zieba2023_JWST4008}), designed to measure thermal beaming effects, similar to those seen for Solar System asteroids or other airless bodies. These data will distinguish between a young, smooth surface texture and an old, rough one. 

\clearpage
\newpage
\backmatter

\section*{Methods}

\subsection*{Observations}

This work uses data collected by the Transiting Exoplanet Survey Satellite \cite[TESS;][]{Ricker2015}, the James Webb Space Telescope \cite[JWST;][]{Gardner2006}, and the Spitzer Space Telescope \cite{Werner2004}. For the TESS and JWST observations, we perform a data reduction and analysis that is outlined below. For the Spitzer observations, we use the reported planet-to-star flux value reported in Ref. \cite{Kreidberg2019}, but we do not include the actual Spitzer time series observations in any of the white or spectroscopic fits.  

\subsubsection*{TESS Observations}

To improve constraints on planetary parameters such as the orbital period and potentially the orbital eccentricity, we used data from the Transiting Exoplanet Survey Satellite (TESS). LHS 3844 (TIC 410153553, T = 11.9 mag) was observed over 5 sectors in August 2018, July 2020, August 2020, July 2023, and August 2023. Table \ref{tab:observations_TESS} lists a summary of the observation dates as accessed from the TESS data release notes\footnote{\url{https://archive.stsci.edu/tess/tess_drn.html}}. The discovery of LHS 3844\,b was first reported by Ref. \cite{Vanderspek2019} using data from TESS Sector 1. Additional ground-based photometry from MEarth South, El Sauce Observatory, and the Cerro Tololo International Observatory (CTIO) at Las Cumbres Observatory was used in the discovery paper to confirm the planetary nature of LHS 3844\,b. However, in this study, we exclusively use the TESS and JWST observations. For Sector 1, TESS data were collected at a 2-minute cadence, while for all other sectors in this study, observations were taken at a 20-second cadence. The light curves were processed by the Science Processing Operations Center \cite[SPOC;][]{Jenkins2016} and made publicly available via the Mikulski Archive for Space Telescopes (MAST). We retrieved the data from all five sectors using the open-source Python package \texttt{lightkurve} \cite{Lightkurve2018}. Both simple aperture photometry (SAP) and systematics-corrected Presearch Data Conditioning SAP (PDCSAP) light curves \cite{Smith2012, Stumpe2012, Stumpe2014} are available. Following Ref. \cite{Vanderspek2019}, we use PDCSAP data for the remainder of this work.

\begin{table}[!h]
\centering
\caption{\textbf{Summary of the TESS observations of LHS 3844\,b.}}
\label{tab:observations_TESS}
\renewcommand{\arraystretch}{1.25}
\begin{tabular}{c|ccccc}
\hline
Sector        & 1                                                              & 27                                                             & 28                                                             & 67                                                             & 68                                                             \\ \hline
start (UTC)   & \begin{tabular}[c]{@{}c@{}}2018-07-25 \\ 19:00:27\end{tabular} & \begin{tabular}[c]{@{}c@{}}2020-07-05 \\ 18:31:16\end{tabular} & \begin{tabular}[c]{@{}c@{}}2020-07-31 \\ 08:15:15\end{tabular} & \begin{tabular}[c]{@{}c@{}}2023-07-01 \\ 03:14:33\end{tabular} & \begin{tabular}[c]{@{}c@{}}2023-07-29 \\ 02:30:32\end{tabular} \\
end (UTC)     & \begin{tabular}[c]{@{}c@{}}2018-08-22 \\ 16:14:51\end{tabular} & \begin{tabular}[c]{@{}c@{}}2020-07-30 \\ 03:21:15\end{tabular} & \begin{tabular}[c]{@{}c@{}}2020-08-25 \\ 14:17:14\end{tabular} & \begin{tabular}[c]{@{}c@{}}2023-07-28 \\ 21:26:32\end{tabular} & \begin{tabular}[c]{@{}c@{}}2023-08-25 \\ 15:16:31\end{tabular} \\
cadence (sec) & 120                                                            & 20                                                             & 20                                                             & 20                                                             & 20                                                             \\
N$_{\rm{transits}}$     & 57                                                             & 49                                                             & 46                                                             & 39                                                             & 44                                                             \\ \hline
\end{tabular}
\end{table}

\subsubsection*{JWST MIRI/LRS Observations}

The JWST observations used in this work were conducted as part of GO program \#1846 (PI: Laura Kreidberg, co-PI: Renyu Hu) titled ``A Search for Signatures of Volcanism and Geodynamics on the Hot Rocky Exoplanet LHS 3844\,b'' \cite{Kreidberg2021_go1846}. The program consisted of three eclipses of the planet LHS 3844\,b, with the MIRI Low Resolution Spectroscopy (LRS)
time-series observation mode on JWST \cite{Kendrew2015}. This instrument provides coverage of wavelengths from approximately 5 to 12 \textmu m with a resolving power of R$ \sim 100$. The observations employed the SLITLESSPRISM subarray and the FASTR1 readout pattern. Each of the three observations spanned 2.58 hours and followed an identical setup, consisting of 1887 integrations with 30 groups per integration. A group is a non-destructive read of the detector, and an integration is a set of such groups. Table \ref{tab:observations} lists a summary of each observation. The first observation (\#1) on June 18, 2023 and the third observation (\#3) on May 4, 2024, were successfully completed. However, the second observation (\#2) on August 18, 2023, failed due to a guide-star acquisition issue caused by a masked hot pixel at the guide-star location on the camera. A different guide star was used for the repeat observation on September 30, 2023 (\#102), which was successfully executed.
Notably, no high-gain antenna (HGA) movements occurred during any of the three observations.

\begin{table}[!h]
\centering
\caption{\textbf{Summary of the observations in JWST program GO 1846.} We do not list any information on the failed observation \#2, which occurred on 28. August 2023. The repeat observation is listed as \#102.}
\label{tab:observations}
\renewcommand{\arraystretch}{1.25}
\begin{tabular}{l|cccc}\hline
                 & observation \#1      & observation \#102       & observation \#3        \\\hline
date             & 19. Jun. 2023      & 30. Sep. 2023         & 4. May. 2024  \\
start time (UTC)       & 03:29:45           &  09:18:59             & 00:29:32             \\
end time (UTC)        & 06:04:49           &  11:54:02             & 03:04:35             \\
duration (hours) & 2.58               & 2.58                  & 2.58          \\
Nint             & 1887               & 1887                  & 1887           \\
Ngroups/int      & 30                 & 30                    & 30            \\\hline
\end{tabular}
\end{table}

\subsection*{Data Reduction}

For this work, we performed a data reduction of the TESS observations and two independent data reductions of the JWST data using \texttt{Eureka!} to determine whether varying reduction methods influence the final emission spectra of the planet. These reductions, referred to as Reduction SZ and Reduction ABA, are described below, with a summary in Table \ref{tab:data_reduction}.

\subsubsection*{TESS Data Reduction}

\begin{figure}[!h]
    \centering
    \includegraphics[width=1\linewidth]{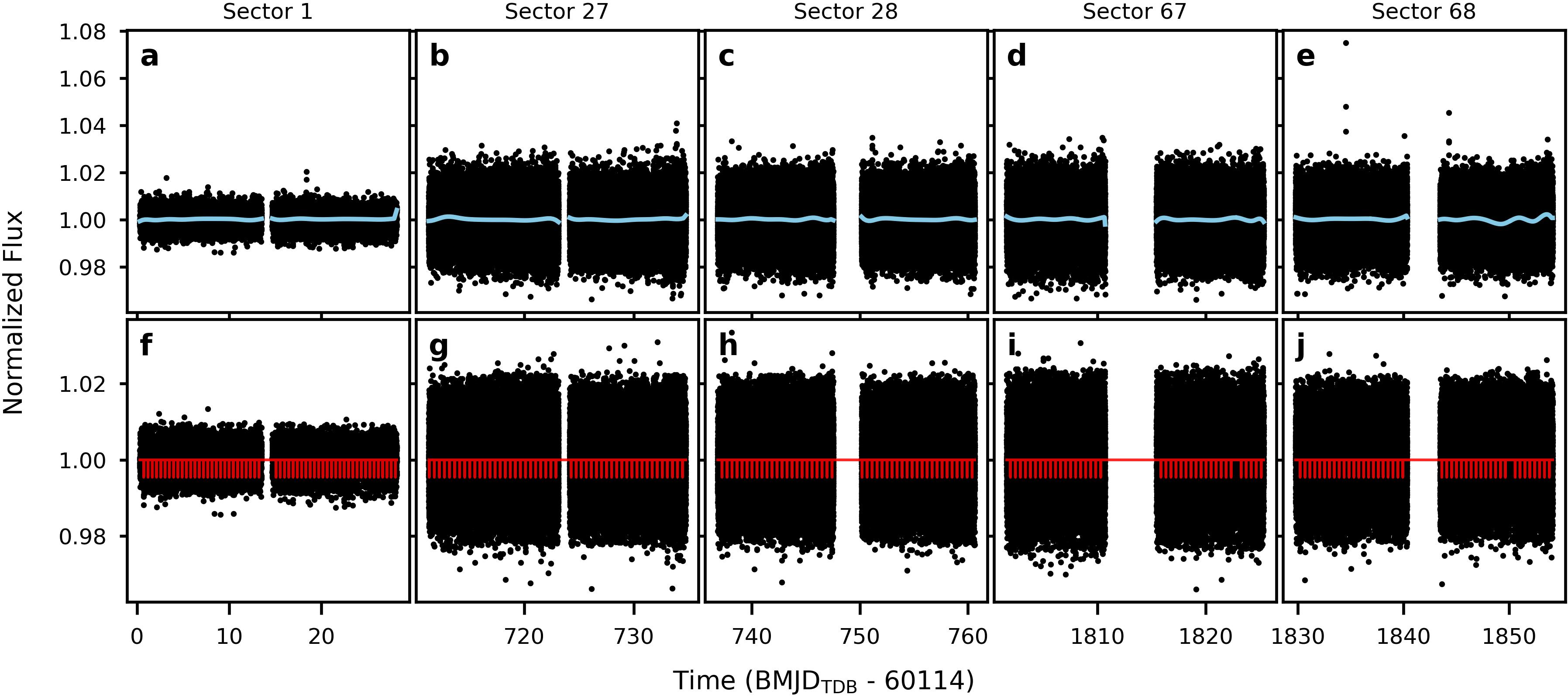}
    \caption{\textbf{TESS light curves of LHS 3844.} \textbf{a--e,} The raw PDCSAP TESS light curves with a spline fit (solid blue line) modeling any stellar variability or instrumental systematics. \textbf{f--j,} The final light curves after dividing the observations by the spline fit. The best-fitting \texttt{batman} transit model for each sector is shown in red. Some transits were not observed within the sector start and end times due to gaps in the data caused by the regular TESS data downlinks or scattered light contamination. The scatter in Sector 1 (panel \textbf{a} and \textbf{f}) is lower due to the longer 2-minute exposure time compared to the 20-second exposures in the other sectors.}
    \label{fig:tess_reduction}
\end{figure}

For the reduction of the TESS data, we followed a similar approach to that described in the LHS 3844\,b discovery paper by Ref. \cite{Vanderspek2019}. Since LHS 3844\,b was one of the first planets detected by the TESS mission, the discovery paper only used data collected during TESS Sector 1. However, in the meantime, LHS 3844 has been reobserved in four additional sectors. Here we apply the same reduction procedure to all five sectors. 
First, we remove all data points with nonzero quality flag values, which typically mark anomalous events. In our case, the majority of the removed data were affected by scattered light from Earth or the Moon. We then iteratively clip outliers greater than 3$\sigma$ relative to the median flux, while masking transits and eclipses of LHS 3844\,b based on the literature ephemeris. Approximately 0.3\% of the data were excluded in this step. 
Following Ref. \cite{Vanderburg2019}, we then fit a cubic spline with breakpoints every 1.5 days to the light curve, excluding transits and eclipses, to mitigate any remaining systematics or stellar variability. The spline is then interpolated for the points during transits and eclipses and divided out from the full light curve. We fit all the TESS data together using a constant baseline and a \texttt{batman} transit model to derive a significantly improved ephemeris (see Fig. \ref{fig:tess_reduction}).

\subsubsection*{JWST Data Reduction SZ}

\begin{figure}
    \centering
    \includegraphics[width=1\linewidth]{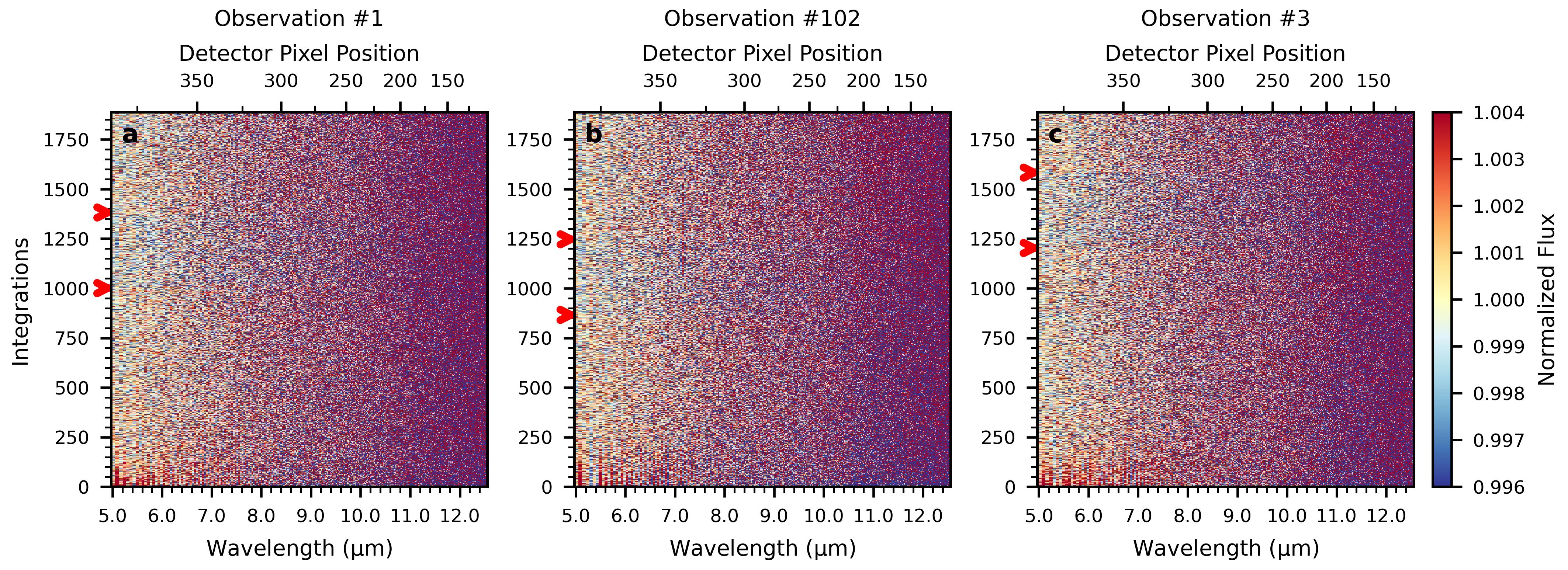}
    \caption{\textbf{2D light curves of the JWST observations.} The plots are based on the SZ \texttt{Eureka!} reduction for the three JWST observations of LHS 3844\,b after completing Stage 3 of the pipeline. Each row in the 2D light curves corresponds to an integration, and each column to a wavelength. The integration number increases upward to a maximum of 1887 for all observations. The wavelengths range from 5.060 to 12.368 \textmu m. The eclipse itself is marginally visible in this raw data, noticeable by the reduced flux observed at the times of eclipse. We marked the times of ingress and egress using red arrows for each observation.}
    \label{fig:2d_lc}
\end{figure}

This primary data reduction used the publicly accessible end-to-end pipeline \texttt{Eureka!} to process the \texttt{uncal.fits} files obtained from MAST into white and spectroscopic light curves. For that, we used \texttt{Eureka!} version \texttt{1.1.2.dev175+gf23df3a7}, \texttt{jwst} package version \texttt{1.15.1}, and CRDS version \texttt{11.18.4} with the CRDS context pmap 1303. The uncalibrated data were processed using software version \texttt{2024\_3a}, with each observation split into three segments. In Table \ref{tab:data_reduction}, we summarize the settings used for all stages of our reduction. In Stages 1 and 2 of \texttt{Eureka!}, which convert the uncalibrated data \texttt{uncal.fits} into count rate frames \texttt{rateints.fits} and then the calibrated data products \texttt{calint.fits}, we retained the default settings with the following exceptions: \texttt{skip\_firstframe} and \texttt{skip\_lastframe} were set to True, and the jump thresholds were set to 6$\sigma$, 5$\sigma$, and 7$\sigma$ for the three observations, respectively. These settings and thresholds were chosen to minimize the median absolute deviation (MAD) of the white light curves in Stage 4, after testing thresholds ranging from 4$\sigma$ to 20$\sigma$. We only found marginal differences when including a correction for the 390Hz noise. We did include the correction (i.e., \texttt{skip\_emicorr} = False) when it reduced the final MAD for the respective observation in Stage 4, which was the case in observation \#1 and \#102. To enhance uncertainty estimation in Stage 3, we skipped photometric calibration in Stage 2 (\texttt{skip\_photom = True}). For absolute stellar flux spectra, however, this parameter was set to \texttt{False} (see ``Absolute Stellar Flux Observation'' in Methods). In Stage 3, \texttt{Eureka!} performed optimal spectral extraction following the method described in Ref. \cite{Horne1986}. The 2D spectra were first rotated to align rows with the dispersion direction and columns with the spatial direction. Pixels flagged in the DQ array were masked, and for background subtraction, we fit and subtracted a constant value from each column. To minimize the MAD, we selected a half-width target aperture of \texttt{spec\_hw} = 4 and a half-width background aperture of \texttt{bw\_hw} = 12 relative to the source position. We used a 5$\sigma$ threshold for outlier rejection during background subtraction (\texttt{p3thresh}) and 4$\sigma$ threshold for time-axis outlier rejection, which was performed twice (\texttt{bg\_thresh}). Finally, in Stage 4 of \texttt{Eureka!}, we clipped data points deviating by more than 4$\sigma$ from a binned version of the data, using a box car filter with a width of 20 integrations. This procedure resulted in the removal of only a small number of data points compared to the total of 1887 integrations per visit, specifically, 1, 3, and 5 data points for the white light curves of the three observations, respectively. After completing Stage 4, we obtained the 2D light curves for each of the three observations (see Fig. \ref{fig:2d_lc}).

\begin{table}[]
\centering
\caption{\textbf{Summary of the data reductions.} Settings listed in round brackets specify the setup for the respective observations in the following order: (observation \#1, observation \#102, observation \#3). We refer to the text and the documentation of \texttt{Eureka!} for the description of these parameters.}
\label{tab:data_reduction}
\renewcommand{\arraystretch}{1.25}
\begin{tabular}{c|cc}
\hline
                         & SZ reduction         & ABA reduction                  \\ \hline
Stage 1                  &                      &                                \\
DP software              & 2024\_3a             & (2023\_1a, 2023\_4a, 2023\_2a) \\
jump rejection threshold & (6,5,7)              & (7,7,7)                        \\
skip\_emicorr            & (False, False, True) & True                           \\
skip\_firstframe         & True                 & False                          \\
skip\_lastframe          & True                 & False                          \\ \hline
Stage 2                  &                      &                                \\ 
skip\_photom             & True                 & True                           \\ \hline
Stage 3                  &                      &                                \\ 
ywindow                  & {[}109, 389{]}       & {[}109, 388{]}                 \\
xwindow                  & {[}13, 64{]}         & {[}11, 62{]}                   \\
bg\_thresh               & {[}4,4{]}            & {[}np.inf,np.inf{]}            \\
spec\_hw                 & 4                    & 3                              \\
bg\_hw                   & 12                   & 12                             \\
bg\_deg                  & 0                    & 0                              \\
bg\_method               & std                  & std                            \\
p3thresh                 & 5                    & 5                              \\
p5thresh                 & 14                   & 10                             \\
p7thresh                 & 14                   & 10                             \\
fittype                  & meddata              & meddata                        \\
median\_thresh           & 5                    & 7                              \\
window\_len              & 3                    & 5                              \\\hline
\end{tabular}
\end{table}

\subsubsection*{JWST Data Reduction ABA}
We performed an additional reduction of the data using version 0.10 of the \texttt{Eureka!} pipeline \cite{Bell2022}. We started from the uncalibrated raw data files (\texttt{uncal.fits}) and ran Stage 1 of \texttt{Eureka!} with the default parameters, except for \texttt{jump\_rejection\_threshold}, which we set to 7$\sigma$. We furthermore set \texttt{skip\_emicorr} to True and did not skip the \texttt{firstframe} and \texttt{lastframe} steps. The \texttt{firstframe} and \texttt{lastframe} steps flag the first and last groups in each integration as bad, excluding them from all subsequent steps. In Stage 2, we skipped the photometric calibration and ran all other steps as default. In Stage 3, we removed rows (here defined as those that align with the dispersion direction) and columns outside the 11--62 and 109--388, respectively, where the signal from the target is negligible. We masked those pixels with odd DQ values, and we recorded the spatial position of the trace on the detector by fitting a Gaussian function to each detector column. We assumed a gain of 3.1 electrons/data number \cite{Kempton2023} and performed column-by-column background subtraction using the mean value of pixels at least 12 rows away from the source, with a 5$\sigma$ threshold for outlier rejection. We performed optimal extraction on a region centered on the trace and with a half-width of 3 pixels. After generating the spectroscopic light curves, we masked values more than 4$\sigma$ away from the rolling median of a 20-integrations-wide box-car filter. Finally, we trimmed the initial 150 data points from all light curves, which were the integrations most affected by the detector's exponential ramp.

\subsection*{Data Analysis}

In the following, we analyze the TESS and JWST observations. For the JWST data, we analyze each reduction individually, producing both white light and spectroscopic light curves. For a better comparison of the emission spectra from the two reductions, we decided on a wavelength range from 5.060 to 12.368 \textmu m, divided into 12 spectroscopic light curves with each bin having a full width of 0.609 \textmu m. This binning scheme, previously utilized in Ref. \cite{Hu2024}, ensures that the shadowed region --- occurring at wavelengths greater than approximately 10.55 \textmu m --- aligns with the transition between two bins. By doing so, we avoid having any spectroscopic light curve that partially overlaps the shadowed region. This shadowed region has been observed in previous MIRI/LRS observations, such as Refs. \cite{Bell2023, Bell2024} but does not necessarily occur in all observations \cite{Welbanks2024}. This feature is characterized by a change of the initial exponential ramp behavior (like a change in ramp sign, amplitude, and timescale) and is probably caused by different illumination histories above and below the 10.55 \textmu m boundary.

\subsubsection*{Data Analysis SZ}

To estimate our free parameters and their associated uncertainties in this data analysis, we employ a dynamic nested sampling approach \cite{Skilling2004, Skilling2006, Higson2019} implemented in the open-source python package \texttt{dynesty} \cite{Speagle2020, Koposov2024}. For all fits, we use the \texttt{'multi'} bounds and set the sampling to \texttt{'auto'}, which automatically selects the sampling method depending on the number of free parameters in our fits. We initialize the dynamic nested sampler with \texttt{nlive\_init} = 1000 live points and a stopping criterion of \texttt{dlogz\_init}=0.01.\\

\begin{figure}
    \centering
    \includegraphics[width=0.8\linewidth]{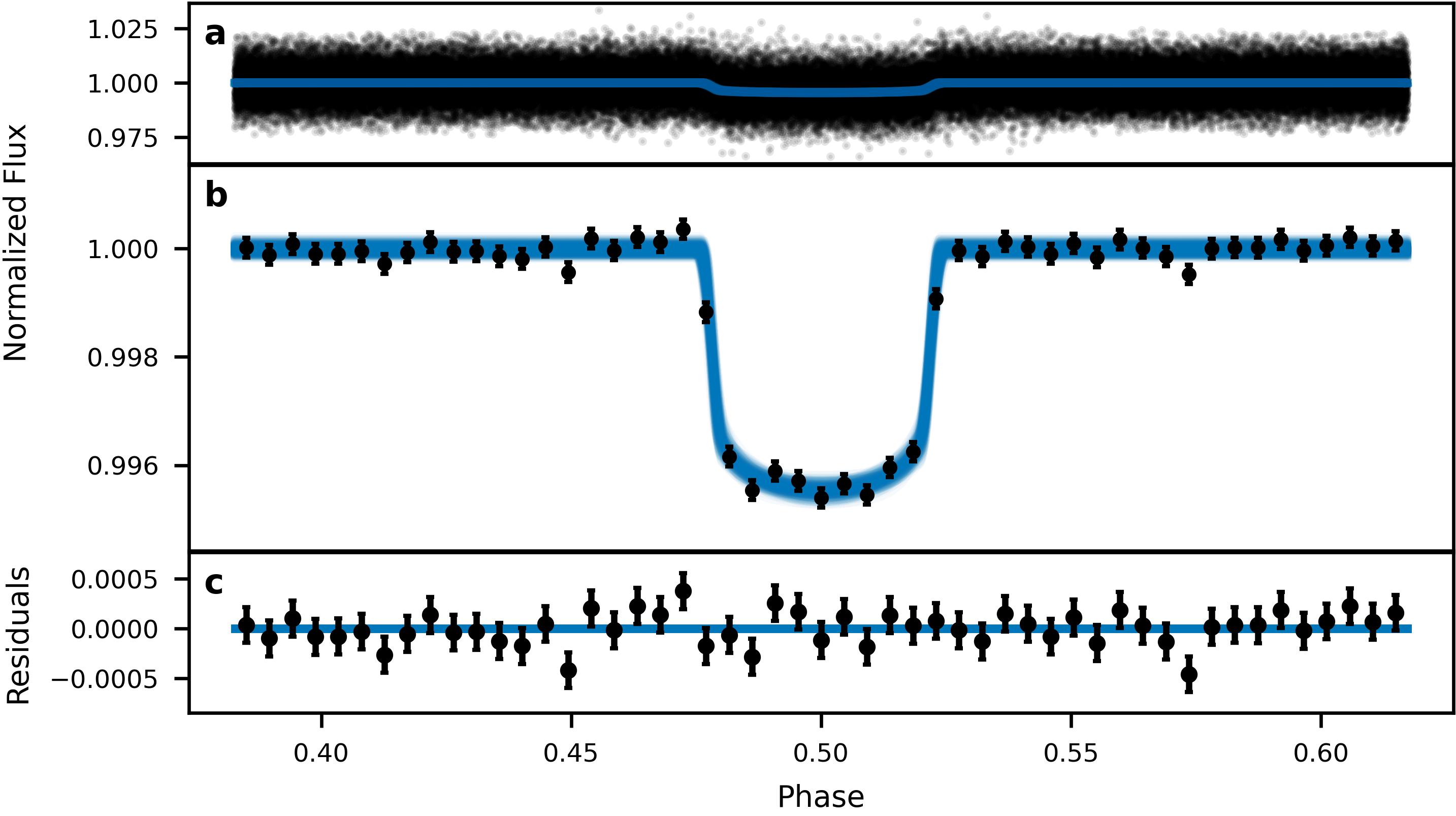}
    \caption{\textbf{Best-fit phase-folded TESS light curve.} \textbf{a,} Detrended phase-folded TESS light curve from all five TESS sectors. The blue solid lines show 100 random draws from the posterior. \textbf{b,} same, but observations have been binned with each bin containing approximately 1,700 integrations. \textbf{c,} The residuals of our best-fitting model. The error bars on the binned data in all panels are 1$\sigma$.}
    \label{fig:tess_phase_folded}
\end{figure}

\noindent\textbf{White light curve fit --- only TESS}\\
\noindent
We fit the five available TESS sectors and increase computational efficiency by exclusively keeping the transits with a transit duration $T_{14}$ and an out-of-transit baseline of $4\times T_{14}$ (two transit durations before and after each transit). Our free parameters are the transit midpoint time $t_0$, the orbital period $P_{\rm{orb}}$, the planet-to-star radius ratio $r_p/r_s$, the semi-major axis to star radius ratio $a/r_s$, and the orbital inclination $i$, and the quadratic limb-darkening parameters $q_1$ and $q_2$, following the reparametrization described in Ref. \cite{Kipping2013}. Additionally to the transit model which we employ using the \texttt{batman} code \cite{Kreidberg2015}, we fit a constant baseline, $c_{\rm{0, sec_i}}$, and a white noise multiplier $\sigma_{\rm{m, sec_i}}$ for each sector. 
Unlike the discovery paper \cite{Vanderspek2019}, we do not impose Gaussian priors on the limb-darkening parameters or the stellar density. Our results are consistent with the values published in Ref. \cite{Vanderspek2019}. A complete list of the best-fit parameters and their uncertainties can be found in Table \ref{tab:white_fits_TESS_and_JWST}, and a phase-folded transit light curve can be found in Figure \ref{fig:tess_phase_folded}.\\

\begin{table}[h]
    \centering
    \caption{\textbf{Best-fit parameter estimates for the TESS-only and the joint white light curve JWST and TESS fits.} We report the median of our posteriors and the 16th and 84th percentiles for the lower and upper 1$\sigma$ bounds. The values in this table are based on the SZ reduction. The adopted fit using the TESS and JWST data is shown in the rightmost column of the table.}
    \renewcommand{\arraystretch}{1.4}
    \footnotesize
    \begin{tabular}{l c c c }
        \hline
        & \begin{tabular}[c]{@{}c@{}}TESS only\\~\\~ \end{tabular}
        & \begin{tabular}[c]{@{}c@{}}JWST + TESS\\$e$ free\\~ \end{tabular}
        & \begin{tabular}[c]{@{}c@{}}\textbf{JWST + TESS}\\\textbf{$e$ = 0}\\\textbf{adopted fit}\end{tabular} \\
        \hline
        free parameters &&& \\
        $t_0$\textsuperscript{a} & $0.725513 \pm 0.000099$ & $0.72551 \pm 0.00010$ & $0.72554 \pm 0.00010$  \\
        $P_{\rm{orb}}$ (days)\textsuperscript{b}& $\left(6.90_{-0.39}^{+0.40}\right)\times10^{-7}$ & $\left(6.87_{-0.38}^{+0.39}\right)\times10^{-7}$ & $\left(6.54_{-0.29}^{+0.30}\right)\times10^{-7}$  \\
        $r_p/r_s$ & $0.06327_{-0.00083}^{+0.00089}$ & $0.06347_{-0.00091}^{+0.00095}$ & $0.06322_{-0.00081}^{+0.00094}$ \\
        $a/r_s$ & $6.98_{-0.48}^{+0.22}$ & $6.63_{-0.51}^{+0.43}$  & $6.75_{-0.51}^{+0.37}$ \\
        $\cos i$ & $0.038_{-0.027}^{+0.031}$ & $0.059_{-0.033}^{+0.027}$ & $0.051_{-0.034}^{+0.029}$ \\
        $q_1$ & $0.160_{-0.078}^{+0.150}$ & $0.198_{-0.090}^{+0.130}$ & $0.24_{-0.10}^{+0.13}$ \\
        $q_2$ & $0.36_{-0.24}^{+0.36}$ & $0.30_{-0.21}^{+0.32}$ & $0.25_{-0.17}^{+0.29}$ \\
        $\sqrt{e}\cos\omega$ & - & $-0.011_{-0.012}^{+0.008}$ & - \\
        $\sqrt{e}\sin\omega$ & - & $0.029_{-0.053}^{+0.063}$ & - \\
        $f_p/f_s$ (ppm) & - &$698 \pm 18$ & $696 \pm 18$ \\
        $AmpCos$ & 1 (fixed) & 1 (fixed) & 1 (fixed) \\
        \hline
        derived parameters &&& \\
        $i$ ($^\circ$) & $87.8_{-1.8}^{+1.5}$ & $86.6_{-1.6}^{+1.9}$ & $87.1_{-1.7}^{+2.0}$ \\
        $R_p$ ($\rm{R}_\oplus$)\textsuperscript{c} & $1.304_{-0.045}^{+0.046}$ & $1.308 \pm 0.046$ & $1.304 \pm 0.045$ \\
        $T_{14}$ (hours) & $0.5245_{-0.0053}^{+0.0065}$ & $0.5305_{-0.0053}^{+0.0071}$ & $0.5296_{-0.0048}^{+0.0066}$  \\
        $e$ & - & \begin{tabular}[c]{@{}c@{}}$0.0020_{-0.0014}^{+0.0067}$\\$<0.017\ (2\sigma)$\\ $<0.039\ (3\sigma)$ \end{tabular} & - \\
        \hline
    \end{tabular}
    \label{tab:white_fits_TESS_and_JWST}

    \vspace{3pt}
    \parbox{\linewidth}{\raggedright \footnotesize 
    \textsuperscript{a} BJD$\rm{_{TBD}}$ - 2458325.0 days.\\
    \textsuperscript{b} $P_{\rm{orb}}$ - 0.462929 days.\\
    \textsuperscript{c} For the calculation of the planetary radius, we use the literature value of $r_s = 0.189 \pm 0.006\ \rm{R}_\odot$ from Ref. \cite{Vanderspek2019}.}

\end{table}

\noindent\textbf{White light curve fit --- only JWST}\\
\noindent
For the white and spectroscopic light curve fits of the SZ reduction, we analyzed the observations using a custom fitting pipeline. We begin with the Stage 4 output files \texttt{Eureka!}, which provide the spectroscopic and white light curves. MIRI/LRS observations exhibit a pronounced initial ramp, typically modeled by one or two exponential terms \cite{Bouwman2023, Bell2024}. To mitigate this effect, we removed the first 45 minutes of each observation, eliminating the majority of the strongest portion of this ramp. This leaves us with 37, 26, and 54 minutes of baseline before the ingress of the eclipse for the three observations, respectively. The after-eclipse baselines are 41, 53, and 25 minutes, respectively. For comparison, the eclipse duration is approximately 31 minutes \cite{Vanderspek2019}. 

Our initial fits are used to determine whether incorporating additional decorrelation parameters, such as the position or width of the spectral trace on the detector, improves our fits. This initial model consists of a \texttt{batman} \cite{Kreidberg2015} eclipse with eclipse time $t_{\rm{sec}}$, eclipse depth $f_p/f_s$, and orbital period $P_{\rm{orb}}$, which we keep the same across the three observations. Furthermore, this model has a cosine term to model the phase curve with amplitude $AmpCos$ that is again shared across the three observations. For any remaining systematic or ramp, we use a quadratic trend in time, which we fit individually to each of the three observations. To account for any excess white noise and the uncertainties in the detector gain, we include a white noise multiplier $\sigma_{\rm{m}}$ as an additional free parameter for each of the three observations.
For our model selection, we use the Bayesian Information Criterion \cite[BIC;][]{Schwarz1978}, which is commonly used in exoplanet science to evaluate model performance by measuring how well a given model describes the data depending on the number of free parameters in the model and the amount of data points. A $\Delta$BIC $>$ 10 corresponds to ``very strong'' evidence in favor of the new, more complex model with more free parameters \cite{Kass1995, Raftery1995}.
We first compare the quadratic trend in time to an exponential one, both having the same number of free parameters, but the BIC strongly favored our original model over the exponential trend ($\Delta$BIC = 79). Notably, the quadratic trend is not intended to capture the phase variation, as this component is already modeled through the $AmpCos$ term in the planetary flux term (see Equation \ref{eq:Fp} and \ref{eq:Fpc}).
We find that adding $c_y$, which is the spatial position of the spectrum on the detector determined by fitting a Gaussian distribution to the spectrum in Stage 3 of \texttt{Eureka!}, improves our fits by $\Delta$BIC = 37.8 and 13.9 for observations 1 and 3, respectively. No significant improvement is found for observation 102. Additionally, we also do not find any improvement in using $c_{\sigma_y}$, which is the width of the spectrum along the spatial direction also determined by fitting a Gaussian distribution to the spectrum in Stage 3, in order to decorrelate against any change in the PSF of the spectrum with time for any of the three observations.

\begin{figure}
    \centering
    \includegraphics[width=1\linewidth]{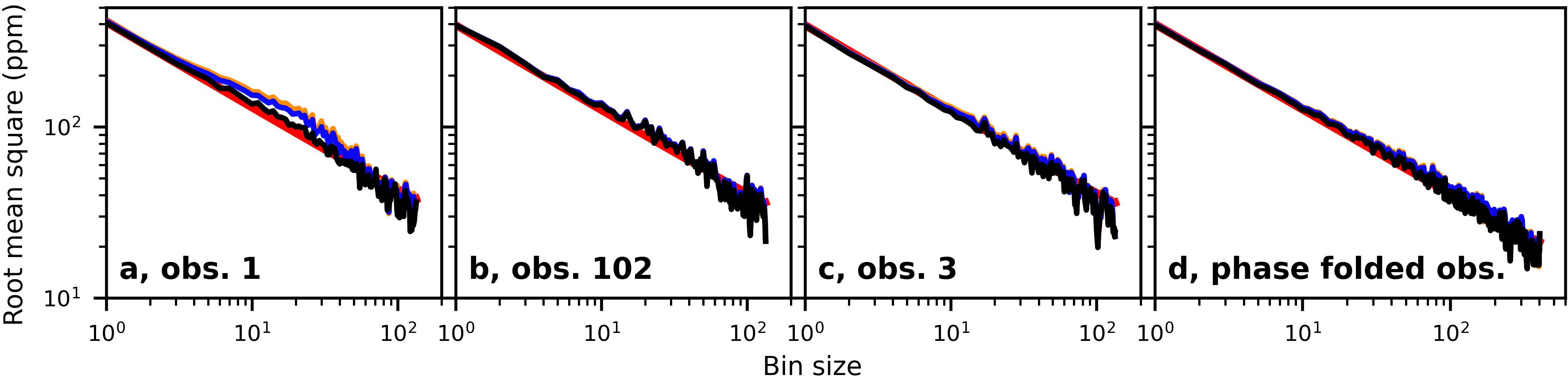}
    \caption{\textbf{Time-averaging plots of the white light curve fit residuals for the three observations (a-c) and the phase-folded data (d).} The plots show the root-mean-square (rms) of the residuals as a function of bin size, where a bin size of one corresponds to unbinned data. The solid red line represents the expected behavior for purely white noise, following the 1/$\sqrt{\rm{bin~size}}$ relationship between rms and bin size. Each subplot presents three different cases: Orange solid line: the behavior of the residuals for our initial fits. This fit does not include any information on the spatial position of the spectral trace or any JWST engineering mnemonics. Excess noise is evident, particularly in observation 1. Blue solid line: The residuals after incorporating decorrelation against the spectral trace position in observations 1 and 3, showing improvement. Black solid line: the final time-averaging plots after including both the spectral position and JWST engineering mnemonics, leading to further reduction of excess noise. Our final residuals show that we are generally consistent with uncorrelated photon shot noise in all three observations.}
    \label{fig:allan}
\end{figure}

We create time-averaging plots to explore correlated noise in the observations by calculating the root-mean-square (rms) of the residuals as a function of bin size \cite{Pont2006, Kipping2025}. By visually inspecting these time-averaging plots, we see excess noise in these initial fits (see Fig. \ref{fig:allan}). Following Refs. \cite{Kirk2024, Wallack2024}, we investigate whether using JWST engineering data can mitigate any access noise. These data are collected by onboard monitoring systems on JWST and can be accessed on the JWST Calibrated Engineering Data Portal\footnote{\url{https://mast.stsci.edu/portal/Mashup/Clients/jwstedb/jwstedb.html}}. We then retrieved all available engineering data (hereafter referred to as ``mnemonics'') that were collected during our three observations. We then filtered those time series, retaining only those that: (1) fully covered the observations, (2) contained floating-point values, and (3) were either associated with the spacecraft attitude control system (prefix: \texttt{SA\_}) or to the MIRI instrument (prefix: \texttt{IMIR\_}). This selection process resulted in approximately 120 mnemonics per observation. To align these mnemonics with our observations, we first applied a barycenter correction to the mnemonics time stamps and resampled them to the observations. We then ran 120 fits for each observation, testing the effect of each mnemonic on the noise properties of the residuals. To determine the most effective mnemonics, we computed the BIC and only kept those that yielded an improvement of at least $\Delta$BIC=10. For observation 1, including \texttt{SA\_ZATTEST4} (Estimated quaternion - 4th Component) led to a $\Delta$BIC of 115.8. For observation 102, using \texttt{SA\_ZPTQAERRX} (Pointing control attitude error calculated from the quaternion error - x axis) improved the BIC by 13.9.  Including \texttt{SA\_ZATTEST4} and \texttt{IMIR\_HK\_SEN05\_TEMP02} (MIR Housekeeping Packet Sensor Temperature) for observation 3 resulted in improvements of $\Delta$BIC of 35.4 and 24.7, respectively. The improvement of the fits is visually shown in the time-averaging plots in Figure \ref{fig:allan}.

For our final white light curve fits of the JWST data, which covers wavelengths from 5.060 to 12.368 \textmu m, we included those 4 mnemonics with their coefficients $c_{\rm{SA, obs_1}}$, $c_{\rm{SA, obs_{102}}}$, $c_{\rm{SA, obs_3}}$ and $c_{\rm{IMIR, obs_3}}$, for the respective observation. As described above, we also used the spatial position of the trace in our fits, with parameters $c_{\rm{y, obs_1}}$ and $c_{\rm{y, obs_3}}$, for observation 1 and 3, respectively. Other systematic parameters include the nine parameters for the second-order polynomial trend for each observation, $c_{\rm{0, obs_i}}$, $c_{\rm{1, obs_i}}$, and $c_{\rm{2, obs_i}}$ and the white noise multiplier $\sigma_{\rm{m, obs_i}}$ for each observation. All the astrophysical parameters are assumed to be identical across the three observations; in this fit, they are the eclipse time $t_{\rm{sec}}$, the eclipse depth $f_p/f_s$, the orbital period $P_{\rm{orb}}$, and the amplitude of the phase curve $AmpCos$. We model the phase curve using this single cosine term that peaks at the eclipse, without accounting for any potential phase curve offset. This choice is motivated by the limited phase coverage and the absence of a hot spot offset in previous studies \cite{Kreidberg2019}. For the JWST-only fit, we fix the planetary eccentricity $e$ to 0. The other orbital parameters are fixed to the literature values presented in Ref. \cite{Vanderspek2019}, that is: planet-to-star radius ratio $r_p/r_s = 0.0635$, the semi-major axis to star radius ratio $a/r_s = 7.109$, and the orbital inclination $i = 88.50^\circ$. 

In sum, the complete model for our observations $F_{\rm{model}}$ is based on the approach of Ref. \cite{Bell2019} and can be described as
\begin{equation}
F_{\rm{model}} (t)= F_{\rm{astro.}} (t) \times F_{\rm{sys.}} (t)
\end{equation}
\noindent
where $F_{\rm{astro.}}$ describes the astrophysical contribution to the flux, and $F_{\rm{sys.}}$ accounts for any contribution due to detector systematics. The astrophysical flux is given by
\begin{equation}
F_{\rm{astro.}} (t) = F_{\rm{s}} (t) + F_{\rm{p}} (t)
\label{eq:astro_notransits}
\end{equation}
\noindent
where $F_{\rm{s}}$ represents the stellar flux, which we assume to be constant during our observations. When simultaneously fitting the JWST and TESS data, the flux is periodically reduced due to the planetary transits. In this case Equation \ref{eq:astro_notransits} turns into:
\begin{equation}
F_{\rm{astro.}} (t) = \left( F_{\rm{s}} (t) \times F_{\rm{transit}} (t)\right) + F_{\rm{p}} (t),
\label{eq:astro_transits}
\end{equation}
\noindent
where $F_{\rm{transit}} = 1$ outside of transits and utilize the \texttt{batman} transit model otherwise. The planetary flux, $F_{\rm{p}}$, is described by 
\begin{equation}
F_{\rm{p}} (t) = F_{\rm{day}} (t) \times F_{\rm{pc}} (t).
\label{eq:Fp}
\end{equation}
\noindent
Here, $F_{\rm{day}}$ is the planet's dayside emission, which is $F_{\rm{day}} = 0$ during the eclipse, and otherwise $F_{\rm{day}} = f_p/f_s$. The phase curve flux, $F_{\rm{pc}}$, is expressed as
\begin{equation}
F_{\rm{pc}} (t) = 1 + \frac{AmpCos}{2} \left(\cos{\phi -1}\right),
\label{eq:Fpc}
\end{equation}
\noindent
where $\phi$ is the orbital phase of the planet. In the case of a circular orbit, this is given by $\phi = 2\pi/P_{\rm{orb}}(t-t_{\rm{sec}})$, with $P_{\rm{orb}}$ being the orbital period and $t_{\rm{sec}}$ being the secondary eclipse time. Finally, the systematics model for an observation is 
\begin{equation}
F_{\rm{sys.}} (t)= c_0 + c_1 (t - \bar{t})+ c_2 (t - \bar{t})^2 + c_{\rm{SA}}  
 \textbf{\rm{SA}}(t) + c_{\rm{IMIR}}  
 \textbf{\rm{IMIR}}(t) + c_{\rm{y}} \textbf{\rm{y}}(t)
\end{equation}
\noindent
where $c_0$, $c_1$, and $c_2$ are the coefficients of the second-order polynomial, $\bar{t}$ is the mean time of an observation, and \textbf{\rm{SA}}(t),  \textbf{\rm{IMIR}}(t) and \textbf{\rm{y}}(t) are arrays containing the information of the SA or IMIR mnemonics, or the y-position of the spectral trace on the detector as a function of time.

\begin{figure}
    \centering
    \includegraphics[width=1\linewidth]{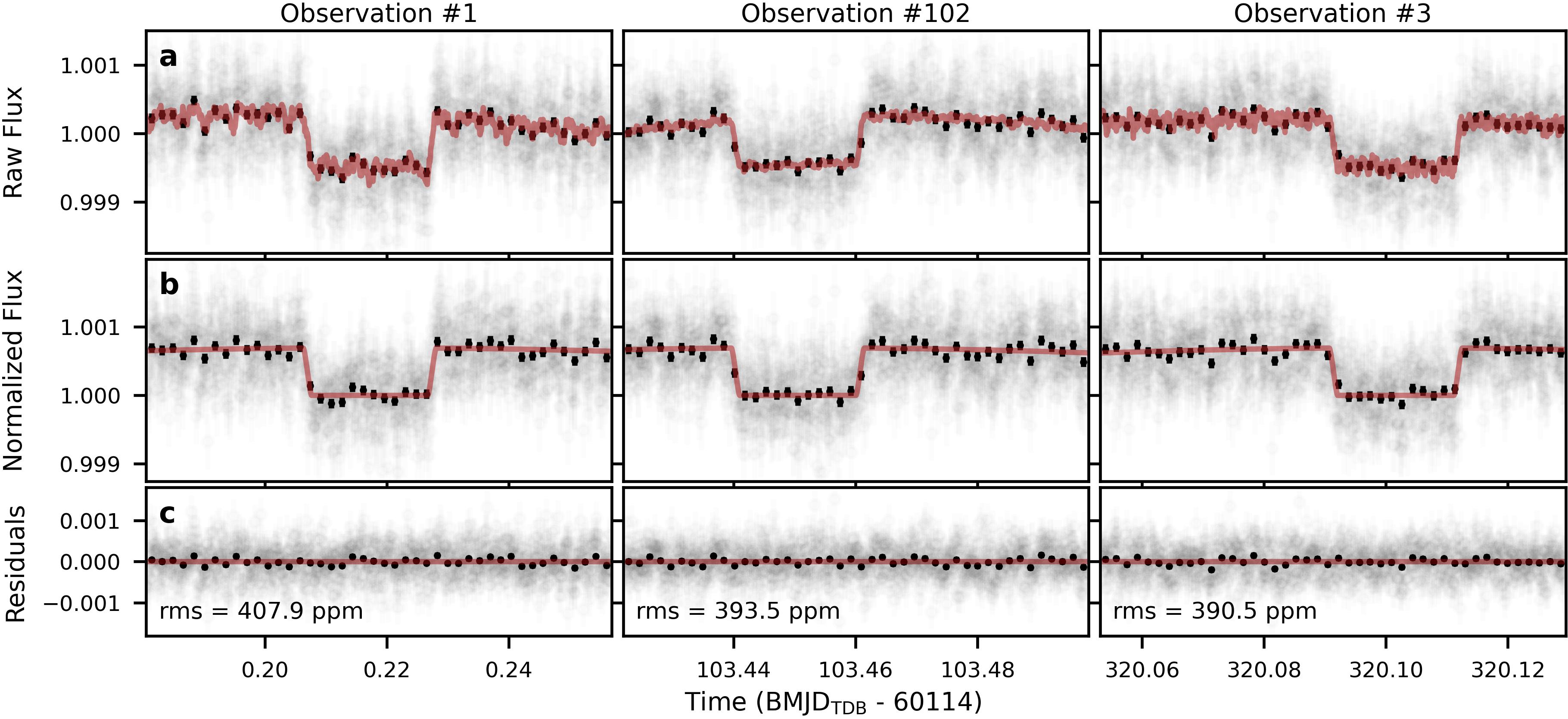}
    \caption{\textbf{Best-fit JWST MIRI/LRS white light curves.} These white light curves cover wavelengths from 5.060 to 12.368 \textmu m and are based on the SZ reduction. Row \textbf{a}: Best-fit full model (astrophysical and systematic) in red with the observed raw flux in grey. Each bin covers approximately 2.5 minutes. Row \textbf{b}: Same as row \textbf{a}, but without the best-fit systematics model, leaving only the astrophysical signal. Row \textbf{c}: Best-fit residuals of the observations. The rms of the residuals for each observation is listed in each panel. The error bars in all panels are 1$\sigma$.}
    \label{fig:white_lc_fits}
\end{figure}

With this initial fit, we find that the phase curve amplitude, $AmpCos$, remains unconstrained within our uniform priors, which we set to U(0, 1). Here, $AmpCos = 1$ would indicate that the full phase curve amplitude is equal to the eclipse depth and there's no night side flux emitted from the planet. When allowing for nonphysical solutions for the planet's amplitude --- such as negative flux (i.e. $AmpCos < 0$) or phase curve amplitudes higher than the eclipse depth (i.e. $AmpCos > 1$) --- we obtain $AmpCos = 1.6_{-12.0}^{+11.8}$. The wide posterior indicates that we are not able to meaningfully contain the planet's phase curve amplitude, which is expected given the limited phase coverage of our observations (see Fig. \ref{fig:white_lc_phasefolded}). 
Based on this, we adopt $AmpCos = 1$ for all other fits. This decision is furthermore supported by the Spitzer observations, which found no significant night side flux in their phase curve data \cite{Kreidberg2019}. 
With this fit that only includes the three JWST observations and sets $AmpCos = 1$, find an eclipse depth of $f_p/f_s = 694 \pm 18$ ppm, an eclipse time of $t_{\rm{sec}}=60114.21718\pm0.00010$ $ \mathrm{BMJD_{TDB}}$. We also find an orbital period of $P_{\rm{orb}} = 0.46292973\pm0.00000023$ days, consistent within 1$\sigma$ with the literature value of $P_{\rm{orb}} = 0.46292913\pm0.00000190$ days \cite{Vanderspek2019}. We list our best-fit parameters and their uncertainties for the JWST-only fits in Table \ref{tab:white_fits_JWST_only}.
We show our best fit to each observation in Figure \ref{fig:white_lc_fits} and the phase-folded light curve in Figure \ref{fig:white_lc_phasefolded}. We find an rms in the residuals of 407.9, 393.5, and 390.5 ppm for observation \#1, \#102, and \#3, respectively. By simulating our observations using the open-source code \texttt{PandExo v3.0} \cite{Batalha2017_pandexo, Batalha2023_pandexo_v3.0}, we determine the expected noise to be 379.2 ppm per integration. This results in our observations being marginally over the photon noise limit at 7\%, 4\%, and 3\% for the three observations, respectively.\\

\begin{table}[h]
    \centering
    \caption{\textbf{Best-fit parameter estimates for the JWST-only fits.} We report the median of our posteriors and the 16th and 84th percentiles for the lower and upper 1$\sigma$ bounds. For these fits, we fixed the following parameters to the literature values: $r_p/r_s = 0.0635$, $a/r_s = 7.109$, and $i = 88.50^\circ$. The values in this table are based on the SZ reduction. The adopted fit using only the JWST data is shown in the rightmost column of the table.}
    \renewcommand{\arraystretch}{1.4}
    \footnotesize
    \begin{tabular}{l c c c }
        \hline
        & \begin{tabular}[c]{@{}c@{}}JWST only\\$AmpCos$, $\mathcal{U}$(0,1)\\with mnemonics\\~   \end{tabular}
        & \begin{tabular}[c]{@{}c@{}}JWST only\\$AmpCos$ = 1                \\without mnemonics\\~    \end{tabular} 
        & \begin{tabular}[c]{@{}c@{}}\textbf{JWST only}\\\textbf{$AmpCos$ = 1}                \\\textbf{with mnemonics}\\\textbf{adopted fit}    \end{tabular}  \\
        \hline
        $t_{\rm{sec}}$\textsuperscript{a} 
        & $0.21718 \pm 0.00010$
        & $0.217145_{-0.000097}^{+0.000100}$ 
        & $0.21718 \pm 0.00010$  \\
        $P_{\rm{orb}}$ (days)\textsuperscript{b} 
        & $\left(7.4 \pm 2.3\right)\times10^{-7}$ 
        & $\left(7.2 \pm 2.2\right)\times10^{-7}$ 
        & $\left(7.3 \pm 2.3\right)\times10^{-7}$ \\
        $f_p/f_s$ (ppm)
        & $694 \pm 18$ 
        & $704 \pm 18$ 
        & $694 \pm 18$ \\
        $AmpCos$ 
        & $0.51_{-0.35}^{+0.33}$\textsuperscript{c} 
        & 1 (fixed) 
        & 1 (fixed) \\
        \hline
    \end{tabular}
    \label{tab:white_fits_JWST_only}

    \vspace{3pt} 
    \parbox{\linewidth}{\raggedright \footnotesize 
    \textsuperscript{a} BJD$\rm{_{TBD}}$ - 2460114.5 days.\\
    \textsuperscript{b} $P_{\rm{orb}}$ - 0.462929 days.\\
    \textsuperscript{c} Note that the posterior for this fit was completely uniform in the prior range of $\mathcal{U}$(0,1). When leaving this parameter completely unconstrained with a wide uniform prior, we retrieved $AmpCos = 1.6_{-12.0}^{+11.8}$. This suggests that we cannot meaningfully contain this parameter with our observations. We therefore set $AmpCos = 1$ in all other fits.   
    }
    
\end{table}

\begin{figure}
    \centering
    \includegraphics[width=1\linewidth]{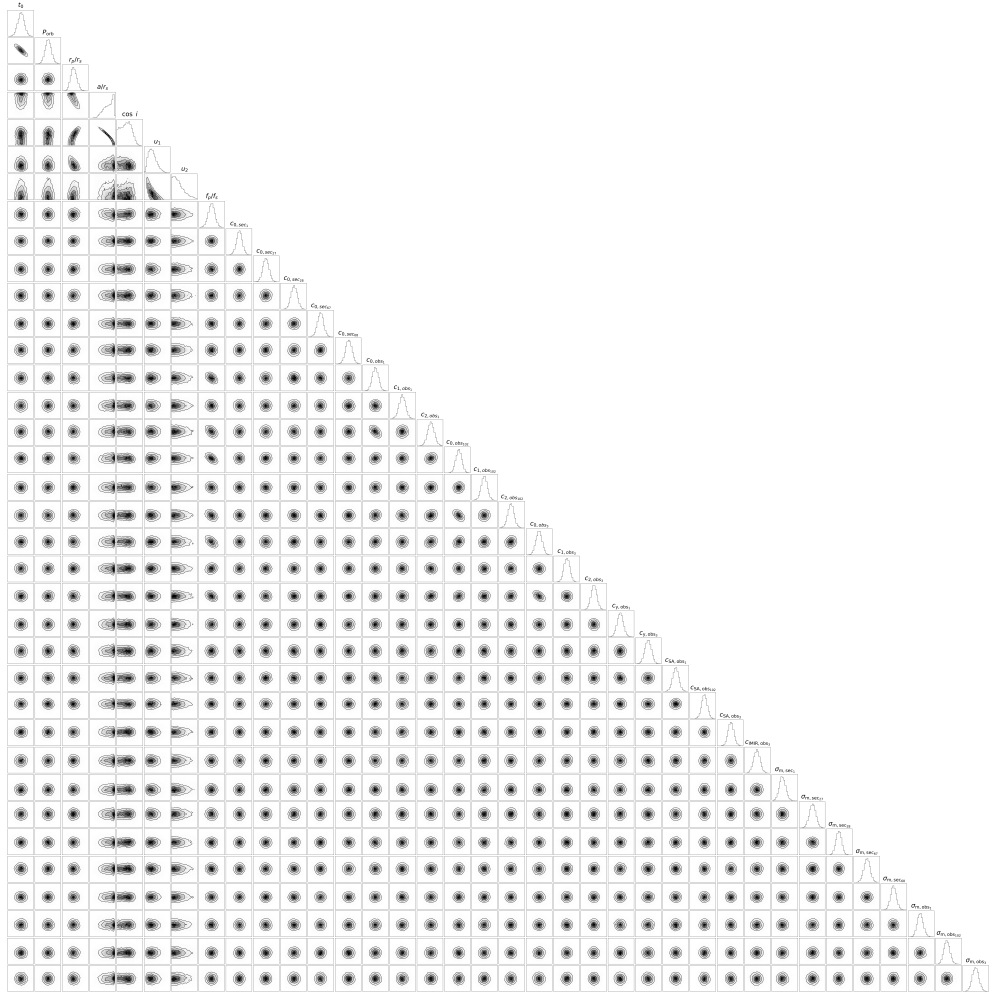}
    \caption{\textbf{Posterior distributions for the circular TESS and white light curve JWST fit.} This fit consists of 36 parameters. We refer to the text in the Methods section for a description of those parameters.}
    \label{fig:corner}
\end{figure}

\noindent\textbf{White light curve fit --- JWST and TESS}\\
\noindent
Finally, we fit both datasets --- TESS and JWST --- jointly. For this analysis, we use the wavelength range from 5.060 to 12.368 \textmu m for all three JWST observations, along with data from all five TESS sectors. We perform two fits: one where we allow the eccentricity to be a free parameter and another where we fix it to $e = 0$. In both of these fits, we account for light travel time that leads to the eclipse arriving approximately 2$\times$3.1 seconds delayed.
In the first fit, our free physical parameters include the transit midpoint time $t_0$, the orbital period $P_{\rm{orb}}$, the planet-to-star radius ratio $r_p/r_s$, the semi-major axis to star radius ratio $a/r_s$, and the cosine of the orbital inclination $\cos i$. Additionally, we fit for the quadratic limb-darkening parameters $q_1$ and $q_2$, (in the reparametrization of Ref. \cite{Kipping2013}), the eclipse depth $f_p/f_s$, and the parameters $\sqrt{e}\sin\omega$ and $\sqrt{e}\cos\omega$, which represent the square root of the eccentricity multiplied by the sine and cosine of the argument of periastron, respectively. We note that the phase curve amplitude parameter, $AmpCos$, is fixed to 1 in these fits, meaning that the full phase amplitude is equal to the eclipse depth. 
For the systematic parameters, we use a constant term in time for each TESS sector, and for the JWST data the same as described above in the JWST-only fits, that is, a quadratic trend in time for each of the three JWST observations, two parameters to account for the position of the spectrum on the trace in observation 1 and 3, and four parameters for the mnemonics. Additionally, each of the eight datasets receives an independent white noise multiplier. In total, this results in 38 free parameters in our fit. We list our results for this fit in Table \ref{tab:white_fits_TESS_and_JWST}. 
Notably, our eccentricity is $e = 0.0020_{-0.0014}^{+0.0067}$, which is consistent with a circular orbit when accounting for the Lucy-Sweeney bias \cite{Lucy1971, Eastman2013, Eastman2019}. We do find upper values of $e <0.017\ (2\sigma)$ and $e <0.039\ (3\sigma)$, which could inform future tidal heating models of LHS 3844\,b. 
Our results for the circular orbit fit, which has two fewer free parameters (36 in total), are also listed in Table \ref{tab:white_fits_TESS_and_JWST}, and a full corner plot is shown in Figure \ref{fig:corner}. The circular and the non-circular orbit fits are consistent with each other within uncertainties. Given the eccentricity is consistent with zero, we adopt the circular fit and use the orbital and planetary values for the following spectroscopic fits. 
Our results are also in agreement with the literature values in Ref. \cite{Vanderspek2019}. However, the discovery paper determined uncertainties in $a/r_s$ and $i$ that are approximately one order of magnitude smaller than ours. This discrepancy, despite our use of more transit observations, is likely in majority due to their adopted prior on the stellar density, which was determined from an empirical relationship between the star's $K_s$ band magnitude and the stellar distance reported by the Gaia mission. In contrast, we do not impose Gaussian priors informed by external data but instead use wide uniform priors.
With our combined dataset, we provide a precise ephemeris that can aid in scheduling future follow-up observations of the planet.
\\

\begin{figure}
    \centering
    \includegraphics[width=1\linewidth]{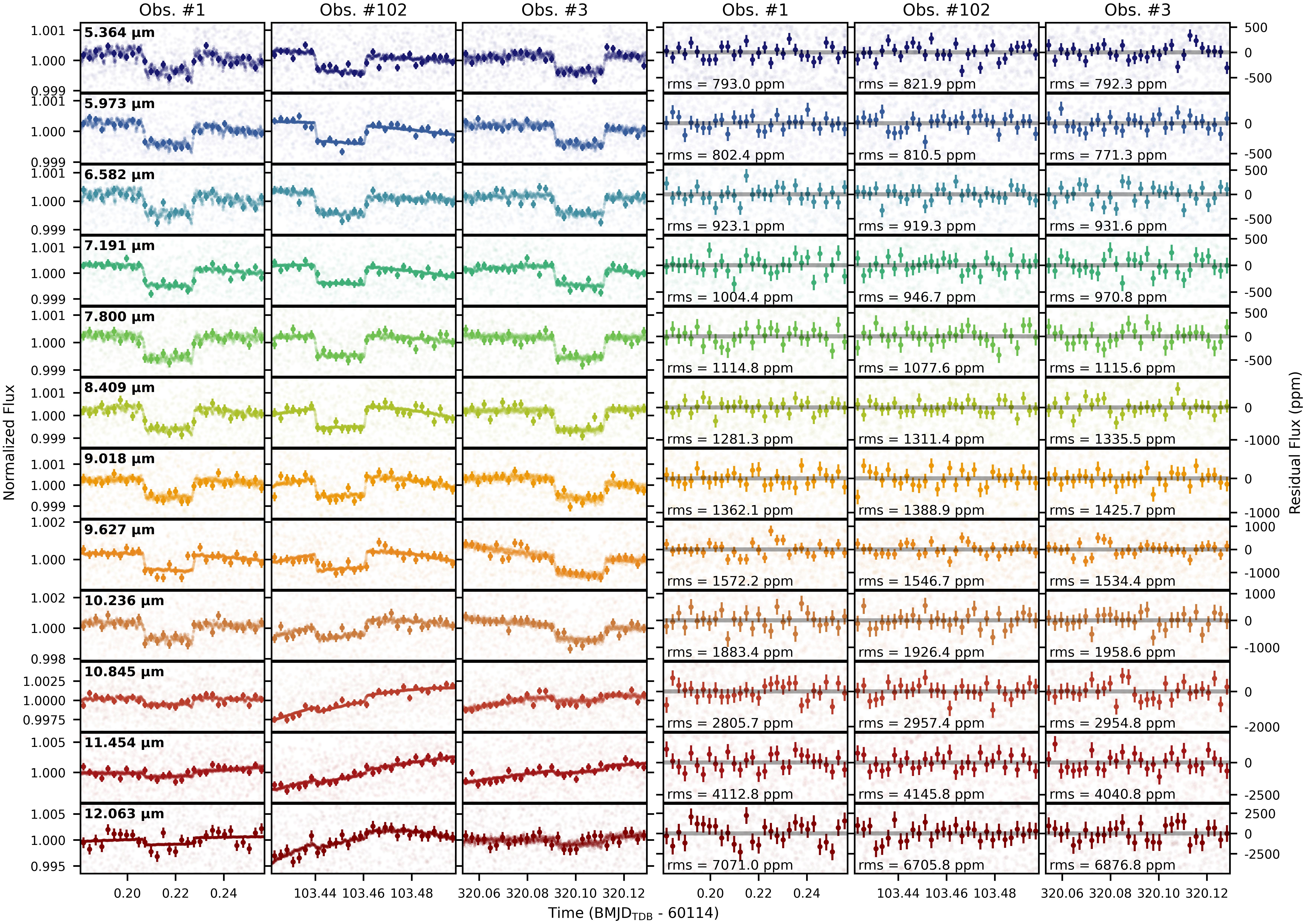}
    \caption{\textbf{Best fits to the JWST spectroscopic light curves and their residuals}. Each row corresponds to a spectroscopic bin, with the mid-wavelength noted in the top left of each subplot. Each column represents a different observation. The first three columns display the binned observations alongside our best-fit astrophysical and systematics model. The remaining three columns show the residuals of the best fit, along with the root mean square (RMS) of the residuals. Time-averaging plots for the unbinned residuals are presented in Figure \ref{fig:allan_specs}. The data and fits in this figure are based on the SZ analysis. All uncertainties are 1$\sigma$.}
    \label{fig:spec_lc}
\end{figure}

\begin{figure}
    \centering
    \includegraphics[width=1\linewidth]{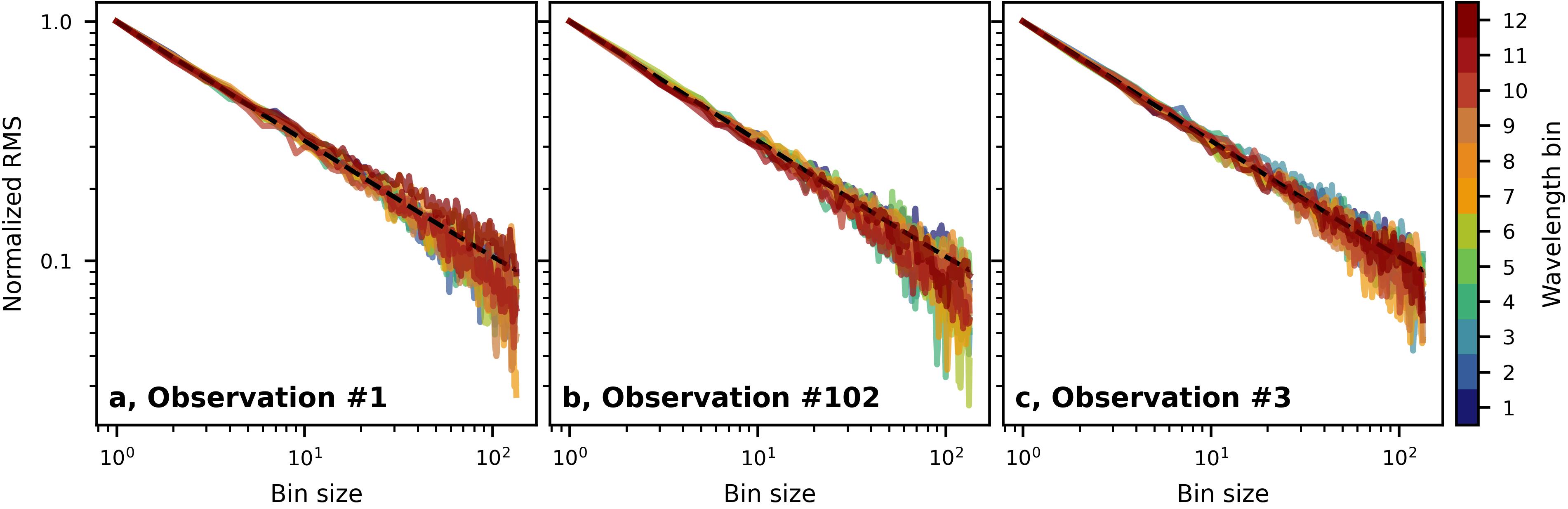}
    \caption{\textbf{Time-averaging plots for all three JWST observations and each spectroscopic light curve}. Each panel (\textbf{a}, \textbf{b}, and \textbf{c}) corresponds to an observation (\#1, \#102, and \#3). Each solid line represents the residuals of a spectroscopic light curve from the SZ analysis as a function of binning. The residuals can be seen in Figure \ref{fig:spec_lc}. The dashed black line shows the expected behavior in the presence of purely Gaussian noise, following the 1/$\sqrt{\rm{bin~size}}$ relationship. The RMS as a function of bin size for each observation is normalized by dividing it by the RMS of the unbinned data (bin size = 1). Each color corresponds to a different wavelength bin.}
    \label{fig:allan_specs}
\end{figure}

\noindent\textbf{Spectroscopic light curve fits}\\
\noindent
We create 12 spectroscopic light curves covering wavelengths from 5.060 to 12.368 \textmu m, with each bin having a full width of 0.609 \textmu m. We also assume a circular orbit for the planet in our fits. We fix the orbital and planetary parameters ($t_0$, $P_{\rm{orb}}$, $r_p/r_s$, $a/r_s$, and $\cos i$) to our circular orbit joint TESS and white light JWST best-fit values, which are listed in Table \ref{tab:white_fits_TESS_and_JWST}.
We adopt the same astrophysical and systematics model as in the white light curve fits, which consists out of a \texttt{batman} eclipse (with eclipse depth $f_p/f_s$), a cosine term to model the phase curve, a quadratic trend in time ($c_{\rm{0, obs_i}}$, $c_{\rm{1, obs_i}}$, and $c_{\rm{2, obs_i}}$), four parameters for the mnemonics ($c_{\rm{SA, obs_1}}$, $c_{\rm{SA, obs_{102}}}$, $c_{\rm{SA, obs_3}}$ and $c_{\rm{IMIR, obs_3}}$), two parameters for the spatial position of the trace ($c_{\rm{y, obs_1}}$ and $c_{\rm{y, obs_3}}$), and the white noise multiplier $\sigma_{\rm{m, obs_i}}$ for each observation. We show the raw flux, best-fit model, and residuals for each of the 12 spectroscopic bins and 3 observations in Figure \ref{fig:spec_lc}. The corresponding time-averaging plot is shown in Figure \ref{fig:allan_specs} and indicates no signs of excessive correlated noise.
This fit involves 19 free parameters when fitting for a common eclipse depth across the three observations in each respective spectroscopic bin (see panel a in Figure \ref{fig:reduction} for the planetary spectrum resulting from this fit). Additionally, we test fitting for a separate eclipse depth per observation (see panel b--d in Figure \ref{fig:reduction}).
When taking a weighted average of the separate eclipse depths per bin, the spectrum remains consistent with the single-depth fit well within $1\sigma$. 
Finally, we also fit the spectroscopic light curves using a Gaussian prior on the orbital and planetary parameters based on TESS data. This resulting spectrum is also consistent with the fixed parameter fit within $1\sigma$ and does not introduce significantly greater uncertainties in the eclipse depths. Therefore, we use the single eclipse depth per wavelength bin with fixed orbital parameters as our final planetary spectrum from the SZ analysis.

\subsubsection*{Data Analysis ABA}
Using \texttt{emcee} as part of \texttt{Eureka!}'s Stage 5, we fitted each light curve with the product of a systematics model and a \texttt{batman} \cite{Kreidberg2015} eclipse model. The systematics model included an exponential ramp and a linear polynomial in time, as well as linear decorrelation against drift in the spatial direction. We also fitted for a white noise parameter in order to scale the data uncertainties according to the scatter of the residuals. The eclipse model, which accounted for light travel time, included a cosine function that peaked around mid-eclipse to model phase curve variations. We fitted all white light curves independently, fixing the planet-to-star radius ratio, orbital period, inclination, and scaled semi-major axis to the values in Ref. \cite{Vanderspek2019} and assuming a circular orbit. Meanwhile, in the spectroscopic light-curve fits, we additionally fixed the mid-transit time and cosine amplitude to the corresponding white light-curve best-fit values. All other orbital and systematic parameters were kept free in all light-curve fits.

\subsubsection*{Final combined planetary spectrum}

\begin{figure}
    \centering
    \includegraphics[width=1\linewidth]{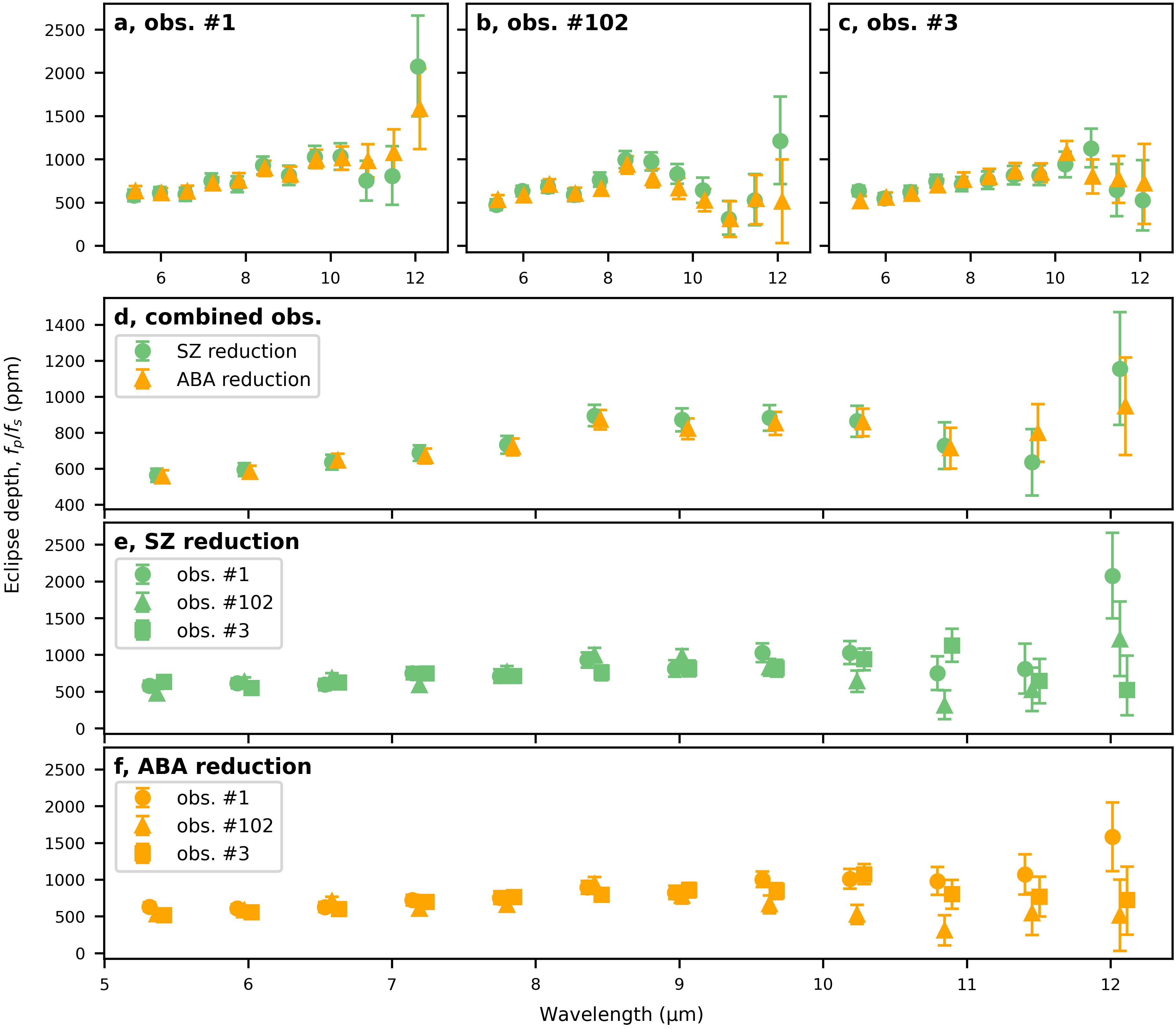}
    \caption{\textbf{Comparison of the planetary spectra from the two reductions}. Panel \textbf{a}--\textbf{c}, The eclipse depth (i.e., planet-to-star flux ratio) as a function of wavelength bin for each individual observation. The green points represent the SZ reduction, while the orange points correspond to the ABA reduction. Panel \textbf{d}, The final eclipse depth as a function of wavelength. Panels \textbf{e} and \textbf{f}, the eclipse depth for the SZ and the ABA reduction for observation \#1 (circular marker),  observation \#102 (triangular marker), and observation \#3 (square marker). A small offset in wavelength is applied to the ABA spectrum in every panel for visual clarity. }
    \label{fig:reduction}
\end{figure}

In panel d of Figure \ref{fig:reduction}, we compare the full planetary spectrum from both data reduction methods and analyses. The measurements remain well within 1$\sigma$ for each individual wavelength bin, indicating strong agreement between the two independent reductions. For the individual observations, 4 out of the 36 measured eclipse depths show discrepancies greater than 1$\sigma$, with the largest disagreement reaching 1.4$\sigma$.
To avoid biasing our results due to pipeline choices, we follow an approach used in previous JWST studies that employ multiple data reductions like Ref. \cite{Greene2023, Zieba2023}. Specifically, we take the mean eclipse depth for each spectroscopic bin. For the uncertainties, we adopt the same approach but additionally compute the standard deviation between the two reductions within a bin and add it to the mean uncertainties in quadrature. Due to the strong agreement between the reductions, the standard deviation typically remains below 10 ppm but reaches approximately 100 ppm for the two reddest bins.
We list the final eclipse depths as a function of wavelength in Table \ref{tab:fpfs}.

\begin{table}[h]
    \centering
    \caption{\textbf{Final measured eclipse depths $f_p/f_s$ as a function of wavelength bin.} 
    $\lambda_\mathrm{mid}$ is the central wavelength of each spectroscopic bin and the bin width was identical for each bin at $\pm0.3045$ \textmu m.}
    \renewcommand{\arraystretch}{1.4}
    \footnotesize
    \begin{tabular}{c c c}
        \hline
        \begin{tabular}[c]{@{}c@{}}$\lambda_\textrm{mid}$\\(µm)\end{tabular} 
        & \begin{tabular}[c]{@{}c@{}}bin width\\(µm)   \end{tabular}    & \begin{tabular}[c]{@{}c@{}}$f_p/f_s$\\(ppm)    \end{tabular} \\
        \hline
5.3645 & 0.3045 & $559.9_{-35.4}^{+34.9}$ \\
5.9735 & 0.3045 & $587.6_{-35.5}^{+35.2}$ \\
6.5825 & 0.3045 & $640.3_{-39.7}^{+40.2}$ \\
7.1915 & 0.3045 & $678.5_{-43.2}^{+42.8}$ \\
7.8005 & 0.3045 & $726.8_{-48.4}^{+49.2}$ \\
8.4095 & 0.3045 & $882.5_{-56.9}^{+58.4}$ \\
9.0185 & 0.3045 & $846.4_{-65.2}^{+65.1}$ \\
9.6275 & 0.3045 & $866.6_{-69.2}^{+69.6}$ \\
10.2365 & 0.3045 & $861.0_{-81.9}^{+81.7}$ \\
10.8455 & 0.3045 & $720.2_{-122.0}^{+122.2}$ \\
11.4545 & 0.3045 & $716.5_{-191.9}^{+190.9}$ \\
12.0635 & 0.3045 & $1050.2_{-310.1}^{+313.0}$ \\
        \hline
    \end{tabular}
    \label{tab:fpfs}
\end{table}

\subsection*{Absolute Stellar Flux --- Observations and modelling}

In order to compare our observations, which are measurements of the planet-to-star flux ratio, to our models, we will have to understand the stellar absolute flux of LHS 3844 in the observed wavelengths. The importance of that was stressed by Ref. \cite{fauchez2025stellar}, which advocates the use of the empirically calibrated stellar flux in rocky exoplanet studies using emission spectroscopy. This is motivated by the observed discrepancy between the stellar flux predicted by stellar models in the mid-infrared as observed by MIRI and the observations for late-type stars \cite{Greene2023, Zieba2023, Ducrot2024}. We, therefore, discuss in the following the absolute flux of LHS 3844 as measured by JWST and Spitzer.

\subsubsection*{Absolute Stellar Flux --- JWST}

\begin{figure}
    \centering
    \includegraphics[width=0.99\linewidth]{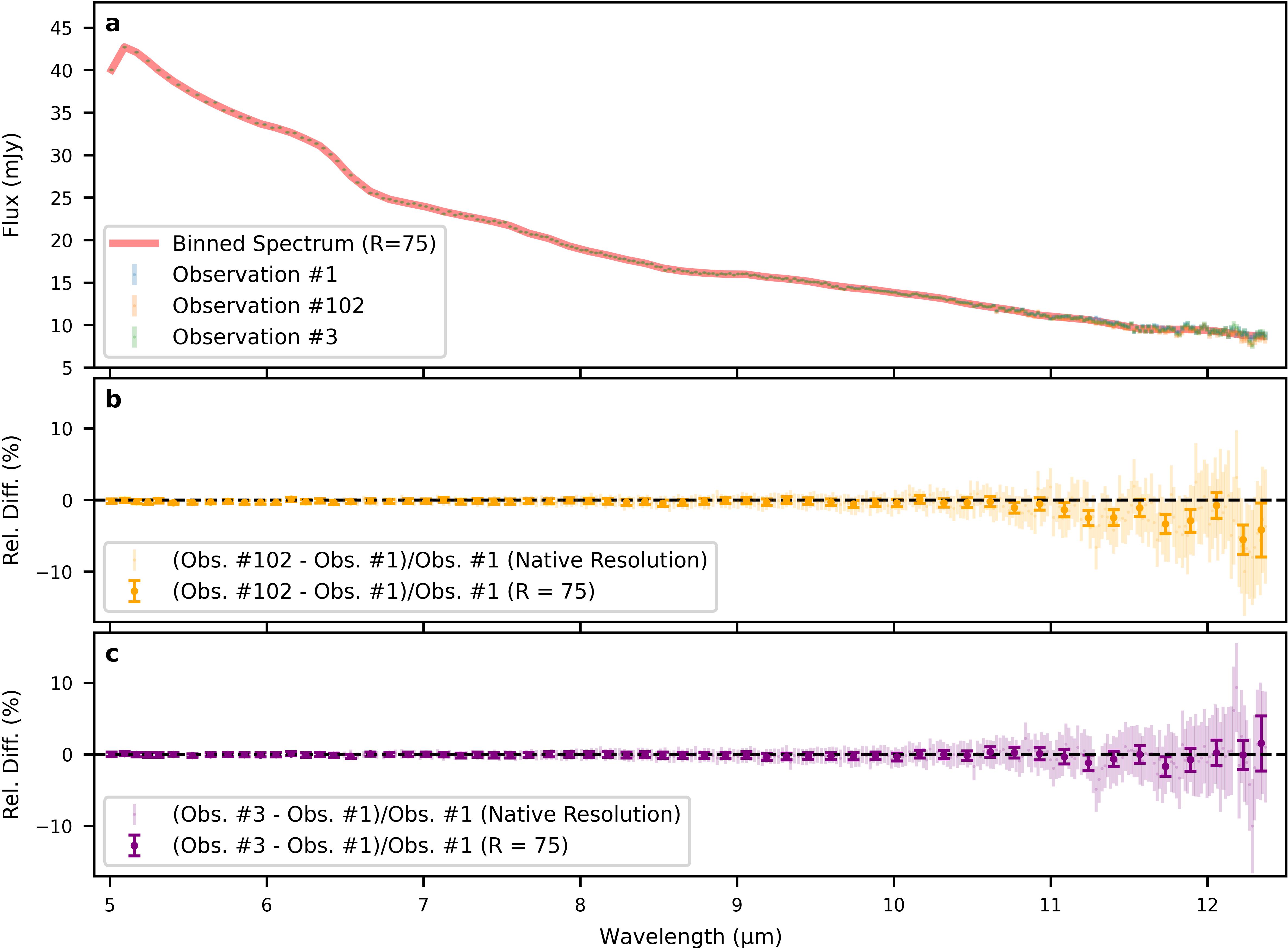}
    \caption{\textbf{Observed stellar spectrum of LHS 3844 by JWST.} \textbf{a}, The stellar flux of LHS 3844 (in units of millijansky) as a function of wavelength. Each of the three observations is shown, with the measurements typically being in agreement and therefore overlapping each other. The combined stellar spectrum of the three observations is shown as a solid red line. \textbf{b}, A comparison of the stellar flux measured in observation \#102 to that in observation \#1. \textbf{c}, Same as panel b, but comparing observation \#3 to observation \#1.}
    \label{fig:stellar_spectrum_1}
\end{figure}

JWST observations provide not only relative flux measurements --- such as eclipse depths --- but also absolute flux densities. This allows us to directly measure the stellar spectrum of LHS 3844 across the three visits and compare it to stellar models. 
To obtain the absolute flux stellar spectrum of LHS 3844 as observed by JWST, we again utilized \texttt{Eureka!}, starting from the same \texttt{uncal.fits} files and package versions as described in Section ``JWST Data Reduction SZ'' of the Methods. In Stages 1 and 2, we retained the same settings as before, with the following exceptions: \texttt{skip\_firstframe} and \texttt{skip\_lastframe} were set to False, and \texttt{skip\_emicorr} was set to True, which led to less noisy final stellar spectra. We. Additionally, in Stage 2, we set \texttt{skip\_flat\_field} and \texttt{skip\_photom} to False. Finally, in Stage 3, we increase the target aperture size to capture more stellar flux. We therefore set \texttt{spec\_hw} = 17 and \texttt{bg\_hw} = 18, so that the majority of stellar flux was included while still retaining enough background pixels in each column for background subtraction. 
We then only extract the stellar flux at the time of the planetary eclipses. Although we use only the stellar flux during eclipse, the planet's contribution is less than 0.1\%, while the absolute flux calibration has an uncertainty of a few percent\footnote{\url{https://jwst-docs.stsci.edu/jwst-calibration-status/miri-calibration-status/miri-lrs-calibration-status}}. Thus, whether we extract the flux only during eclipses or not has little impact. Figure \ref{fig:stellar_spectrum_1} compares the stellar spectrum extracted from each observation, showing no evidence of variability. We therefore take the weighted mean across the three observations to obtain the final stellar spectrum. The bluest wavelength column at 5.011 \textmu m shows a drop in flux compared to the second bluest wavelength column (5.094 \textmu m). We leave out this single column for our final spectra, as it might be an artifact of the spectrum being calibrated to hotter stars than LHS 3844. 

\begin{figure}
    \centering
    \includegraphics[width=0.99\linewidth]{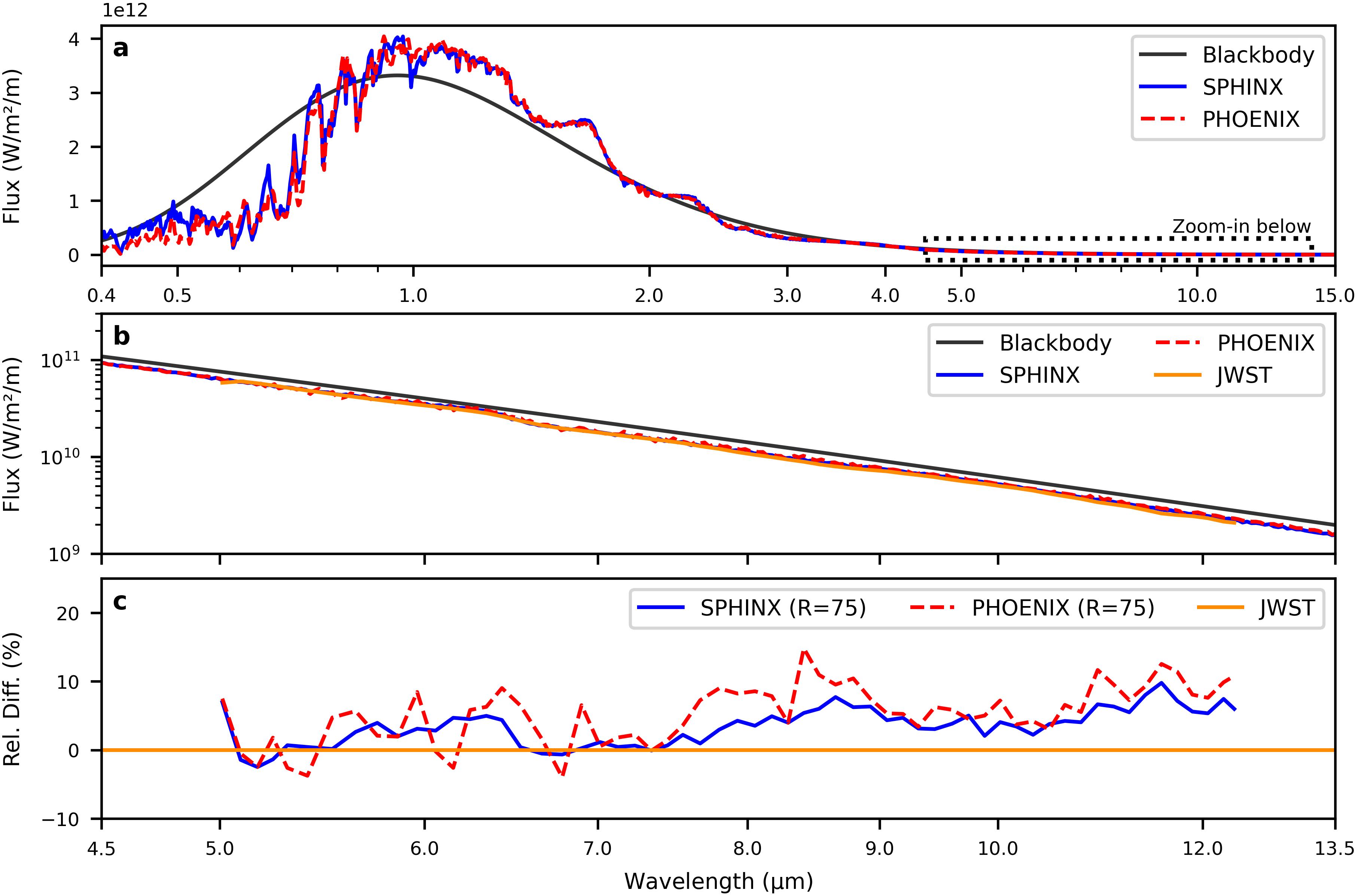}
    \caption{\textbf{Comparison of the JWST stellar spectrum to simulated stellar models.} \textbf{a}, Comparison of a black body spectrum, a PHOENIX model, and a SPHINX model using the stellar parameters published in Ref. \cite{Vanderspek2019}. \textbf{b}, A zoomed-in view of planet a into the MIRI/LRS wavelength range. The obtained stellar spectrum, converted from milliJanskys to W/m$^2$/m, is shown as a solid orange line. \textbf{c}, A comparison of the PHOENIX and SPHINX models to the observed spectrum, expressed as a percentage.}
    \label{fig:stellar_spectrum_2}
\end{figure}

The stellar flux is initially in units of millijansky (mJy) after extracting it from \texttt{Eureka!}, which we convert into $\rm{W/m^2/m}$ for comparison with stellar models. This conversion between flux density $S_\nu$ (in millijansky) to flux density $F_\lambda$ (in $\rm{W/m^2/m}$) with a given wavelength array $wvl$ (in meters) is performed using the following relation:
\begin{equation}
F_\lambda \, [\mathrm{W/m^2/m}] = \frac{S_\nu \, [\mathrm{mJy}] \cdot c \cdot 10^{-29}}{wvl^2} \left( \frac{d_*}{R_*} \right)^2,
\end{equation}
\noindent
where $c$ is the speed of light. The parameter $d_*$ represents the distance between us and the host star LHS 3844, which is $d_* = 14.867 \pm 0.004$ pc following the Gaia EDR3 release (the star's catalog name is Gaia DR3 6385548541499112448) \cite{BailerJones2021a, BailerJones2021b}. For the stellar radius, we adopt the value reported in the LHS 3844\,b discovery paper of $R_* = 0.189 \pm 0.006\ \rm{R_\odot}$ \cite{Vanderspek2019}. In Figure \ref{fig:stellar_spectrum_2}, we show a comparison of a PHOENIX (\cite{Allard2014}, accessed using the python package pysynphot \cite{STScI2013}) and SPHINX model to the JWST observations (see ``SPHINX'' model in Methods). Both models generally overestimate the flux in the MIRI/LRS wavelength range, with greater disagreement at longer wavelengths. On average, the flux predicted by the SPHINX model is 3.5\% higher than the observed flux, while the PHOENIX model overestimates it by about 5.5\%. These percentages are slightly higher than the absolute flux calibration uncertainty, which is estimated to be only a few percent. The SPHINX model shows better agreement, which is expected since it was specifically designed to better describe the stellar spectra of M-dwarfs. A qualitatively similar agreement has also been observed at longer wavelengths for TRAPPIST-1. The overestimation of flux at 12.8 \textmu m and 15 \textmu m is approximately 13\% for the PHOENIX stellar model and about 7\% for SPHINX using JWST \cite{Greene2023, Zieba2023, Ducrot2024}.

\subsubsection*{Absolute Stellar Flux --- Spitzer}

LHS 3844 was observed by Spitzer in February of 2019 as part of Program 14204 (PI: Laura Kreidberg; \cite{Kreidberg2018sptz.prop14204K}). The observations were conducted using Channel 2 of Spitzer InfraRed Array Camera (IRAC), which is a photometric bandpass covering the wavelength range from 4 to 5 µm \cite{Fazio2004}. We refer to the LHS 3844 Spitzer publication for any further information on the observational setup \cite{Kreidberg2019}.
In order to extract the flux of the star, we use a photometry aperture with a radius of 3.5 pixels to capture the majority of the flux hitting the detector. After subtracting the background, we measure a median electron number of 76,834 electrons per each of the 2-second exposure. This converts into 38,417 electrons per second. Following the Spitzer IRAC Instrument Handbook \cite{IRAC2021}, 1 µJy at 4.5 \textmu m produces 0.73 electrons/second. So this finally results in 38,417/0.73 µJy = 52,626 µJy and therefore $52.626 \pm 0.190$ mJy for the star, assuming only Poisson noise for the uncertainties.
The prediction by the SPHINX stellar model in the Spitzer Channel 2 wavelengths (4--5 \textmu m) is 56.81 mJy (and 56.20 mJy for PHOENIX). We therefore correct the SPHINX stellar model in all plots and calculations by scaling it down by (56.81 mJy$-$52.626 mJy)/56.81 mJy = 7.36\% in the Spitzer wavelengths.

\subsection*{SPHINX model}
Stellar model uncertainties in M-dwarfs remain significant even at longer wavelengths. For this analysis, we use a SPHINX model computed specifically for LHS 3844. An interpolated spectrum from the publicly available SPHINX grid \cite{Iyer2023}, yields uncertainties of approximately 1$\%$ in the MIRI/LRS bandpass when compared to the LHS 3844-specific model. This is consistent with Ref. \cite{Iyer2023}, which demonstrates that biases arising from stellar model assumptions can lead to spectral variability. Furthermore, Ref. \cite{fauchez2025stellar} suggest that stellar contamination may be significant at these wavelengths. To address this, we use the LHS 3844-specific model to mitigate both interpolation errors and the potential omission of key physical effects from the stellar models. The parameters for the high-resolution SPHINX model ($R \sim 100,000$) computed, were using $T_{\rm{eff}}$ of 3036 K, log$g_*$ of 5.06 (cgs), assuming solar metallicity and C/O, following Ref. \cite{Vanderspek2019}.

\section*{Significance calculation}

In order to calculate the rejection significance between the observations and a planetary emission model, we perform a  $\chi^2$ test as follows: 
First, we compute the expected flux of a given model within each wavelength bin of interest. The model flux is then binned using throughput functions of Spitzer IRAC Channel~2 and JWST MIRI/LRS. We compute the $\chi^2$ statistic between the binned model and the observations, with the number of degrees of freedom as the number of bins minus one. The corresponding $p$-value is obtained by integrating the $\chi^2$ distribution from the calculated $\chi^2$ value to infinity (i.e., using the survival function). Finally, we convert this resulting $p$-value into an equivalent Gaussian significance by calculating the inverse of the cumulative distribution function, which yields the corresponding $\sigma$ level between our observations and the model.

\subsection*{Emission modelling for LHS 3844\,b}

We simulated emission spectra for LHS 3844\,b to compare them to our measured eclipse depth in the Spitzer and the MIRI/LRS wavelengths. The models include surface and atmospheric planetary spectra. 
For the creation of all planet-to-star flux spectra, we use the same stellar model. That is a SPHINX model as described in Section ``SPHINX model'' with the following two exceptions: 1) within the Spitzer wavelengths, we scale the stellar model down by 7.36\% as determined in Section ``Absolute Stellar Flux - Spitzer'' of the Methods; 2) in the JWST wavelengths, we use the observed stellar flux as shown in Figure \ref{fig:stellar_spectrum_1} and discussed in Section ``Absolute Stellar Flux - JWST''.

First demonstrated by Ref. \cite{hu2012} by building on the legacy of characterizing rocky surfaces in our Solar System, it was shown that bare rock exoplanets are expected to exhibit distinctive spectral features in the mid-infrared. Different types of silicate surfaces on rocky exoplanets can be distinguished spectroscopically by analyzing the prominent Si-O vibrational features, specifically between 7 \textmu m and 13  \textmu m. These features provide insight into the silica content of a planetary surface, allowing for classification into ultramafic, mafic, or felsic types, categories that correspond to surfaces with increasing SiO$_2$ content. Such spectral diagnostics enable us to constrain both the past and present geology of rocky exoplanets by analyzing their emission spectra. These key wavelengths are accessible to JWST, particularly through the MIRI Low Resolution Spectroscopy (LRS) instrument, covering wavelengths approximately from 5 to 12 \textmu m. Therefore, thanks to JWST, a fundamental question can be explored: What is the surface composition of rocky exoplanets and what is the geological history of planets outside of our solar system? Observations of planets orbiting small M dwarfs are especially promising for addressing this question. Due to their high planet-to-star radius ratios, these systems have deeper secondary eclipse depths and, therefore, greater emission signals, an observational advantage often referred to as the ``M dwarf opportunity.'' Hot, molten planets, however, are generally expected to exhibit spectra consistent with blackbody emission, with any surface features strongly attenuated \cite[e.g.,][]{hu2012}. The optimal regime for surface characterization, therefore, is a rocky M-dwarf exoplanet that is hot enough to emit strongly in the infrared, but not so hot that its dayside becomes molten.

\subsubsection*{Planetary albedo and brightness temperature ratio}

Following methods outlined in Ref. \cite{ParkCoy2024}, we fit for the ``brightness temperature ratio'' $\mathcal{R}$, defined as the disk-averaged brightness temperature compared to that expected of a zero-albedo blackbody: $\mathcal{R} = T_p/T_{\rm{max}}$ with $T_{\rm{max}}= T_{\rm{eff}}/\sqrt{a/r_s}\left(\frac{2}{3}\right)^{1/4}$ where $T_{\rm{eff}}$ is the effective temperature of the star and $a/r_s$ is the semi-major axis to star radius ratio.
The MIRI/LRS wavelength range captures $\sim30\%$ of LHS 3844\,b's thermal emission. Combining the MIRI/LRS observations with the Spitzer data increases the coverage to $\sim50\%$ of the distribution, thereby providing more stringent constraints on the planet's Bond albedo than Spitzer alone. For the planetary and orbital parameters ($r_p/r_s$ and $a/r_s$), we use the posteriors from our adopted joint Spitzer and JWST white light curve fit. We furthermore use the observed stellar flux, assume Gaussian priors on stellar parameters from Ref. \cite{Vanderspek2019}, and fix the stellar metallicity to solar. We find that $\mathcal{R}=0.93^{+0.04}_{-0.05}$ using the white light eclipse depth and $\mathcal{R}=0.96^{+0.03}_{-0.04}$ fitting the eclipse spectrum combined with the measured Spitzer eclipse depth (see Tab. \ref{tab:T_AB_R}).  Both values are consistent with the Spitzer value of $\mathcal{R}=0.995\pm0.034$ from Ref. \cite{ParkCoy2024} within $2\,\sigma$.

Our comparably large uncertainty on $\mathcal{R}$, despite the additional data, is primarily due to the higher uncertainties in $a/r_s$. In our analysis, we exclusively used the JWST and TESS data to determine this orbital parameter, whereas Ref. \cite{Vanderspek2019} used a Gaussian prior on the stellar density derived from empirical relationships.
Assuming that the planet is indeed airless, this $\mathcal{R}$ corresponds to a best-fit Bond albedo of $A_{B}=0.14^{+0.13}_{-0.14}$, which is consistent with estimates of the Moon and Mercury of 0.11 and 0.088, respectively \cite{mallama2002,buratti1996,mallama2017spherical}. We note that this approach neglects the effects of infrared thermal beaming due to surface roughness (e.g., Ref.  \cite{emery98}, \cite{rozitis2011}, \cite{delbo2015}, \cite{davidsson2015}, \cite{wohlfarth2023}) which would cause the planet to appear hotter at opposition and may lead to a slight underestimation of the true Bond albedo.

\begin{table}[!h]
\centering
\caption{\textbf{Derived planetary parameters from the planets' emission.} $T_p$ is the planet's brightness temperature, $A_B$ the Bond albedo and $\mathcal{R}$ the brightness ratio.}
\label{tab:T_AB_R}
\renewcommand{\arraystretch}{1.25}
\begin{tabular}{l|cc}\hline
              & \begin{tabular}[c]{@{}c@{}}JWST (MIRI/LRS)\\1 bin\end{tabular}          
              & \begin{tabular}[c]{@{}c@{}}JWST (MIRI/LRS) and Spitzer \\12 bins + 1 bin\end{tabular} \\\hline
$T_p$ (K)     & $984.9^{+16.4}_{-17.4}$ & $999.6^{+15.5}_{-14.3}$ \\
$A_B$         & $0.22^{+0.15}_{-0.14}$  & $0.14^{+0.13}_{-0.14}$  \\
$\mathcal{R}$ & $0.93^{+0.04}_{-0.05}$  & $0.96^{+0.03}_{-0.04}$   \\\hline
\end{tabular}
\end{table}

\subsubsection*{Bare rock surfaces --- darkening agents and space weathering}

We model LHS 3844\,b's emission spectrum using the model developed in Ref. \cite{lyu2024}, which simulates the 3D temperature profile (latitude, longitude, and depth) of an airless planet. The model takes into account instellation, reflection, thermal emission, and heat conduction in the subsurface. The model's radiative transfer assumes a surface made of regolith, like the Moon and Mercury, whose bi-directional reflectance and emission are modeled following Ref. \cite{hapke2012}. Following Ref. \cite{lyu2024}, we assume LHS 3844\,b is in 1:1 synchronous rotation and set tidal heating to zero.

\begin{figure}[h]
\centering
\includegraphics[width=0.8\textwidth]{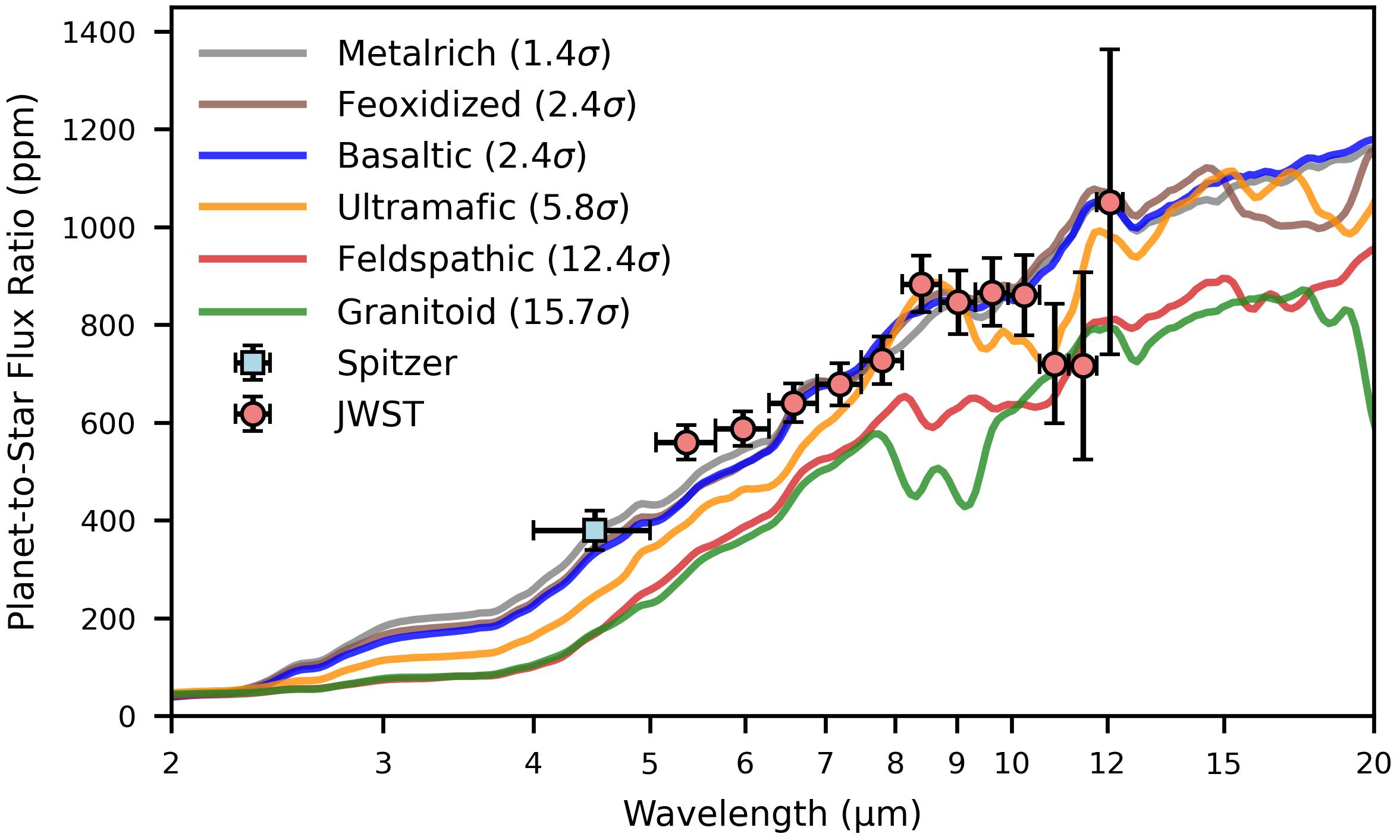}
\caption{\textbf{The measured planet-to-star flux ratio as a function of wavelength compared to a range of fresh powdered surfaces.} Same as Figure \ref{fig:surface} but showing the models presented in Ref. \cite{lyu2024}.}
\label{fig:fresh_powder}
\end{figure}

First, we simulate spectra for different geologically fresh surfaces, using data of powdered materials that were compiled in Ref. \cite{hu2012}. Comparing the observations with the fresh powdered surfaces, we find that the metal-rich surface (a 50\% nanophase hematite and 50\% basalt mixture, \cite{hu2012}) is consistent with the observations ($<2\sigma$), Feoxidized and basaltic surfaces are disfavored ($2-3\sigma$), and the others (utramafic, feldspathic, and granitoid) are ruled out ($>5\sigma$) (see Fig. \ref{fig:fresh_powder}) (we refer to Ref. \cite{hu2012} for a more detailed explanation of these surfaces).

In the Solar System, several agents reduce surface albedo and enhance spectral emissivity, most notably the production of nanophase iron through space weathering and also carbon. Nanophase iron (npFe) is a collective term for submicroscopic metallic iron particles ranging from a few to several hundred nanometers \cite{denevi2023}. It forms in mineral rims and agglutinates and is well documented on the Moon and inferred on Mercury, asteroids, and dwarf planets \cite{hapke2001, domingue2014, pieters2016,denevi2023}. Mercury's proximity to the Sun suggests enhanced space weathering, which may also affect LHS 3844\,b. Carbon darkens C-type asteroid Ryugu \cite{yada2022} and contributes to Mercury's low-reflectance material. While cometary or micrometeoritic delivery has been proposed \cite{syal2015}, the geological context of Mercury favors an endogenous origin from a primordial flotation crust \cite{peplowski2016}, though exogenic sources may remain relevant for LHS 3844\,b. Other, less widespread darkening agents include ilmenite, sulfur, and volcanic or impact glasses.

We next investigate how these darkening agents modify the surfaces. We use the same approach as in Ref. \cite{lyu2024} and simulate space weathering with iron nanoparticles or additional carbon, adding up to $\sim5$\% darkening material to the fresh surface. For iron nanoparticles, this represents the extreme case where all surface iron (assuming an iron abundance similar to that of Earth's crust, Mars, or the Moon) is converted to metallic nanophase particles. For carbon, this limit approximates the amount inferred in Mercury's darkest regions \cite{Klima2018}. While particle size can affect near-infrared spectral slopes, we adopt a first-order approximation with only a single size of particles, similar to Ref. \cite{lyu2024}.

\begin{figure}
    \centering
    \includegraphics[width=0.999\linewidth]{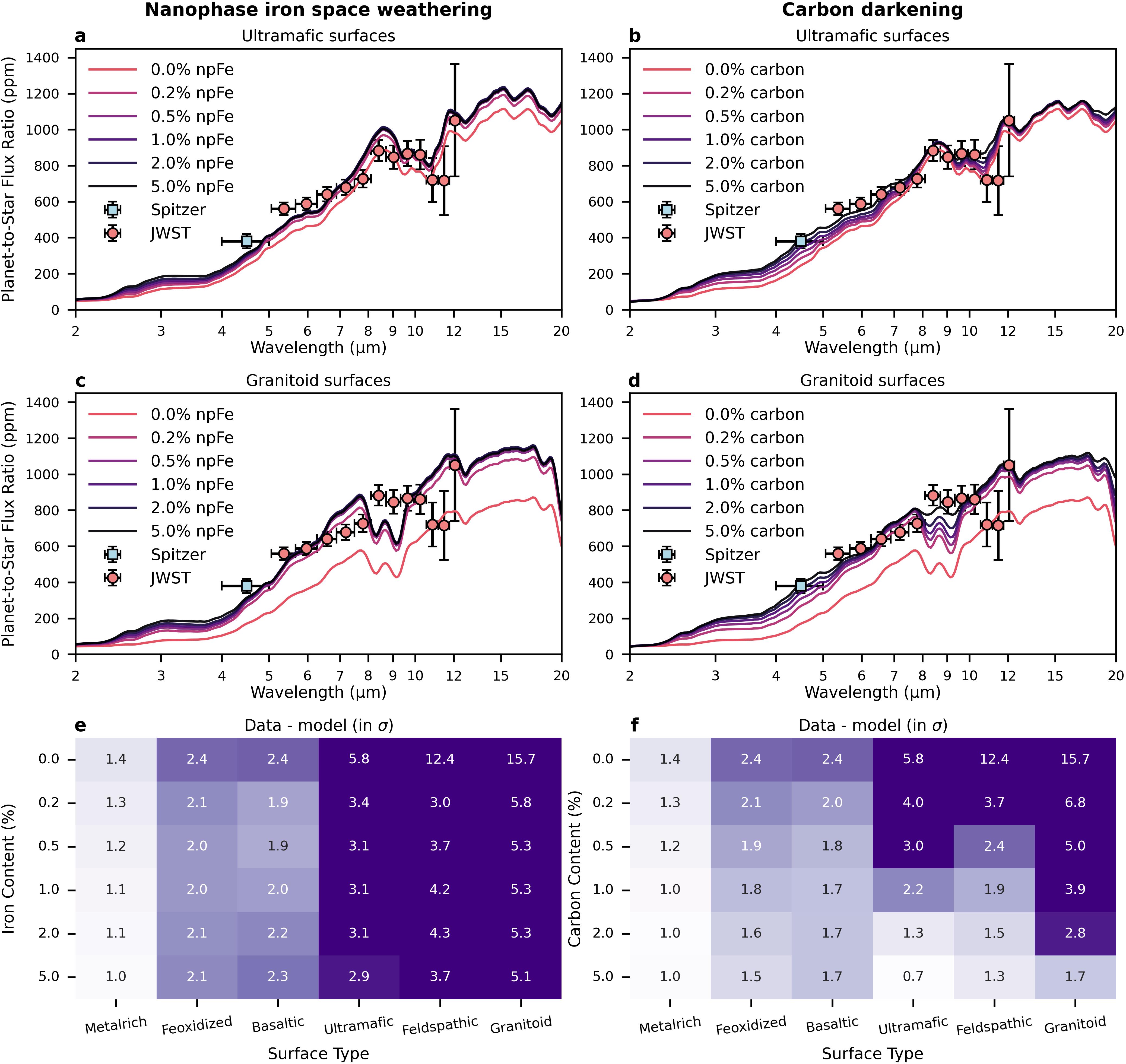}
    \caption{\textbf{A comparison between the observations and a range of weathered powdered surfaces.} The observations have been taken using Spitzer (blue square-shaped marker) and JWST (red circle-shaped markers). The models are from Ref. \cite{lyu2024}. The effect of nanophase iron (npFe) space weathering on ultramafic (granitoid) surfaces can be seen in panel \textbf{a} (\textbf{b}). In panels \textbf{a} and \textbf{c}, we show models with no weathering (i.e., a fresh, powdered surface) for an ultramafic and granitoid surface, up to our maximum considered weathering. The darkest model has the highest weathering at 5\% neFe. Analogous to panel \textbf{b} (\textbf{d}) for carbon darkening.  Panels \textbf{c} and \textbf{f}, show the difference between the model-predicted flux and the observed flux in terms of sigmas. All fresh surfaces are shown in Figure \ref{fig:fresh_powder}.}
    \label{fig:weathering}
\end{figure}

Figure \ref{fig:weathering} shows the impact of the darkening agents on ultramafic and granitoid surfaces and a rejection significance plot for all surfaces from Ref. \cite{lyu2024}. Both iron nanoparticles and carbon lower the surface’s albedo and thus improve the match to the observations. Space weathering via iron nanoparticles better preserves the surface’s infrared spectral features, especially noticeable for granitoid surfaces around 8--10 \textmu m (see panel c in Fig. \ref{fig:weathering}), whereas all spectral features get increasingly smoothed by large amounts of carbon (see panel d in Fig. \ref{fig:weathering}). 

We therefore find that moderate (0.5\%) to extreme (5\%) amounts of space weathering by iron nanoparticles typically improves the fit between the powdered surfaces and the observations. As high levels of carbon darkening make the predicted planetary emission as a function of wavelength comparable to that of a low albedo black body, we also find good agreement between our observations and those models. Together, these results indicate that our observations are well fit by an old, darkened surface.

\subsubsection*{Bare Rock surfaces --- A New Spectral Library}

We modeled a variety of emission spectra spanning different igneous rock compositions and surface textures (solid slab, coarsely crushed, and powdered) using the recently published spectral library in Ref. \cite{Paragas2025} and its implementation into version 6.3 of the open-source retrieval code PLATON. More details on PLATON can be found in Refs. \cite{Zhang2019, Zhang2020, Zhang2025, Paragas2025}. This library uses measurements of hemispherical reflectance rather than bidirectional reflectance. These models assign a single temperature to the dayside of the planet that is calculated using energy balance with a 1D correction factor to account for how the incident stellar energy is redistributed across the dayside. Additional modeling details can be found in Ref. \cite{Paragas2025}.

A number of surface compositions and textures are consistent ($<2\sigma$)  with the measured emission spectrum (see Fig. \ref{fig:textures_vs_sio2}). Notably, the observations cannot be matched by powders because their high reflectances consistently underpredict the planetary flux. The data prefer either slab or coarsely crushed textures of Fe-oxidized, ultramafic ($\leq$45\% SiO$_2$), mafic (45--52\% SiO$_2$), or intermediate  (52--63\% SiO$_2$) compositions. There is a very slight preference for ultramafic samples overall, which is driven by the two data points around 11 \textmu m. This feature is consistent with where one might expect the Si-O stretching feature of an ultramafic rock (rich in olivine and/or pyroxene with a generally low SiO$_2$ content) to occur in wavelength space. However, the uncertainties in the spectrum are quite high longward of 10 \textmu m, so additional observations are necessary to robustly pin the surface composition down.

\subsubsection*{Bare Rock surfaces --- RELAB comparison}

In addition to the above analyses, we searched the RELAB Spectral Database\footnote{\url{https://sites.brown.edu/relab/relab-spectral-database/}} for possible geological materials that match the observed emission spectrum.  We followed the methods outlined in Refs. \cite{ParkCoy2024,hammond25} to compute model emission spectra from laboratory-measured bidirectional reflectance spectra by first converting to single scattering albedo and subsequently using a 0-D energy balance model. 

\begin{figure}
    \centering
    \includegraphics[width=0.8\linewidth]{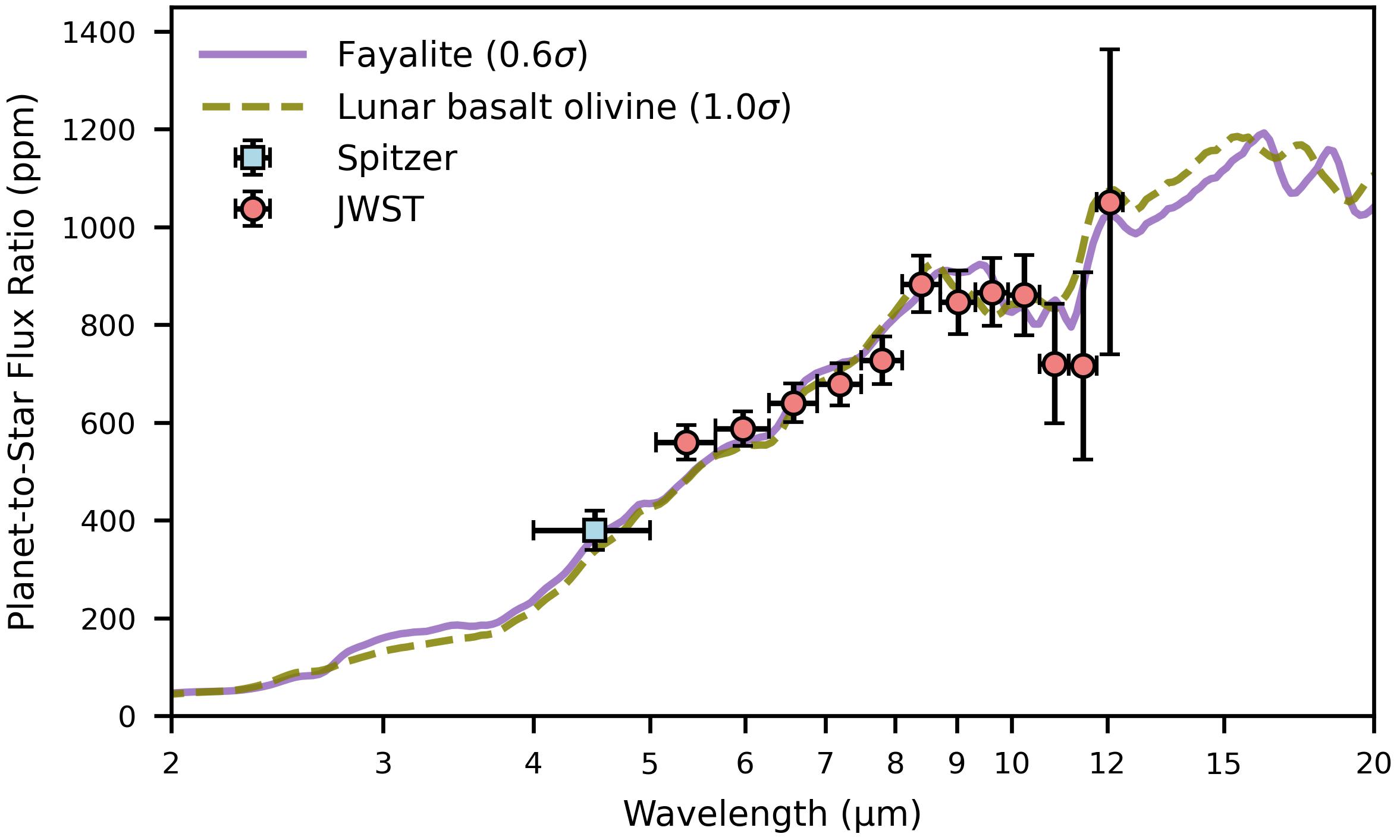}
    \caption{\textbf{The measured planet-to-star flux ratio as a function of wavelength compared to a range of surfaces from the RELAB database.} Same as Figure \ref{fig:surface} but showing a selection of RELAB models.}
    \label{fig:RELAB}
\end{figure}

Similar to the above analyses, we find that pure olivine or olivine-rich rocks and meteorites generally provide the best match to the shape of the measured emission spectrum between 8--12 \textmu m.
The measured wavelength-dependent emissivity in this region of the spectrum qualitatively resembles that of olivine-rich regions on the Moon (e.g., Ref. \cite{sprague1992}) and Mars (e.g., Ref. \cite{ruff2022}).  
In Fig. \ref{fig:RELAB}, we show a selection of models among multiple olivine-rich surfaces, including fayalite ($0.6\sigma$, Sample ID: DD-MDD-101) and an olivine-rich lunar basalt sample ($1.0\sigma$, Sample ID: LR-CMP-169) which fit the data well. 

\subsubsection*{CO$_2$/O$_2$/SO$_2$ atmospheres}

LHS 3844\,b orbits extremely close to its host star (at about 0.006 AU) and currently experiences roughly 4000$\times$ more XUV flux than modern-day Earth \cite{diamond21}, making the existence of a permanent thick atmosphere difficult.  According to Ref. \cite{tian2009}, a Venus-like (92 bar CO$_2$) atmosphere could be lost to thermal escape in as little as 30 Myr, and thus extremely vigorous outgassing would be required to sustain a thick atmosphere. In addition, the Spitzer phase curve confidently rules out atmospheres $\gtrsim10$ bar, as thicker atmospheres should exhibit detectable heat redistribution to the nightside \cite{Kreidberg2019}. 

However, transient outgassing events have the ability to produce thin atmospheres that may still be detectable within JWST's capabilities.  For example, the potentially carbon-rich and time-variable atmosphere of 55 Cancri e is difficult to explain without such transient outgassing \cite{heng2023}.  Recent modeling has also suggested that N$_2$/O$_2$-dominated atmospheres may prevent significant thermal loss even for extremely high XUV fluxes \cite{nakayama22}, giving some hope for the presence of a secondary atmosphere on LHS 3844\,b. 

MIRI/LRS has great potential for detecting or providing stringent upper limits on a variety of common outgassed species, including SO$_2$, H$_2$O, and CH$_4$ \cite{whittaker22}.  Of these, SO$_2$ shows the most distinct spectral features amenable to calculation of upper-limits on concentration. H$_2$O and CH$_4$ are also highly susceptible to photolytic destruction and escape to space. SO$_2$ is the third-most outgassed species on modern Earth and is expected to be a major contributor for rocky planets with oxidized mantles \cite{LICHTENBERG2025}. SO$_2$ has also been proposed as a tracer of tidal heating on rocky exoplanets \cite{seligman2024} (inspired by its ubiquity in Io's volcanism). Given LHS 3844\,b's extremely close-in orbit, a small but non-zero eccentricity could lead to significant tidal heating that promotes volcanic activity.

Here, we perform grid forward modeling of potential atmospheric scenarios for LHS 3844\,b using the radiative-convective equilibrium code \texttt{HELIOS 3.0} \cite{malik2017helios,Malik2019,whittaker22}.  We focus on two scenarios indicative of recent outgassing events; trace SO$_2$ with a CO$_2$-dominated background and trace SO$_2$ with an O$_2$-dominated background. We follow the approach of Ref. \cite{whittaker22}, using a low-resolution ($R=250$) SPHINX stellar spectrum to calculate the initial temperature-pressure profile, then using post-processing mode with a high-resolution ($R=4000$) SPHINX stellar spectrum to calculate the emergent planetary flux.  We then used the observed stellar spectrum to convert the emergent \texttt{HELIOS} flux (in W/m$^{2}$) to an eclipse spectrum.
We model pressures of [0.01, 0.1, 1, 10, 100, 1,000] mbar for SO$_2$ mixing ratios of [1, 10, 100, 1,000, 10,000, 100,000] ppm ($0.0001-10\%$).  Below this pressure range, SO$_2$ absorption features begin to disappear almost completely, and atmospheres above this pressure range are highly inconsistent with the data for our given SO$_2$ fractions. For the underlying rock, we do not assume any particular surface composition, but instead use a blackbody curve. For that, we use the brightness temperature we have determined in our joint Spitzer and JWST fit, $999.6^{+15.5}_{-14.3}$ K (see Fig. \ref{fig:surface}). To avoid biasing the search for CO$_2$ or SO$_2$ signatures by the exact assumption of the underlying emission, we implement a grid-like sampling approach of the temperature and perform a $\chi^2$ minimization. The results of our sampling across different surface pressures and relative SO$_2$ contents are shown in panels e and f of Figure \ref{fig:atmosphere}, with some exemplary spectra in panels a--d.

We confidently rule out CO$_2$-dominated atmospheres $\geq100\,$mbar (5$\sigma$) and any atmospheres thicker than 1 mbar with appreciable amounts of SO$_2$ ($>1\%$) (3$\sigma$).  This is consistent with the Spitzer phase curve that suggested a thin or no atmosphere.

\subsubsection*{Constraints on the crust}

The spectra of LHS 3844\,b are comparatively dark and broadly consistent with an (ultra)mafic silicate mineralogy rich in pyroxene and olivine. While additional data are needed to confirm specific mineralogical features, several implications for the crust and its evolution naturally follow.

The Christiansen feature (CF; emissivity maximum near 8 \textmu m) of silicates shifts systematically with SiO$_2$ content, from shorter wavelengths ($\sim7.5$--8.5 \textmu m) for felsic to longer wavelengths ($\sim8.5$--9.0 \textmu m) for mafic materials \cite{Salisbury1989,Walter1989,hu2012}. Although its exact position cannot be resolved with the present spectral resolution, the lower planet-to-star flux at 7.8 \textmu m relative to 8.5 \textmu m implies that the CF lies above 7.8 \textmu m, ruling out a very SiO$_2$-rich feldspathic surface. This is generally tentative and can be clearer seen in Figure \ref{fig:textures_vs_sio2}. Together with spectral fits, this supports an (ultra)mafic composition, dominated by pyroxene and olivine.

The SiO$_2$ content of crust is a rough index of magma-production environment. Partial melting of solid planetary mantle usually forms low-SiO$_2$ magmas that crystallize to form pyroxene-rich crusts (pyroxene has stochiometric (Mg+Fe)/Si = 1) \cite{bvtp1981,pc2009}. SiO$_2$-poor mantles (i.e., higher (Mg+Fe)/Si)), and/or high mantle temperatures, increase the olivine abundance of crust (olivine has stochiometric (Mg+Fe)/Si = 2). Partial melting of olivine-rich planetary mantle produces melts that are richer in SiO$_2$ than the mantle (this melt buoyantly ascends to form the crust), and a correspondingly SiO$_2$-impoverished residuum/restite which remains in the mantle. As a result, the median magma from the modern mantle of Earth is rich in pyroxene and has little olivine, even though Earth's upper mantle is mostly olivine. In other words, the median magma from the modern mantle of Earth is more SiO$_2$-rich than the mantle, while still overall being SiO$_2$-poor.  However, in the limit of 100\% melting, the SiO$_2$ content of the mantle and melt are identical. Thus, higher melt fractions for olivine-rich mantles lead to a correspondingly lower SiO$_2$ content in the melt. ``Mafic" (magnesium and iron rich, rich in pyroxene and/or olivine) igneous rocks usually appear black. High-SiO$_2$ magmas ($>57$ wt\% SiO$_2$) form from a range of processes, including partial melting of previously produced lower-SiO$_2$ crusts, and crystallize to form olivine-absent, commonly white, gray, or pink ``felsic" rocks. 

A feldspathic crust without space weathering would have a high albedo, inconsistent with the data, and there is no evidence for high-Si ash, which also reflects strongly. If LHS 3844\,b’s dayside is indeed Si-poor, then processes analogous to those producing Earth’s high-Si continental crust were likely inefficient. For a thick crust, partial melting at its base would generate Si-rich magmas \cite{Lee2025}. Whether such melts could ascend depends on the ability of dense, warm but subsolidus basalt (and therefore more viscous) to deform or founder, allowing buoyant high-Si magmas to reach the surface. As olivine is a proxy for high (Mg+Fe)/Si, any high olivine content of LHS 3844\,b's surface could in turn correspond to a high Mg/Si mantle. The star is old, so its Mg/Si is high according to \cite{adibekyan2015}. A high Mg/Si mantle has lower viscosity (e.g. \cite{spaargaren2020}), allowing volcanic activity to continue for longer than would be the case for a low Mg/Si.

The dip in emission near 11 \textmu m is unexpectedly strong, and its origin remains uncertain; however, given the comparably large error bars at the red end of MIRI/LRS, it can very likely be just a statistical coincidence. One possibility is that it is a silicate spectral feature, though more evidence is needed to constrain the band’s shape and depth. This wavelength range overlaps the Reststrahlen bands and transparency features of silicates and may correspond to the pronounced transparency feature band of an olivine-rich, low-SiO$_2$ crust. Laboratory and simulation studies of lunar and Mercury analogs show that spectral bands can shift and deepen at higher temperatures \cite{donaldsonHanna2017,Ferrari2020,Prem2022}. Given the extreme conditions on LHS 3844\,b, temperature effects not captured in current laboratory data cannot be excluded.

The darkness of the surface can be explained by space weathering, a low-SiO$_2$ crust, and/or an ancient carbon flotation crust \cite{peplowski2016,keppler2019}. The spectral fits of nanophase iron appear to be most in line with the measurements. 

Laboratory experiments and thermodynamic databases can reproduce the observed diversity of Solar System igneous crustal compositions \cite{bvtp1981, young2003, wilson1989, winter2014, taylormclennan2009}. There is no evidence for exotic planetary compositions (e.g., carbides) in white dwarf data \cite{Doyle2023}, so this terrestrial work is a useful touchstone for interpreting exoplanet data. However, the mapping between crust composition and planetary tectonic mode (i.e., the tectonic drivers of magmatism) is still uncertain.

\bmhead{Acknowledgments} 
We thank Matej Malik for his contributions to this project. 
This work is based on observations made with the NASA/ESA/CSA James Webb Space Telescope. The data were obtained from the Mikulski Archive for Space Telescopes at the Space Telescope Science Institute, which is operated by the Association of Universities for Research in Astronomy, Inc., under NASA contract NAS 5-03127 for JWST. These observations are associated with program \#1846.
This paper includes data collected with the TESS mission, obtained from the MAST data archive at the Space Telescope Science Institute (STScI). Funding for the TESS mission is provided by the NASA Explorer Program. STScI is operated by the Association of Universities for Research in Astronomy, Inc., under NASA contract NAS 5-26555.
This research has made use of the NASA Exoplanet Archive, which is operated by the California Institute of Technology, under contract with the National Aeronautics and Space Administration under the Exoplanet Exploration Program.\\
S.Z. was supported by NASA through the NASA Hubble Fellowship grant \#HST-HF2-51570.001-A awarded by the Space Telescope Science Institute, which is operated by the Association of Universities for Research in Astronomy, Incorporated, under NASA contract NAS5- 26555.
X.L. and D.D.B.K. acknowledge support from NSFC research grants \#42250410318 and \#12473064.
B.P.C. acknowledges support from a NASA grant awarded to the Illinois/NASA Space Grant Consortium.
E.A.W acknowledges support from the Eugene V. Cota-Robles Award and the Nathan P. Myhrvold Graduate Fellowship.
This research utilizes spectra acquired by Martha Schaefer and Carle M. Pieters with the NASA RELAB facility at Brown University.
Part of this research was carried out at the Jet Propulsion Laboratory, California Institute of Technology, under a contract with the National Aeronautics and Space Administration (80NM0018D0004). Part of the high-performance computing resources used in this investigation were provided by funding from the JPL Information and Technology Solutions Directorate.

\section*{Author Contributions}
All authors played a significant role in the creation of this manuscript. It includes steps starting with the writing and submission of the JWST GO proposals, up to the submission of this manuscript. Some of those contributions are listed here.
S.Z. led the paper writing on this manuscript.
L.K. is the principal investigator and R.H. is the co-principal investigator of JWST program GO1846.
Other co-Is of JWST program GO1846 are EK, DDBK, CM, LS, and EW.
SZ and ABA have contributed to the manuscript with a data reduction and analysis of the JWST observations.
BPC, AI, XL, and KP have contributed with models of the planet's or stellar emission.
Additionally, KW and EK have contributed significantly to the writing of this manuscript.
RH, EW, HK, and RW have provided significant feedback to the manuscript.

\section*{Data availability}
The JWST data used in this publication were collected as part of the GO program 1846 and have the default proprietary period of one year. All observations (\#1, \#102, and \#3) are therefore publicly accessible on the Mikulski Archive for Space Telescopes (MAST) already.

\section*{Code availability}
We used the following codes, resources, and \texttt{python} packages to reduce, analyze, and interpret our data: \texttt{numpy}\cite{numpy2020}, \texttt{matplotlib}\cite{matplotlib2007}, \texttt{astropy}\cite{astropy2022}, \texttt{batman}\cite{Kreidberg2015}, \texttt{Eureka!}\cite{Bell2022}, \texttt{jwst}\cite{Bushouse2022}, \texttt{emcee}\cite{ForemanMackey2013},  \texttt{dynesty}\cite{Speagle2020, Koposov2023}, \texttt{pysynphot} \cite{STScI2013}, \texttt{PLATON} \cite{Zhang2019, Zhang2020, Zhang2025, Paragas2025}.
Everything used to make this paper can be obtained for unrestricted further use by contacting the lead author.

\section*{Competing Interests}
The authors declare no competing interests.


\begin{thebibliography}{100}
\expandafter\ifx\csname url\endcsname\relax
  \def\url#1{\burl{#1}}\fi
\expandafter\ifx\csname urlprefix\endcsname\relax\def\urlprefix{URL }\fi
\providecommand{\bibinfo}[2]{#2}
\providecommand{\eprint}[2][]{\url{#2}}
\providecommand{\doi}[1]{\url{https://doi.org/#1}}
\bibcommenthead

\bibitem{ParkCoy2024}
\bibinfo{author}{{Coy}, B.~P.} \emph{et~al.}
\newblock \bibinfo{title}{{Population-level Hypothesis Testing with Rocky Planet Emission Data: A Tentative Trend in the Brightness Temperatures of M-Earths}}.
\newblock \emph{\bibinfo{journal}{Astrophys. J.}} \textbf{\bibinfo{volume}{987}}, \bibinfo{pages}{22} (\bibinfo{year}{2025}).

\bibitem{Kreidberg2025}
\bibinfo{author}{{Kreidberg}, L.} \& \bibinfo{author}{{Stevenson}, K.~B.}
\newblock \bibinfo{title}{{A first look at rocky exoplanets with JWST}}.
\newblock \emph{\bibinfo{journal}{Proc. Nat. Acad. Sci. USA}} \textbf{\bibinfo{volume}{122}}, \bibinfo{pages}{e2416190122} (\bibinfo{year}{2025}).

\bibitem{hu2012}
\bibinfo{author}{{Hu}, R.}, \bibinfo{author}{{Ehlmann}, B.~L.} \& \bibinfo{author}{{Seager}, S.}
\newblock \bibinfo{title}{{Theoretical Spectra of Terrestrial Exoplanet Surfaces}}.
\newblock \emph{\bibinfo{journal}{Astrophys. J.}} \textbf{\bibinfo{volume}{752}}, \bibinfo{pages}{7} (\bibinfo{year}{2012}).

\bibitem{Vanderspek2019}
\bibinfo{author}{{Vanderspek}, R.} \emph{et~al.}
\newblock \bibinfo{title}{{TESS Discovery of an Ultra-short-period Planet around the Nearby M Dwarf LHS 3844}}.
\newblock \emph{\bibinfo{journal}{Astrophys.~J.~Lett.}} \textbf{\bibinfo{volume}{871}}, \bibinfo{pages}{L24} (\bibinfo{year}{2019}).

\bibitem{Kreidberg2019}
\bibinfo{author}{{Kreidberg}, L.} \emph{et~al.}
\newblock \bibinfo{title}{{Absence of a thick atmosphere on the terrestrial exoplanet LHS 3844b}}.
\newblock \emph{\bibinfo{journal}{Nature}} \textbf{\bibinfo{volume}{573}}, \bibinfo{pages}{87--90} (\bibinfo{year}{2019}).

\bibitem{whittaker22}
\bibinfo{author}{{Whittaker}, E.~A.} \emph{et~al.}
\newblock \bibinfo{title}{{The Detectability of Rocky Planet Surface and Atmosphere Composition with the JWST: The Case of LHS 3844b}}.
\newblock \emph{\bibinfo{journal}{Astron. J.}} \textbf{\bibinfo{volume}{164}}, \bibinfo{pages}{258} (\bibinfo{year}{2022}).

\bibitem{lyu2024}
\bibinfo{author}{{Lyu}, X.} \emph{et~al.}
\newblock \bibinfo{title}{{Super-Earth LHS3844b is Tidally Locked}}.
\newblock \emph{\bibinfo{journal}{Astrophys. J.}} \textbf{\bibinfo{volume}{964}}, \bibinfo{pages}{152} (\bibinfo{year}{2024}).

\bibitem{Bell2022}
\bibinfo{author}{{Bell}, T.} \emph{et~al.}
\newblock \bibinfo{title}{{Eureka!: An End-to-End Pipeline for JWST Time-Series Observations}}.
\newblock \emph{\bibinfo{journal}{J.~Open~Source~Softw.}} \textbf{\bibinfo{volume}{7}}, \bibinfo{pages}{4503} (\bibinfo{year}{2022}).

\bibitem{Greene2023}
\bibinfo{author}{{Greene}, T.~P.} \emph{et~al.}
\newblock \bibinfo{title}{{Thermal emission from the Earth-sized exoplanet TRAPPIST-1 b using JWST}}.
\newblock \emph{\bibinfo{journal}{Nature}} \textbf{\bibinfo{volume}{618}}, \bibinfo{pages}{39--42} (\bibinfo{year}{2023}).

\bibitem{Zieba2023}
\bibinfo{author}{{Zieba}, S.} \emph{et~al.}
\newblock \bibinfo{title}{{No thick carbon dioxide atmosphere on the rocky exoplanet TRAPPIST-1 c}}.
\newblock \emph{\bibinfo{journal}{Nature}} \textbf{\bibinfo{volume}{620}}, \bibinfo{pages}{746--749} (\bibinfo{year}{2023}).

\bibitem{fauchez2025stellar}
\bibinfo{author}{{Fauchez}, T.~J.} \emph{et~al.}
\newblock \bibinfo{title}{{Stellar Models Also Limit Exoplanet Atmosphere Studies in Emission}}.
\newblock \emph{\bibinfo{journal}{Astrophys. J.}} \textbf{\bibinfo{volume}{989}}, \bibinfo{pages}{170} (\bibinfo{year}{2025}).

\bibitem{Paragas2025}
\bibinfo{author}{{Paragas}, K.} \emph{et~al.}
\newblock \bibinfo{title}{{A New Spectral Library for Modeling the Surfaces of Hot, Rocky Exoplanets}}.
\newblock \emph{\bibinfo{journal}{Astrophys. J.}} \textbf{\bibinfo{volume}{981}}, \bibinfo{pages}{130} (\bibinfo{year}{2025}).

\bibitem{Campbell1983}
\bibinfo{author}{{Campbell}, I.~H.} \& \bibinfo{author}{{Taylor}, S.~R.}
\newblock \bibinfo{title}{{No water, no granites-no oceans, no continents}}.
\newblock \emph{\bibinfo{journal}{Geophys. Res. Lett.}} \textbf{\bibinfo{volume}{10}}, \bibinfo{pages}{1061--1064} (\bibinfo{year}{1983}).

\bibitem{Taylor1989}
\bibinfo{author}{{Taylor}, S.~R.}
\newblock \bibinfo{title}{{Growth of Planetary Crusts}}.
\newblock \emph{\bibinfo{journal}{Tectonophysics}} \textbf{\bibinfo{volume}{161}}, \bibinfo{pages}{147--156} (\bibinfo{year}{1989}).

\bibitem{mansfield2019}
\bibinfo{author}{{Mansfield}, M.} \emph{et~al.}
\newblock \bibinfo{title}{{Identifying Atmospheres on Rocky Exoplanets through Inferred High Albedo}}.
\newblock \emph{\bibinfo{journal}{Astrophys. J.}} \textbf{\bibinfo{volume}{886}}, \bibinfo{pages}{141} (\bibinfo{year}{2019}).

\bibitem{hammond25}
\bibinfo{author}{{Hammond}, M.} \emph{et~al.}
\newblock \bibinfo{title}{{Reliable Detections of Atmospheres on Rocky Exoplanets with Photometric JWST Phase Curves}}.
\newblock \emph{\bibinfo{journal}{Astrophys.~J.~Lett.}} \textbf{\bibinfo{volume}{978}}, \bibinfo{pages}{L40} (\bibinfo{year}{2025}).

\bibitem{Lee2025}
\bibinfo{author}{{Lee}, C.-T.} \emph{et~al.}
\newblock \bibinfo{title}{{Crustal thickness effects on chemical differentiation and hydrology on Mars}}.
\newblock \emph{\bibinfo{journal}{Earth~Planet.~Sci.~Lett.}} \textbf{\bibinfo{volume}{651}}, \bibinfo{pages}{119155} (\bibinfo{year}{2025}).

\bibitem{Phillips2025}
\bibinfo{author}{Phillips, M.~S.} \emph{et~al.}
\newblock \bibinfo{title}{Widespread ancient anorthosites in the lower crust of {Mars}}.
\newblock \emph{\bibinfo{journal}{Commun.~Earth~Environ.}} \textbf{\bibinfo{volume}{6}}, \bibinfo{pages}{1026} (\bibinfo{year}{2025}).
\newblock \urlprefix\url{https://doi.org/10.1038/s43247-025-03004-7}.

\bibitem{Widemann2023}
\bibinfo{author}{{Widemann}, T.} \emph{et~al.}
\newblock \bibinfo{title}{{Venus Evolution Through Time: Key Science Questions, Selected Mission Concepts and Future Investigations}}.
\newblock \emph{\bibinfo{journal}{Space Sci. Rev.}} \textbf{\bibinfo{volume}{219}}, \bibinfo{pages}{56} (\bibinfo{year}{2023}).

\bibitem{cassidy1975}
\bibinfo{author}{{Cassidy}, W.} \& \bibinfo{author}{{Hapke}, B.}
\newblock \bibinfo{title}{{Effects of Darkening Processes on Surfaces of Airless Bodies}}.
\newblock \emph{\bibinfo{journal}{Icarus}} \textbf{\bibinfo{volume}{25}}, \bibinfo{pages}{371--383} (\bibinfo{year}{1975}).

\bibitem{Pieters2000}
\bibinfo{author}{{Pieters}, C.~M.} \emph{et~al.}
\newblock \bibinfo{title}{{Space weathering on airless bodies: Resolving a mystery with lunar samples}}.
\newblock \emph{\bibinfo{journal}{Meteorit.~Planet.~Sci.}} \textbf{\bibinfo{volume}{35}}, \bibinfo{pages}{1101--1107} (\bibinfo{year}{2000}).

\bibitem{hapke2001}
\bibinfo{author}{{Hapke}, B.}
\newblock \bibinfo{title}{{Space weathering from Mercury to the asteroid belt}}.
\newblock \emph{\bibinfo{journal}{J.~Geophys.~Res.}} \textbf{\bibinfo{volume}{106}}, \bibinfo{pages}{10039--10074} (\bibinfo{year}{2001}).

\bibitem{pieters2016}
\bibinfo{author}{{Pieters}, C.~M.} \& \bibinfo{author}{{Noble}, S.~K.}
\newblock \bibinfo{title}{{Space weathering on airless bodies}}.
\newblock \emph{\bibinfo{journal}{J.~Geophys.~Res.~Planets}} \textbf{\bibinfo{volume}{121}}, \bibinfo{pages}{1865--1884} (\bibinfo{year}{2016}).

\bibitem{Noble2007}
\bibinfo{author}{{Noble}, S.~K.}, \bibinfo{author}{{Pieters}, C.~M.} \& \bibinfo{author}{{Keller}, L.~P.}
\newblock \bibinfo{title}{{An experimental approach to understanding the optical effects of space weathering}}.
\newblock \emph{\bibinfo{journal}{Icarus}} \textbf{\bibinfo{volume}{192}}, \bibinfo{pages}{629--642} (\bibinfo{year}{2007}).

\bibitem{Fleischer1954}
\bibinfo{author}{{Fleischer}, M.}
\newblock \bibinfo{title}{{The abundance and distribution of the chemical elements in the earth's crust}}.
\newblock \emph{\bibinfo{journal}{J.~Chem.~Educ.}} \textbf{\bibinfo{volume}{31}}, \bibinfo{pages}{446} (\bibinfo{year}{1954}).

\bibitem{Klima2018}
\bibinfo{author}{{Klima}, R.~L.}, \bibinfo{author}{{Denevi}, B.~W.}, \bibinfo{author}{{Ernst}, C.~M.}, \bibinfo{author}{{Murchie}, S.~L.} \& \bibinfo{author}{{Peplowski}, P.~N.}
\newblock \bibinfo{title}{{Global Distribution and Spectral Properties of Low-Reflectance Material on Mercury}}.
\newblock \emph{\bibinfo{journal}{Geophys. Res. Lett.}} \textbf{\bibinfo{volume}{45}}, \bibinfo{pages}{2945--2953} (\bibinfo{year}{2018}).

\bibitem{diamond21}
\bibinfo{author}{{Diamond-Lowe}, H.} \emph{et~al.}
\newblock \bibinfo{title}{{The High-energy Spectrum of the Nearby Planet-hosting Inactive Mid-M Dwarf LHS 3844}}.
\newblock \emph{\bibinfo{journal}{Astron. J.}} \textbf{\bibinfo{volume}{162}}, \bibinfo{pages}{10} (\bibinfo{year}{2021}).

\bibitem{nakayama22}
\bibinfo{author}{{Nakayama}, A.}, \bibinfo{author}{{Ikoma}, M.} \& \bibinfo{author}{{Terada}, N.}
\newblock \bibinfo{title}{{Survival of Terrestrial N$_{2}$-O$_{2}$ Atmospheres in Violent XUV Environments through Efficient Atomic Line Radiative Cooling}}.
\newblock \emph{\bibinfo{journal}{Astrophys. J.}} \textbf{\bibinfo{volume}{937}}, \bibinfo{pages}{72} (\bibinfo{year}{2022}).

\bibitem{chatterjee2024}
\bibinfo{author}{{Chatterjee}, R.~D.} \& \bibinfo{author}{{Pierrehumbert}, R.~T.}
\newblock \bibinfo{title}{{Novel Physics of Escaping Secondary Atmospheres May Shape the Cosmic Shoreline}}.
\newblock \emph{\bibinfo{journal}{Astrophys. J.}} \textbf{\bibinfo{volume}{998}}, \bibinfo{pages}{236} (\bibinfo{year}{2026}).

\bibitem{Hu2013}
\bibinfo{author}{{Hu}, R.}, \bibinfo{author}{{Seager}, S.} \& \bibinfo{author}{{Bains}, W.}
\newblock \bibinfo{title}{{Photochemistry in Terrestrial Exoplanet Atmospheres. II. H$_{2}$S and SO$_{2}$ Photochemistry in Anoxic Atmospheres}}.
\newblock \emph{\bibinfo{journal}{Astrophys. J.}} \textbf{\bibinfo{volume}{769}}, \bibinfo{pages}{6} (\bibinfo{year}{2013}).

\bibitem{LICHTENBERG2025}
\bibinfo{author}{{Lichtenberg}, T.} \& \bibinfo{author}{{Miguel}, Y.}
\newblock \bibinfo{title}{{Super-Earths and Earth-like Exoplanets}}.
\newblock \emph{\bibinfo{journal}{Treatise on Geochemistry}} \textbf{\bibinfo{volume}{7}}, \bibinfo{pages}{51--112} (\bibinfo{year}{2025}).

\bibitem{Kaltenegger2010}
\bibinfo{author}{{Kaltenegger}, L.}, \bibinfo{author}{{Henning}, W.~G.} \& \bibinfo{author}{{Sasselov}, D.~D.}
\newblock \bibinfo{title}{{Detecting Volcanism on Extrasolar Planets}}.
\newblock \emph{\bibinfo{journal}{Astron. J.}} \textbf{\bibinfo{volume}{140}}, \bibinfo{pages}{1370--1380} (\bibinfo{year}{2010}).

\bibitem{malik2017helios}
\bibinfo{author}{{Malik}, M.} \emph{et~al.}
\newblock \bibinfo{title}{{HELIOS: An Open-source, GPU-accelerated Radiative Transfer Code for Self-consistent Exoplanetary Atmospheres}}.
\newblock \emph{\bibinfo{journal}{Astron. J.}} \textbf{\bibinfo{volume}{153}}, \bibinfo{pages}{56} (\bibinfo{year}{2017}).

\bibitem{Malik2019}
\bibinfo{author}{{Malik}, M.} \emph{et~al.}
\newblock \bibinfo{title}{{Analyzing Atmospheric Temperature Profiles and Spectra of M Dwarf Rocky Planets}}.
\newblock \emph{\bibinfo{journal}{Astrophys. J.}} \textbf{\bibinfo{volume}{886}}, \bibinfo{pages}{142} (\bibinfo{year}{2019}).

\bibitem{Powell2024}
\bibinfo{author}{{Powell}, D.}, \bibinfo{author}{{Wordsworth}, R.} \& \bibinfo{author}{{{\"O}berg}, K.}
\newblock \bibinfo{title}{{Nightside Clouds on Tidally Locked Terrestrial Planets Mimic Atmosphere-free Scenarios}}.
\newblock \emph{\bibinfo{journal}{Astrophys.~J.~Lett.}} \textbf{\bibinfo{volume}{974}}, \bibinfo{pages}{L4} (\bibinfo{year}{2024}).

\bibitem{Barker1979}
\bibinfo{author}{{Barker}, E.~S.}
\newblock \bibinfo{title}{{Detection of SO$_{2}$ in the UV spectrum of Venus}}.
\newblock \emph{\bibinfo{journal}{Geophys. Res. Lett.}} \textbf{\bibinfo{volume}{6}}, \bibinfo{pages}{117--120} (\bibinfo{year}{1979}).

\bibitem{Encrenaz2018}
\bibinfo{author}{Encrenaz, T.} \& \bibinfo{author}{Coustenis, A.}
\newblock \emph{\bibinfo{title}{Composition and Chemistry of the Atmospheres of Terrestrial Planets: Venus, the Earth, Mars, and Titan}}, \bibinfo{pages}{187--214} (\bibinfo{publisher}{Springer International Publishing}, \bibinfo{address}{Cham}, \bibinfo{year}{2018}).
\newblock \urlprefix\url{https://doi.org/10.1007/978-3-319-55333-7_45}.

\bibitem{Oyama1979}
\bibinfo{author}{{Oyama}, V.~I.}, \bibinfo{author}{{Carle}, G.~C.}, \bibinfo{author}{{Woeller}, F.} \& \bibinfo{author}{{Pollack}, J.~B.}
\newblock \bibinfo{title}{{Venus Lower Atmospheric Composition: Analysis by Gas Chromatography}}.
\newblock \emph{\bibinfo{journal}{Science}} \textbf{\bibinfo{volume}{203}}, \bibinfo{pages}{802--805} (\bibinfo{year}{1979}).

\bibitem{Bezard1993}
\bibinfo{author}{{Bezard}, B.} \emph{et~al.}
\newblock \bibinfo{title}{{The abundance of sulfur dioxide below the clouds of Venus}}.
\newblock \emph{\bibinfo{journal}{Geophys. Res. Lett.}} \textbf{\bibinfo{volume}{20}}, \bibinfo{pages}{1587--1590} (\bibinfo{year}{1993}).

\bibitem{Taylor2014}
\bibinfo{author}{Taylor, F.~W.} \& \bibinfo{author}{Hunten, D.~M.}
\newblock \bibinfo{title}{ in \textit{Chapter 14 - venus: Atmosphere}} \bibinfo{edition}{Third edition} edn, (eds \bibinfo{editor}{Spohn, T.}, \bibinfo{editor}{Breuer, D.} \& \bibinfo{editor}{Johnson, T.~V.}) \emph{\bibinfo{booktitle}{Encyclopedia of the Solar System (Third Edition)}} \bibinfo{pages}{305--322} (\bibinfo{publisher}{Elsevier}, \bibinfo{address}{Boston}, \bibinfo{year}{2014}).
\newblock \urlprefix\url{https://www.sciencedirect.com/science/article/pii/B9780124158450000141}.

\bibitem{dePater2010}
\bibinfo{author}{{de Pater}, I.} \& \bibinfo{author}{{Lissauer}, J.~J.}
\newblock \emph{\bibinfo{title}{{Planetary Sciences}}} \bibinfo{edition}{2} edn (\bibinfo{publisher}{Cambridge University Press}, \bibinfo{address}{Cambridge}, \bibinfo{year}{2010}).

\bibitem{Seinfeld2006}
\bibinfo{author}{Seinfeld, J.~H.} \& \bibinfo{author}{Pandis, S.~N.}
\newblock \emph{\bibinfo{title}{Atmospheric chemistry and physics: from air pollution to climate change.}}  (\bibinfo{publisher}{John Wiley \& Sons, Inc.}, \bibinfo{address}{Hoboken}, \bibinfo{year}{2006}).

\bibitem{Brimblecombe1989}
\bibinfo{author}{Brimblecombe, P.} \& \bibinfo{author}{Lein, A.}
\newblock \emph{\bibinfo{title}{Evolution of the Global Biogeochemical Sulphur Cycle}}  (\bibinfo{publisher}{Wiley}, \bibinfo{address}{Germany}, \bibinfo{year}{1989}).

\bibitem{Loftus2019}
\bibinfo{author}{{Loftus}, K.}, \bibinfo{author}{{Wordsworth}, R.~D.} \& \bibinfo{author}{{Morley}, C.~V.}
\newblock \bibinfo{title}{{Sulfate Aerosol Hazes and SO$_{2}$ Gas as Constraints on Rocky Exoplanets{\textquoteright} Surface Liquid Water}}.
\newblock \emph{\bibinfo{journal}{Astrophys. J.}} \textbf{\bibinfo{volume}{887}}, \bibinfo{pages}{231} (\bibinfo{year}{2019}).

\bibitem{Lellouch1990}
\bibinfo{author}{{Lellouch}, E.}, \bibinfo{author}{{Belton}, M.}, \bibinfo{author}{{de Pater}, I.}, \bibinfo{author}{{Gulkis}, S.} \& \bibinfo{author}{{Encrenaz}, T.}
\newblock \bibinfo{title}{{Io's atmosphere from microwave detection of SO$_{2}$}}.
\newblock \emph{\bibinfo{journal}{Nature}} \textbf{\bibinfo{volume}{346}}, \bibinfo{pages}{639--641} (\bibinfo{year}{1990}).

\bibitem{Lellouch1992}
\bibinfo{author}{{Lellouch}, E.} \emph{et~al.}
\newblock \bibinfo{title}{{The structure, stability, and global distribution of Io's atmosphere}}.
\newblock \emph{\bibinfo{journal}{Icarus}} \textbf{\bibinfo{volume}{98}}, \bibinfo{pages}{271--295} (\bibinfo{year}{1992}).

\bibitem{Tsang2016}
\bibinfo{author}{{Tsang}, C. C.~C.}, \bibinfo{author}{{Spencer}, J.~R.}, \bibinfo{author}{{Lellouch}, E.}, \bibinfo{author}{{Lopez-Valverde}, M.~A.} \& \bibinfo{author}{{Richter}, M.~J.}
\newblock \bibinfo{title}{{The collapse of Io's primary atmosphere in Jupiter eclipse}}.
\newblock \emph{\bibinfo{journal}{J.~Geophys.~Res.~Planets}} \textbf{\bibinfo{volume}{121}}, \bibinfo{pages}{1400--1410} (\bibinfo{year}{2016}).

\bibitem{Lopes2023}
\bibinfo{editor}{{Lopes}, R. M.~C.}, \bibinfo{editor}{{de Kleer}, K.} \& \bibinfo{editor}{{Keane}, J.~T.} (eds).
\newblock \emph{\bibinfo{title}{{Io: A New View of Jupiter's Moon}}}, Vol. \bibinfo{volume}{468} of \emph{\bibinfo{series}{Astrophysics and Space Science Library}} (\bibinfo{year}{2023}).

\bibitem{Pearl1979}
\bibinfo{author}{{Pearl}, J.} \emph{et~al.}
\newblock \bibinfo{title}{{Identification of gaseous SO$_{2}$ and new upper limits for other gases on Io}}.
\newblock \emph{\bibinfo{journal}{Nature}} \textbf{\bibinfo{volume}{280}}, \bibinfo{pages}{755--758} (\bibinfo{year}{1979}).

\bibitem{Wordsworth2015}
\bibinfo{author}{{Wordsworth}, R.}
\newblock \bibinfo{title}{{Atmospheric Heat Redistribution and Collapse on Tidally Locked Rocky Planets}}.
\newblock \emph{\bibinfo{journal}{Astrophys. J.}} \textbf{\bibinfo{volume}{806}}, \bibinfo{pages}{180} (\bibinfo{year}{2015}).

\bibitem{Paragas2025_JWST7953}
\bibinfo{author}{{Paragas}, K.} \emph{et~al.}
\newblock \bibinfo{title}{{Exo-Geology: Surface Spectral Features from a Rocky Exoplanet}}.
\newblock \bibinfo{howpublished}{JWST Proposal. Cycle 4, ID. \#7953} (\bibinfo{year}{2025}).

\bibitem{Zieba2023_JWST4008}
\bibinfo{author}{{Zieba}, S.} \emph{et~al.}
\newblock \bibinfo{title}{{The search for regolith on the airless exoplanet LHS 3844 b}}.
\newblock \bibinfo{howpublished}{JWST Proposal. Cycle 2, ID. \#4008} (\bibinfo{year}{2023}).

\bibitem{Ricker2015}
\bibinfo{author}{{Ricker}, G.~R.} \emph{et~al.}
\newblock \bibinfo{title}{{Transiting Exoplanet Survey Satellite (TESS)}}.
\newblock \emph{\bibinfo{journal}{J.~Astron.~Telesc.~Instrum.~Syst.}} \textbf{\bibinfo{volume}{1}}, \bibinfo{pages}{014003} (\bibinfo{year}{2015}).

\bibitem{Gardner2006}
\bibinfo{author}{{Gardner}, J.~P.} \emph{et~al.}
\newblock \bibinfo{title}{{The James Webb Space Telescope}}.
\newblock \emph{\bibinfo{journal}{Space Sci. Rev.}} \textbf{\bibinfo{volume}{123}}, \bibinfo{pages}{485--606} (\bibinfo{year}{2006}).

\bibitem{Werner2004}
\bibinfo{author}{{Werner}, M.~W.} \emph{et~al.}
\newblock \bibinfo{title}{{The Spitzer Space Telescope Mission}}.
\newblock \emph{\bibinfo{journal}{Astrophys. J. Suppl.}} \textbf{\bibinfo{volume}{154}}, \bibinfo{pages}{1--9} (\bibinfo{year}{2004}).

\bibitem{Jenkins2016}
\bibinfo{author}{{Jenkins}, J.~M.} \emph{et~al.}
\newblock \bibinfo{editor}{{Chiozzi}, G.} \& \bibinfo{editor}{{Guzman}, J.~C.} (eds) \emph{\bibinfo{title}{{The TESS science processing operations center}}}.
\newblock (eds \bibinfo{editor}{{Chiozzi}, G.} \& \bibinfo{editor}{{Guzman}, J.~C.}) \emph{\bibinfo{booktitle}{Software and Cyberinfrastructure for Astronomy IV}}, Vol. \bibinfo{volume}{9913} of \emph{\bibinfo{series}{Society of Photo-Optical Instrumentation Engineers (SPIE) Conference Series}}, \bibinfo{pages}{99133E} (\bibinfo{year}{2016}).

\bibitem{Lightkurve2018}
\bibinfo{author}{{Lightkurve Collaboration}} \emph{et~al.}
\newblock \bibinfo{title}{{Lightkurve: Kepler and TESS time series analysis in Python}}.
\newblock \bibinfo{howpublished}{Astrophysics Source Code Library} (\bibinfo{year}{2018}).
\newblock \bibinfo{eprint}{{\href{https://arxiv.org/abs/1812.013}{{ascl:1812.013}}}}.

\bibitem{Smith2012}
\bibinfo{author}{{Smith}, J.~C.} \emph{et~al.}
\newblock \bibinfo{title}{{Kepler Presearch Data Conditioning II - A Bayesian Approach to Systematic Error Correction}}.
\newblock \emph{\bibinfo{journal}{Publ.~Astron.~Soc.~Pac.}} \textbf{\bibinfo{volume}{124}}, \bibinfo{pages}{1000} (\bibinfo{year}{2012}).

\bibitem{Stumpe2012}
\bibinfo{author}{{Stumpe}, M.~C.} \emph{et~al.}
\newblock \bibinfo{title}{{Kepler Presearch Data Conditioning I{\textemdash}Architecture and Algorithms for Error Correction in Kepler Light Curves}}.
\newblock \emph{\bibinfo{journal}{Publ.~Astron.~Soc.~Pac.}} \textbf{\bibinfo{volume}{124}}, \bibinfo{pages}{985} (\bibinfo{year}{2012}).

\bibitem{Stumpe2014}
\bibinfo{author}{{Stumpe}, M.~C.} \emph{et~al.}
\newblock \bibinfo{title}{{Multiscale Systematic Error Correction via Wavelet-Based Bandsplitting in Kepler Data}}.
\newblock \emph{\bibinfo{journal}{Publ.~Astron.~Soc.~Pac.}} \textbf{\bibinfo{volume}{126}}, \bibinfo{pages}{100} (\bibinfo{year}{2014}).

\bibitem{Kreidberg2021_go1846}
\bibinfo{author}{{Kreidberg}, L.} \emph{et~al.}
\newblock \bibinfo{title}{{A Search for Signatures of Volcanism and Geodynamics on the Hot Rocky Exoplanet LHS 3844b}}.
\newblock \bibinfo{howpublished}{JWST Proposal. Cycle 1, ID. \#1846} (\bibinfo{year}{2021}).

\bibitem{Kendrew2015}
\bibinfo{author}{{Kendrew}, S.} \emph{et~al.}
\newblock \bibinfo{title}{{The Mid-Infrared Instrument for the James Webb Space Telescope, IV: The Low-Resolution Spectrometer}}.
\newblock \emph{\bibinfo{journal}{Publ.~Astron.~Soc.~Pac.}} \textbf{\bibinfo{volume}{127}}, \bibinfo{pages}{623} (\bibinfo{year}{2015}).

\bibitem{Vanderburg2019}
\bibinfo{author}{{Vanderburg}, A.} \emph{et~al.}
\newblock \bibinfo{title}{{TESS Spots a Compact System of Super-Earths around the Naked-eye Star HR 858}}.
\newblock \emph{\bibinfo{journal}{Astrophys.~J.~Lett.}} \textbf{\bibinfo{volume}{881}}, \bibinfo{pages}{L19} (\bibinfo{year}{2019}).

\bibitem{Horne1986}
\bibinfo{author}{{Horne}, K.}
\newblock \bibinfo{title}{{An optimal extraction algorithm for CCD spectroscopy.}}
\newblock \emph{\bibinfo{journal}{Publ.~Astron.~Soc.~Pac.}} \textbf{\bibinfo{volume}{98}}, \bibinfo{pages}{609--617} (\bibinfo{year}{1986}).

\bibitem{Kempton2023}
\bibinfo{author}{{Kempton}, E. M.~R.} \emph{et~al.}
\newblock \bibinfo{title}{{A reflective, metal-rich atmosphere for GJ 1214b from its JWST phase curve}}.
\newblock \emph{\bibinfo{journal}{Nature}} \textbf{\bibinfo{volume}{620}}, \bibinfo{pages}{67--71} (\bibinfo{year}{2023}).

\bibitem{Hu2024}
\bibinfo{author}{{Hu}, R.} \emph{et~al.}
\newblock \bibinfo{title}{{A secondary atmosphere on the rocky exoplanet 55 Cancri e}}.
\newblock \emph{\bibinfo{journal}{Nature}} \textbf{\bibinfo{volume}{630}}, \bibinfo{pages}{609--612} (\bibinfo{year}{2024}).

\bibitem{Bell2023}
\bibinfo{author}{{Bell}, T.~J.} \emph{et~al.}
\newblock \bibinfo{title}{{A First Look at the JWST MIRI/LRS Phase Curve of WASP-43b}}.
\newblock \emph{\bibinfo{journal}{arXiv e-prints}} \bibinfo{pages}{arXiv:2301.06350} (\bibinfo{year}{2023}).

\bibitem{Bell2024}
\bibinfo{author}{{Bell}, T.~J.} \emph{et~al.}
\newblock \bibinfo{title}{{Nightside clouds and disequilibrium chemistry on the hot Jupiter WASP-43b}}.
\newblock \emph{\bibinfo{journal}{Nat.~Astron.}} \textbf{\bibinfo{volume}{8}}, \bibinfo{pages}{879--898} (\bibinfo{year}{2024}).

\bibitem{Welbanks2024}
\bibinfo{author}{{Welbanks}, L.} \emph{et~al.}
\newblock \bibinfo{title}{{A high internal heat flux and large core in a warm Neptune exoplanet}}.
\newblock \emph{\bibinfo{journal}{Nature}} \textbf{\bibinfo{volume}{630}}, \bibinfo{pages}{836--840} (\bibinfo{year}{2024}).

\bibitem{Skilling2004}
\bibinfo{author}{Skilling, J.}
\newblock \bibinfo{title}{Nested sampling}.
\newblock \emph{\bibinfo{journal}{AIP Conference Proceedings}} \textbf{\bibinfo{volume}{735}}, \bibinfo{pages}{395--405} (\bibinfo{year}{2004}).
\newblock \urlprefix\url{https://doi.org/10.1063/1.1835238}.

\bibitem{Skilling2006}
\bibinfo{author}{Skilling, J.}
\newblock \bibinfo{title}{{Nested sampling for general Bayesian computation}}.
\newblock \emph{\bibinfo{journal}{Bayesian~Anal.}} \textbf{\bibinfo{volume}{1}}, \bibinfo{pages}{833 -- 859} (\bibinfo{year}{2006}).
\newblock \urlprefix\url{https://doi.org/10.1214/06-BA127}.

\bibitem{Higson2019}
\bibinfo{author}{{Higson}, E.}, \bibinfo{author}{{Handley}, W.}, \bibinfo{author}{{Hobson}, M.} \& \bibinfo{author}{{Lasenby}, A.}
\newblock \bibinfo{title}{{Dynamic nested sampling: an improved algorithm for parameter estimation and evidence calculation}}.
\newblock \emph{\bibinfo{journal}{Stat.~Comput.}} \textbf{\bibinfo{volume}{29}}, \bibinfo{pages}{891--913} (\bibinfo{year}{2019}).

\bibitem{Speagle2020}
\bibinfo{author}{{Speagle}, J.~S.}
\newblock \bibinfo{title}{{DYNESTY: a dynamic nested sampling package for estimating Bayesian posteriors and evidences}}.
\newblock \emph{\bibinfo{journal}{Mon.~Not.~R.~Astron.~Soc.}} \textbf{\bibinfo{volume}{493}}, \bibinfo{pages}{3132--3158} (\bibinfo{year}{2020}).

\bibitem{Koposov2024}
\bibinfo{author}{{Koposov}, S.} \emph{et~al.}
\newblock \bibinfo{title}{{joshspeagle/dynesty: v2.1.4}} (\bibinfo{year}{2024}).

\bibitem{Kipping2013}
\bibinfo{author}{{Kipping}, D.~M.}
\newblock \bibinfo{title}{{Efficient, uninformative sampling of limb darkening coefficients for two-parameter laws}}.
\newblock \emph{\bibinfo{journal}{Mon.~Not.~R.~Astron.~Soc.}} \textbf{\bibinfo{volume}{435}}, \bibinfo{pages}{2152--2160} (\bibinfo{year}{2013}).

\bibitem{Kreidberg2015}
\bibinfo{author}{{Kreidberg}, L.}
\newblock \bibinfo{title}{{batman: BAsic Transit Model cAlculatioN in Python}}.
\newblock \emph{\bibinfo{journal}{Publ.~Astron.~Soc.~Pac.}} \textbf{\bibinfo{volume}{127}}, \bibinfo{pages}{1161} (\bibinfo{year}{2015}).

\bibitem{Bouwman2023}
\bibinfo{author}{{Bouwman}, J.} \emph{et~al.}
\newblock \bibinfo{title}{{Spectroscopic Time Series Performance of the Mid-infrared Instrument on the JWST}}.
\newblock \emph{\bibinfo{journal}{Publ.~Astron.~Soc.~Pac.}} \textbf{\bibinfo{volume}{135}}, \bibinfo{pages}{038002} (\bibinfo{year}{2023}).

\bibitem{Schwarz1978}
\bibinfo{author}{{Schwarz}, G.}
\newblock \bibinfo{title}{{Estimating the Dimension of a Model}}.
\newblock \emph{\bibinfo{journal}{Ann.~Stat.}} \textbf{\bibinfo{volume}{6}}, \bibinfo{pages}{461--464} (\bibinfo{year}{1978}).

\bibitem{Kass1995}
\bibinfo{author}{Kass, R.~E.} \& \bibinfo{author}{Raftery, A.~E.}
\newblock \bibinfo{title}{Bayes factors}.
\newblock \emph{\bibinfo{journal}{J.~Am.~Stat.~Assoc.}} \textbf{\bibinfo{volume}{90}}, \bibinfo{pages}{773--795} (\bibinfo{year}{1995}).

\bibitem{Raftery1995}
\bibinfo{author}{Raftery, A.~E.}
\newblock \bibinfo{title}{Bayesian model selection in social research}.
\newblock \emph{\bibinfo{journal}{Sociol.~Methodol.}} \textbf{\bibinfo{volume}{25}}, \bibinfo{pages}{111--163} (\bibinfo{year}{1995}).
\newblock \urlprefix\url{http://www.jstor.org/stable/271063}.

\bibitem{Pont2006}
\bibinfo{author}{{Pont}, F.}, \bibinfo{author}{{Zucker}, S.} \& \bibinfo{author}{{Queloz}, D.}
\newblock \bibinfo{title}{{The effect of red noise on planetary transit detection}}.
\newblock \emph{\bibinfo{journal}{Mon.~Not.~R.~Astron.~Soc.}} \textbf{\bibinfo{volume}{373}}, \bibinfo{pages}{231--242} (\bibinfo{year}{2006}).

\bibitem{Kipping2025}
\bibinfo{author}{{Kipping}, D.}
\newblock \bibinfo{title}{{Exoplaneteers Keep Calling Plots ``Allan Variance'' Plots When They Aren't}}.
\newblock \emph{\bibinfo{journal}{arXiv e-prints}} \bibinfo{pages}{arXiv:2504.13238} (\bibinfo{year}{2025}).

\bibitem{Kirk2024}
\bibinfo{author}{{Kirk}, J.} \emph{et~al.}
\newblock \bibinfo{title}{{JWST/NIRCam Transmission Spectroscopy of the Nearby Sub-Earth GJ 341b}}.
\newblock \emph{\bibinfo{journal}{Astron. J.}} \textbf{\bibinfo{volume}{167}}, \bibinfo{pages}{90} (\bibinfo{year}{2024}).

\bibitem{Wallack2024}
\bibinfo{author}{{Wallack}, N.~L.} \emph{et~al.}
\newblock \bibinfo{title}{{JWST COMPASS: A NIRSpec/G395H Transmission Spectrum of the Sub-Neptune TOI-836c}}.
\newblock \emph{\bibinfo{journal}{Astron. J.}} \textbf{\bibinfo{volume}{168}}, \bibinfo{pages}{77} (\bibinfo{year}{2024}).

\bibitem{Bell2019}
\bibinfo{author}{{Bell}, T.~J.} \emph{et~al.}
\newblock \bibinfo{title}{{Mass loss from the exoplanet WASP-12b inferred from Spitzer phase curves}}.
\newblock \emph{\bibinfo{journal}{Mon.~Not.~R.~Astron.~Soc.}} \textbf{\bibinfo{volume}{489}}, \bibinfo{pages}{1995--2013} (\bibinfo{year}{2019}).

\bibitem{Batalha2017_pandexo}
\bibinfo{author}{{Batalha}, N.~E.} \emph{et~al.}
\newblock \bibinfo{title}{{PandExo: A Community Tool for Transiting Exoplanet Science with JWST \& HST}}.
\newblock \emph{\bibinfo{journal}{Publ.~Astron.~Soc.~Pac.}} \textbf{\bibinfo{volume}{129}}, \bibinfo{pages}{064501} (\bibinfo{year}{2017}).

\bibitem{Batalha2023_pandexo_v3.0}
\bibinfo{author}{Batalha, N.} \emph{et~al.}
\newblock \bibinfo{title}{natashabatalha/pandexo: Pandexo 3.0} (\bibinfo{year}{2023}).
\newblock \urlprefix\url{https://doi.org/10.5281/zenodo.8377201}.

\bibitem{Lucy1971}
\bibinfo{author}{{Lucy}, L.~B.} \& \bibinfo{author}{{Sweeney}, M.~A.}
\newblock \bibinfo{title}{{Spectroscopic binaries with circular orbits.}}
\newblock \emph{\bibinfo{journal}{Astron. J.}} \textbf{\bibinfo{volume}{76}}, \bibinfo{pages}{544--556} (\bibinfo{year}{1971}).

\bibitem{Eastman2013}
\bibinfo{author}{{Eastman}, J.}, \bibinfo{author}{{Gaudi}, B.~S.} \& \bibinfo{author}{{Agol}, E.}
\newblock \bibinfo{title}{{EXOFAST: A Fast Exoplanetary Fitting Suite in IDL}}.
\newblock \emph{\bibinfo{journal}{Publ.~Astron.~Soc.~Pac.}} \textbf{\bibinfo{volume}{125}}, \bibinfo{pages}{83} (\bibinfo{year}{2013}).

\bibitem{Eastman2019}
\bibinfo{author}{{Eastman}, J.~D.} \emph{et~al.}
\newblock \bibinfo{title}{{EXOFASTv2: A public, generalized, publication-quality exoplanet modeling code}}.
\newblock \emph{\bibinfo{journal}{arXiv e-prints}} \bibinfo{pages}{arXiv:1907.09480} (\bibinfo{year}{2019}).

\bibitem{Ducrot2024}
\bibinfo{author}{{Ducrot}, E.} \emph{et~al.}
\newblock \bibinfo{title}{{Combined analysis of the 12.8 and 15 {\ensuremath{\mu}}m JWST/MIRI eclipse observations of TRAPPIST-1 b}}.
\newblock \emph{\bibinfo{journal}{Nat.~Astron.}} \textbf{\bibinfo{volume}{9}}, \bibinfo{pages}{358--369} (\bibinfo{year}{2025}).

\bibitem{BailerJones2021a}
\bibinfo{author}{{Bailer-Jones}, C.~A.~L.}, \bibinfo{author}{{Rybizki}, J.}, \bibinfo{author}{{Fouesneau}, M.}, \bibinfo{author}{{Demleitner}, M.} \& \bibinfo{author}{{Andrae}, R.}
\newblock \bibinfo{title}{{Estimating Distances from Parallaxes. V. Geometric and Photogeometric Distances to 1.47 Billion Stars in Gaia Early Data Release 3}}.
\newblock \emph{\bibinfo{journal}{Astron. J.}} \textbf{\bibinfo{volume}{161}}, \bibinfo{pages}{147} (\bibinfo{year}{2021}).

\bibitem{BailerJones2021b}
\bibinfo{author}{{Bailer-Jones}, C.~A.~L.}, \bibinfo{author}{{Rybizki}, J.}, \bibinfo{author}{{Fouesneau}, M.}, \bibinfo{author}{{Demleitner}, M.} \& \bibinfo{author}{{Andrae}, R.}
\newblock \bibinfo{title}{{VizieR Online Data Catalog: Distances to 1.47 billion stars in Gaia EDR3 (Bailer-Jones+, 2021)}}.
\newblock \bibinfo{howpublished}{VizieR On-line Data Catalog: I/352. Originally published in: 2021AJ....161..147B} (\bibinfo{year}{2021}).

\bibitem{Allard2014}
\bibinfo{author}{{Allard}, F.}
\newblock \bibinfo{editor}{{Booth}, M.}, \bibinfo{editor}{{Matthews}, B.~C.} \& \bibinfo{editor}{{Graham}, J.~R.} (eds) \emph{\bibinfo{title}{{The BT-Settl Model Atmospheres for Stars, Brown Dwarfs and Planets}}}.
\newblock (eds \bibinfo{editor}{{Booth}, M.}, \bibinfo{editor}{{Matthews}, B.~C.} \& \bibinfo{editor}{{Graham}, J.~R.}) \emph{\bibinfo{booktitle}{Exploring the Formation and Evolution of Planetary Systems}}, Vol. \bibinfo{volume}{299} of \emph{\bibinfo{series}{IAU Symposium}}, \bibinfo{pages}{271--272} (\bibinfo{year}{2014}).

\bibitem{STScI2013}
\bibinfo{author}{{STScI Development Team}}.
\newblock \bibinfo{title}{{pysynphot: Synthetic photometry software package}}.
\newblock \bibinfo{howpublished}{Astrophysics Source Code Library, record ascl:1303.023} (\bibinfo{year}{2013}).

\bibitem{Kreidberg2018sptz.prop14204K}
\bibinfo{author}{{Kreidberg}, L.} \emph{et~al.}
\newblock \bibinfo{title}{{A Test for the Existence of An Atmosphere on a Terrestrial Exoplanet Orbiting a Small Star}}.
\newblock \bibinfo{howpublished}{Spitzer Proposal ID \#14204} (\bibinfo{year}{2018}).

\bibitem{Fazio2004}
\bibinfo{author}{{Fazio}, G.~G.} \emph{et~al.}
\newblock \bibinfo{title}{{The Infrared Array Camera (IRAC) for the Spitzer Space Telescope}}.
\newblock \emph{\bibinfo{journal}{Astrophys. J. Suppl.}} \textbf{\bibinfo{volume}{154}}, \bibinfo{pages}{10--17} (\bibinfo{year}{2004}).

\bibitem{IRAC2021}
\bibinfo{author}{{IRAC Instrument Team}} \& \bibinfo{author}{{IRAC Instrument Support Team}}.
\newblock \bibinfo{title}{{IRAC Instrument Handbook}}.
\newblock \bibinfo{howpublished}{NASA IPAC DataSet, IRSA486} (\bibinfo{year}{2021}).

\bibitem{Iyer2023}
\bibinfo{author}{{Iyer}, A.~R.}, \bibinfo{author}{{Line}, M.~R.}, \bibinfo{author}{{Muirhead}, P.~S.}, \bibinfo{author}{{Fortney}, J.~J.} \& \bibinfo{author}{{Gharib-Nezhad}, E.}
\newblock \bibinfo{title}{{The SPHINX M-dwarf Spectral Grid. I. Benchmarking New Model Atmospheres to Derive Fundamental M-dwarf Properties}}.
\newblock \emph{\bibinfo{journal}{Astrophys. J.}} \textbf{\bibinfo{volume}{944}}, \bibinfo{pages}{41} (\bibinfo{year}{2023}).

\bibitem{mallama2002}
\bibinfo{author}{{Mallama}, A.}, \bibinfo{author}{{Wang}, D.} \& \bibinfo{author}{{Howard}, R.~A.}
\newblock \bibinfo{title}{{Photometry of Mercury from SOHO/LASCO and Earth. The Phase Function from 2 to 170 deg.}}
\newblock \emph{\bibinfo{journal}{Icarus}} \textbf{\bibinfo{volume}{155}}, \bibinfo{pages}{253--264} (\bibinfo{year}{2002}).

\bibitem{buratti1996}
\bibinfo{author}{Buratti, B.~J.}, \bibinfo{author}{Hillier, J.~K.} \& \bibinfo{author}{Wang, M.}
\newblock \bibinfo{title}{The lunar opposition surge: Observations by clementine}.
\newblock \emph{\bibinfo{journal}{Icarus}} \textbf{\bibinfo{volume}{124}}, \bibinfo{pages}{490--499} (\bibinfo{year}{1996}).

\bibitem{mallama2017spherical}
\bibinfo{author}{{Mallama}, A.}
\newblock \bibinfo{title}{{The Spherical Bolometric Albedo of Planet Mercury}}.
\newblock \emph{\bibinfo{journal}{arXiv e-prints}} \bibinfo{pages}{arXiv:1703.02670} (\bibinfo{year}{2017}).

\bibitem{emery98}
\bibinfo{author}{Emery, J.} \emph{et~al.}
\newblock \bibinfo{title}{Mercury: Thermal modeling and mid-infrared (5--12 $\mu$m) observations}.
\newblock \emph{\bibinfo{journal}{Icarus}} \textbf{\bibinfo{volume}{136}}, \bibinfo{pages}{104--123} (\bibinfo{year}{1998}).

\bibitem{rozitis2011}
\bibinfo{author}{{Rozitis}, B.} \& \bibinfo{author}{{Green}, S.~F.}
\newblock \bibinfo{title}{{Directional characteristics of thermal-infrared beaming from atmosphereless planetary surfaces - a new thermophysical model}}.
\newblock \emph{\bibinfo{journal}{Mon.~Not.~R.~Astron.~Soc.}} \textbf{\bibinfo{volume}{415}}, \bibinfo{pages}{2042--2062} (\bibinfo{year}{2011}).

\bibitem{delbo2015}
\bibinfo{author}{Delbo, M.}, \bibinfo{author}{Mueller, M.}, \bibinfo{author}{Emery, J.~P.}, \bibinfo{author}{Rozitis, B.} \& \bibinfo{author}{Capria, M.~T.}
\newblock \bibinfo{title}{ in \textit{Asteroid thermophysical modeling}} (eds \bibinfo{editor}{Michel, P.}, \bibinfo{editor}{Demeo, F.~E.} \& \bibinfo{editor}{Bottke, W.~F.}) \emph{\bibinfo{booktitle}{Asteroids IV}} \bibinfo{pages}{107--128} (\bibinfo{publisher}{The University of Arizona Press}, \bibinfo{address}{Tucson}, \bibinfo{year}{2015}).

\bibitem{davidsson2015}
\bibinfo{author}{{Davidsson}, B. J.~R.} \emph{et~al.}
\newblock \bibinfo{title}{{Interpretation of thermal emission. I. The effect of roughness for spatially resolved atmosphereless bodies}}.
\newblock \emph{\bibinfo{journal}{Icarus}} \textbf{\bibinfo{volume}{252}}, \bibinfo{pages}{1--21} (\bibinfo{year}{2015}).

\bibitem{wohlfarth2023}
\bibinfo{author}{{Wohlfarth}, K.}, \bibinfo{author}{{W{\"o}hler}, C.}, \bibinfo{author}{{Hiesinger}, H.} \& \bibinfo{author}{{Helbert}, J.}
\newblock \bibinfo{title}{{An advanced thermal roughness model for airless planetary bodies. Implications for global variations of lunar hydration and mineralogical mapping of Mercury with the MERTIS spectrometer}}.
\newblock \emph{\bibinfo{journal}{Astron. \& Astrophys.}} \textbf{\bibinfo{volume}{674}}, \bibinfo{pages}{A69} (\bibinfo{year}{2023}).

\bibitem{hapke2012}
\bibinfo{author}{Hapke, B.}
\newblock \emph{\bibinfo{title}{Theory of Reflectance and Emittance Spectroscopy}} \bibinfo{edition}{Second} edn (\bibinfo{publisher}{{Cambridge University Press}}, \bibinfo{address}{{Cambridge}}, \bibinfo{year}{2012}).

\bibitem{denevi2023}
\bibinfo{author}{Denevi, B.~W.} \emph{et~al.}
\newblock \emph{\bibinfo{title}{14 Space Weathering At The Moon}}, \bibinfo{pages}{611--650} (\bibinfo{publisher}{De Gruyter}, \bibinfo{address}{Berlin, Boston}, \bibinfo{year}{2023}).
\newblock \urlprefix\url{https://doi.org/10.1515/9781501519895-017}.

\bibitem{domingue2014}
\bibinfo{author}{Domingue, D.~L.} \emph{et~al.}
\newblock \bibinfo{title}{Mercury's {{Weather-Beaten Surface}}: {{Understanding Mercury}} in the {{Context}} of {{Lunar}} and {{Asteroidal Space Weathering Studies}}}.
\newblock \emph{\bibinfo{journal}{Space Sci. Rev.}} \textbf{\bibinfo{volume}{181}}, \bibinfo{pages}{121--214} (\bibinfo{year}{2014}).

\bibitem{yada2022}
\bibinfo{author}{{Yada}, T.} \emph{et~al.}
\newblock \bibinfo{title}{{Preliminary analysis of the Hayabusa2 samples returned from C-type asteroid Ryugu}}.
\newblock \emph{\bibinfo{journal}{Nat.~Astron.}} \textbf{\bibinfo{volume}{6}}, \bibinfo{pages}{214--220} (\bibinfo{year}{2021}).

\bibitem{syal2015}
\bibinfo{author}{{Syal}, M.~B.}, \bibinfo{author}{{Schultz}, P.~H.} \& \bibinfo{author}{{Riner}, M.~A.}
\newblock \bibinfo{title}{{Darkening of Mercury's surface by cometary carbon}}.
\newblock \emph{\bibinfo{journal}{Nat.~Geosci.}} \textbf{\bibinfo{volume}{8}}, \bibinfo{pages}{352--356} (\bibinfo{year}{2015}).

\bibitem{peplowski2016}
\bibinfo{author}{Peplowski, P.~N.} \emph{et~al.}
\newblock \bibinfo{title}{Remote sensing evidence for an ancient carbon-bearing crust on mercury}.
\newblock \emph{\bibinfo{journal}{Nat.~Geosci.}} \textbf{\bibinfo{volume}{9}}, \bibinfo{pages}{273--276} (\bibinfo{year}{2016}).

\bibitem{Zhang2019}
\bibinfo{author}{{Zhang}, M.}, \bibinfo{author}{{Chachan}, Y.}, \bibinfo{author}{{Kempton}, E. M.~R.} \& \bibinfo{author}{{Knutson}, H.~A.}
\newblock \bibinfo{title}{{Forward Modeling and Retrievals with PLATON, a Fast Open-source Tool}}.
\newblock \emph{\bibinfo{journal}{Publ.~Astron.~Soc.~Pac.}} \textbf{\bibinfo{volume}{131}}, \bibinfo{pages}{034501} (\bibinfo{year}{2019}).

\bibitem{Zhang2020}
\bibinfo{author}{{Zhang}, M.}, \bibinfo{author}{{Chachan}, Y.}, \bibinfo{author}{{Kempton}, E. M.~R.}, \bibinfo{author}{{Knutson}, H.~A.} \& \bibinfo{author}{{Chang}, W.~H.}
\newblock \bibinfo{title}{{PLATON II: New Capabilities and a Comprehensive Retrieval on HD 189733b Transit and Eclipse Data}}.
\newblock \emph{\bibinfo{journal}{Astrophys. J.}} \textbf{\bibinfo{volume}{899}}, \bibinfo{pages}{27} (\bibinfo{year}{2020}).

\bibitem{Zhang2025}
\bibinfo{author}{{Zhang}, M.} \emph{et~al.}
\newblock \bibinfo{title}{{Retrievals on NIRCam Transmission and Emission Spectra of HD 189733b with PLATON 6, a GPU Code for the JWST Era}}.
\newblock \emph{\bibinfo{journal}{Astron. J.}} \textbf{\bibinfo{volume}{169}}, \bibinfo{pages}{38} (\bibinfo{year}{2025}).

\bibitem{sprague1992}
\bibinfo{author}{Sprague, A.} \emph{et~al.}
\newblock \bibinfo{title}{The moon: Mid-infrared (7.5- to 11.4-$\mu$m) spectroscopy of selected regions}.
\newblock \emph{\bibinfo{journal}{Icarus}} \textbf{\bibinfo{volume}{100}}, \bibinfo{pages}{73--84} (\bibinfo{year}{1992}).
\newblock \urlprefix\url{https://www.sciencedirect.com/science/article/pii/0019103592900194}.

\bibitem{ruff2022}
\bibinfo{author}{{Ruff}, S.~W.}, \bibinfo{author}{{Hamilton}, V.~E.}, \bibinfo{author}{{Rogers}, A.~D.}, \bibinfo{author}{{Edwards}, C.~S.} \& \bibinfo{author}{{Horgan}, B. H.~N.}
\newblock \bibinfo{title}{{Olivine and carbonate-rich bedrock in Gusev crater and the Nili Fossae region of Mars may be altered ignimbrite deposits}}.
\newblock \emph{\bibinfo{journal}{Icarus}} \textbf{\bibinfo{volume}{380}}, \bibinfo{pages}{114974} (\bibinfo{year}{2022}).

\bibitem{tian2009}
\bibinfo{author}{{Tian}, F.}
\newblock \bibinfo{title}{{Thermal Escape from Super Earth Atmospheres in the Habitable Zones of M Stars}}.
\newblock \emph{\bibinfo{journal}{Astrophys. J.}} \textbf{\bibinfo{volume}{703}}, \bibinfo{pages}{905--909} (\bibinfo{year}{2009}).

\bibitem{heng2023}
\bibinfo{author}{{Heng}, K.}
\newblock \bibinfo{title}{{The Transient Outgassed Atmosphere of 55 Cancri e}}.
\newblock \emph{\bibinfo{journal}{Astrophys.~J.~Lett.}} \textbf{\bibinfo{volume}{956}}, \bibinfo{pages}{L20} (\bibinfo{year}{2023}).

\bibitem{seligman2024}
\bibinfo{author}{{Seligman}, D.~Z.} \emph{et~al.}
\newblock \bibinfo{title}{{Potential Melting of Extrasolar Planets by Tidal Dissipation}}.
\newblock \emph{\bibinfo{journal}{Astrophys. J.}} \textbf{\bibinfo{volume}{961}}, \bibinfo{pages}{22} (\bibinfo{year}{2024}).

\bibitem{Salisbury1989}
\bibinfo{author}{Salisbury, J.~W.} \& \bibinfo{author}{Walter, L.~S.}
\newblock \bibinfo{title}{Thermal infrared (2.5–13.5 µm) spectroscopic remote sensing of igneous rock types on particulate planetary surfaces}.
\newblock \emph{\bibinfo{journal}{J.~Geophys.~Res.~Solid~Earth}} \textbf{\bibinfo{volume}{94}}, \bibinfo{pages}{9192--9202} (\bibinfo{year}{1989}).

\bibitem{Walter1989}
\bibinfo{author}{Walter, L.~S.} \& \bibinfo{author}{Salisbury, J.~W.}
\newblock \bibinfo{title}{Spectral characterization of igneous rocks in the 8- to 12-µm region}.
\newblock \emph{\bibinfo{journal}{J.~Geophys.~Res.~Solid~Earth}} \textbf{\bibinfo{volume}{94}}, \bibinfo{pages}{9203--9213} (\bibinfo{year}{1989}).

\bibitem{bvtp1981}
\bibinfo{author}{{Basaltic Volcanism Study Project}}.
\newblock \emph{\bibinfo{title}{{Basaltic Volcanism on the Terrestrial Planets}}}  (\bibinfo{year}{1981}).

\bibitem{pc2009}
\bibinfo{author}{{Taylor}, S.~R.} \& \bibinfo{author}{{McLennan}, S.}
\newblock \emph{\bibinfo{title}{{Planetary Crusts: Their Composition, Origin and Evolution}}}  (\bibinfo{year}{2009}).

\bibitem{adibekyan2015}
\bibinfo{author}{{Adibekyan}, V.} \emph{et~al.}
\newblock \bibinfo{title}{{From stellar to planetary composition: Galactic chemical evolution of Mg/Si mineralogical ratio}}.
\newblock \emph{\bibinfo{journal}{Astron. \& Astrophys.}} \textbf{\bibinfo{volume}{581}}, \bibinfo{pages}{L2} (\bibinfo{year}{2015}).

\bibitem{spaargaren2020}
\bibinfo{author}{{Spaargaren}, R.~J.}, \bibinfo{author}{{Ballmer}, M.~D.}, \bibinfo{author}{{Bower}, D.~J.}, \bibinfo{author}{{Dorn}, C.} \& \bibinfo{author}{{Tackley}, P.~J.}
\newblock \bibinfo{title}{{The influence of bulk composition on the long-term interior-atmosphere evolution of terrestrial exoplanets}}.
\newblock \emph{\bibinfo{journal}{Astron. \& Astrophys.}} \textbf{\bibinfo{volume}{643}}, \bibinfo{pages}{A44} (\bibinfo{year}{2020}).

\bibitem{donaldsonHanna2017}
\bibinfo{author}{{Donaldson Hanna}, K.~L.} \emph{et~al.}
\newblock \bibinfo{title}{{Effects of varying environmental conditions on emissivity spectra of bulk lunar soils: Application to Diviner thermal infrared observations of the Moon}}.
\newblock \emph{\bibinfo{journal}{Icarus}} \textbf{\bibinfo{volume}{283}}, \bibinfo{pages}{326--342} (\bibinfo{year}{2017}).

\bibitem{Ferrari2020}
\bibinfo{author}{{Ferrari}, S.} \emph{et~al.}
\newblock \bibinfo{title}{{Thermal infrared emissivity of felsic-rich to mafic-rich analogues of hot planetary regoliths}}.
\newblock \emph{\bibinfo{journal}{Earth~Planet.~Sci.~Lett.}} \textbf{\bibinfo{volume}{534}}, \bibinfo{pages}{116089} (\bibinfo{year}{2020}).

\bibitem{Prem2022}
\bibinfo{author}{{Prem}, P.}, \bibinfo{author}{{Greenhagen}, B.~T.}, \bibinfo{author}{{Donaldson Hanna}, K.~L.}, \bibinfo{author}{{Shirley}, K.~A.} \& \bibinfo{author}{{Glotch}, T.~D.}
\newblock \bibinfo{title}{{Modeling Thermal Emission under Lunar Surface Environmental Conditions}}.
\newblock \emph{\bibinfo{journal}{Planet.~Sci.~J.}} \textbf{\bibinfo{volume}{3}}, \bibinfo{pages}{180} (\bibinfo{year}{2022}).

\bibitem{keppler2019}
\bibinfo{author}{Keppler, H.} \& \bibinfo{author}{Golabek, G.}
\newblock \bibinfo{title}{Graphite floatation on a magma ocean and the fate of carbon during core formation}.
\newblock \emph{\bibinfo{journal}{Geochem.~Perspect.~Lett.}} \textbf{\bibinfo{volume}{11}}, \bibinfo{pages}{12--17} (\bibinfo{year}{2019}).

\bibitem{young2003}
\bibinfo{author}{{Young}, D.~A.}
\newblock \emph{\bibinfo{title}{{Mind over Magma. The Story of Igneous Petrology}}}  (\bibinfo{publisher}{Princeton University Press}, \bibinfo{year}{2003}).

\bibitem{wilson1989}
\bibinfo{editor}{Wilson, M.} (ed.) \emph{\bibinfo{title}{Igneous Petrogenesis: A Global Tectonic Approach}}  (\bibinfo{publisher}{Springer}, \bibinfo{address}{Dordrecht}, \bibinfo{year}{1989}).

\bibitem{winter2014}
\bibinfo{author}{Winter, J.~D.}
\newblock \emph{\bibinfo{title}{Principles of igneous and metamorphic petrology: Pearson new international edition}} \bibinfo{edition}{2} edn.
\newblock Pearson custom library (\bibinfo{publisher}{Pearson Education}, \bibinfo{address}{London, England}, \bibinfo{year}{2013}).

\bibitem{taylormclennan2009}
\bibinfo{author}{{Taylor}, S.~R.} \& \bibinfo{author}{{McLennan}, S.}
\newblock \emph{\bibinfo{title}{{Planetary Crusts: Their Composition, Origin and Evolution}}}  (\bibinfo{publisher}{Cambridge Catalogue}, \bibinfo{year}{2009}).

\bibitem{Doyle2023}
\bibinfo{author}{{Doyle}, A.~E.} \emph{et~al.}
\newblock \bibinfo{title}{{New Chondritic Bodies Identified in Eight Oxygen-bearing White Dwarfs}}.
\newblock \emph{\bibinfo{journal}{Astrophys. J.}} \textbf{\bibinfo{volume}{950}}, \bibinfo{pages}{93} (\bibinfo{year}{2023}).

\bibitem{numpy2020}
\bibinfo{author}{{Harris}, C.~R.} \emph{et~al.}
\newblock \bibinfo{title}{{Array programming with NumPy}}.
\newblock \emph{\bibinfo{journal}{Nature}} \textbf{\bibinfo{volume}{585}}, \bibinfo{pages}{357--362} (\bibinfo{year}{2020}).

\bibitem{matplotlib2007}
\bibinfo{author}{{Hunter}, J.~D.}
\newblock \bibinfo{title}{{Matplotlib: A 2D Graphics Environment}}.
\newblock \emph{\bibinfo{journal}{Comput.~Sci.~Eng.}} \textbf{\bibinfo{volume}{9}}, \bibinfo{pages}{90--95} (\bibinfo{year}{2007}).

\bibitem{astropy2022}
\bibinfo{author}{{Astropy Collaboration}} \emph{et~al.}
\newblock \bibinfo{title}{{The Astropy Project: Sustaining and Growing a Community-oriented Open-source Project and the Latest Major Release (v5.0) of the Core Package}}.
\newblock \emph{\bibinfo{journal}{Astrophys. J.}} \textbf{\bibinfo{volume}{935}}, \bibinfo{pages}{167} (\bibinfo{year}{2022}).

\bibitem{Bushouse2022}
\bibinfo{author}{{Bushouse}, H.} \emph{et~al.}
\newblock \bibinfo{title}{{JWST Calibration Pipeline}}.
\newblock \bibinfo{howpublished}{Zenodo} (\bibinfo{year}{2022}).
\newblock \urlprefix\url{http://dx.doi.org/10.5281/zenodo.7325378}.

\bibitem{ForemanMackey2013}
\bibinfo{author}{{Foreman-Mackey}, D.}, \bibinfo{author}{{Hogg}, D.~W.}, \bibinfo{author}{{Lang}, D.} \& \bibinfo{author}{{Goodman}, J.}
\newblock \bibinfo{title}{{emcee: The MCMC Hammer}}.
\newblock \emph{\bibinfo{journal}{Publ.~Astron.~Soc.~Pac.}} \textbf{\bibinfo{volume}{125}}, \bibinfo{pages}{306} (\bibinfo{year}{2013}).

\bibitem{Koposov2023}
\bibinfo{author}{{Koposov}, S.} \emph{et~al.}
\newblock \bibinfo{title}{{joshspeagle/dynesty: v2.1.0}}.
\newblock \bibinfo{howpublished}{Zenodo} (\bibinfo{year}{2023}).
\newblock \urlprefix\url{http://dx.doi.org/10.5281/zenodo.7600689}.

\end{thebibliography}

\end{document}